\documentclass[aip,jmp,amsmath,amssymb,reprint,onecolumn]{revtex4-2}
\usepackage{dcolumn}
\usepackage{bm}

\usepackage[T1]{fontenc}
\usepackage{mathptmx}
\usepackage{etoolbox}

\usepackage[english]{babel}

\usepackage[a4paper,margin=30mm]{geometry}
\usepackage{amsmath,amssymb,mathtools,mathrsfs}
\AtBeginDocument{\renewcommand{\hbar}{\hslash}}
\usepackage{graphicx}
\usepackage{microtype}
\usepackage{hyperref}
\hypersetup{hidelinks}

\newcommand{\dd}{\mathrm d}
\newcommand{\ii}{\mathrm i}
\newcommand{\e}{\mathrm e}
\newcommand{\EE}{\mathbb E}

\newcommand{\cR}{\mathcal R}

\newcommand{\cL}{\mathcal L}
\newcommand{\cT}{\mathcal T}
\newcommand{\GamL}{\Gamma_{\mathrm L}}
\newcommand{\ellL}{\ell_{\mathrm L}}
\newcommand{\erfc}{\operatorname{erfc}}

\begin{document}

\setlength{\abovedisplayskip}{5pt}
\setlength{\belowdisplayskip}{5pt}

\title{Reflection Resonances in the One-Dimensional Anderson Localization: Finite-Length Statistics, Wigner Time Delay, and Boundary Eigenfunctions}

\author{Yan V. Fyodorov}
\email{yan.fyodorov@kcl.ac.uk}
\affiliation{ Department of Mathematics, King's College London, Strand, London, WC2R 2LS, United Kingdom}


\begin{abstract}
We study reflection-resonance poles $Z_j=E_j-\ii\Gamma_j$, $\Gamma_j>0$,
of a finite one-dimensional disordered sample of length $L$, coupled at one
end to a semi-infinite lead, in the regime $L\gg\ellL\gg k^{-1}$, where
$\ellL$ is the localization length and $k=\sqrt{E}$. The key step is to
relate the resonance density to the reflection coefficient at the complex
energy $E+\ii\Gamma$, corresponding to uniform absorption. Exact finite-chain
Kac--Rice and Poincar\'e--Lelong identities reduce pole counting to a
finite-length diffusion of the reflected intensity. For the perfect contact transparency $\cT=1$ density crosses over
from the localization-controlled $\Gamma^{-1}$ law to the broad-resonance
$\Gamma^{-2}$ law at $\GamL=k/\ellL$, in agreement with the recent
$L\to\infty$ result of Fyodorov and Meibohm. Finite length cuts off the
$\Gamma^{-1}$ regime at
$\Gamma_{\rm ultra}=\frac{\e^{1-\gamma_{\rm E}}}{2}\GamL\e^{-L/\ellL}$.
We derive the ultranarrow resonances crossover shape as an explicit moving front; for contact
transparency $0<\cT<1$, this scale shifts to $\cT\Gamma_{\rm ultra}$.

The same reflection process yields the finite-length Wigner time-delay
statistics and, in the weak-absorption limit, the Comtet--Texier distribution.
We show that at eigenvalues of the corresponding closed Dirichlet sample the
Wigner delay is inversely proportional to the squared boundary derivative of
the normalized eigenfunction. This quantity also gives the eigenvalue
response to displacement of the Dirichlet boundary at the lead-contact end,
and hence the force exerted by the eigenmode on that boundary; we obtain its
finite-length distribution. Finally, the finite-transparency resonance
density obeys the single-channel Moldauer--Simonius sum rule, linking its
perfect-coupling divergence to the $\Gamma^{-2}$ tail. Direct lattice and
spectral computations test the crossover and front constant.
\end{abstract}

\maketitle

\section{Introduction}
\label{sec:introduction}

\subsection{Reflection resonances and extreme events in a localized sample}

The spectrum of an isolated disordered Hamiltonian is real.  Attaching a lead
turns its discrete eigenvalues into poles of the scattering matrix in the
lower half of the complex-energy plane.  We write a pole as
$Z_j=E_j-\ii\Gamma_j$, with $\Gamma_j>0$.  In our convention the metastable
amplitude contains $\e^{-\ii E_jt-\Gamma_jt}$, so its survival probability
decays as $\e^{-2\Gamma_jt}$.  General background on scattering resonances may
be found in Refs.~\cite{Exner2012,DyatlovZworski2019}. 

\begin{figure}[t]
\centering
\includegraphics[width=\textwidth]{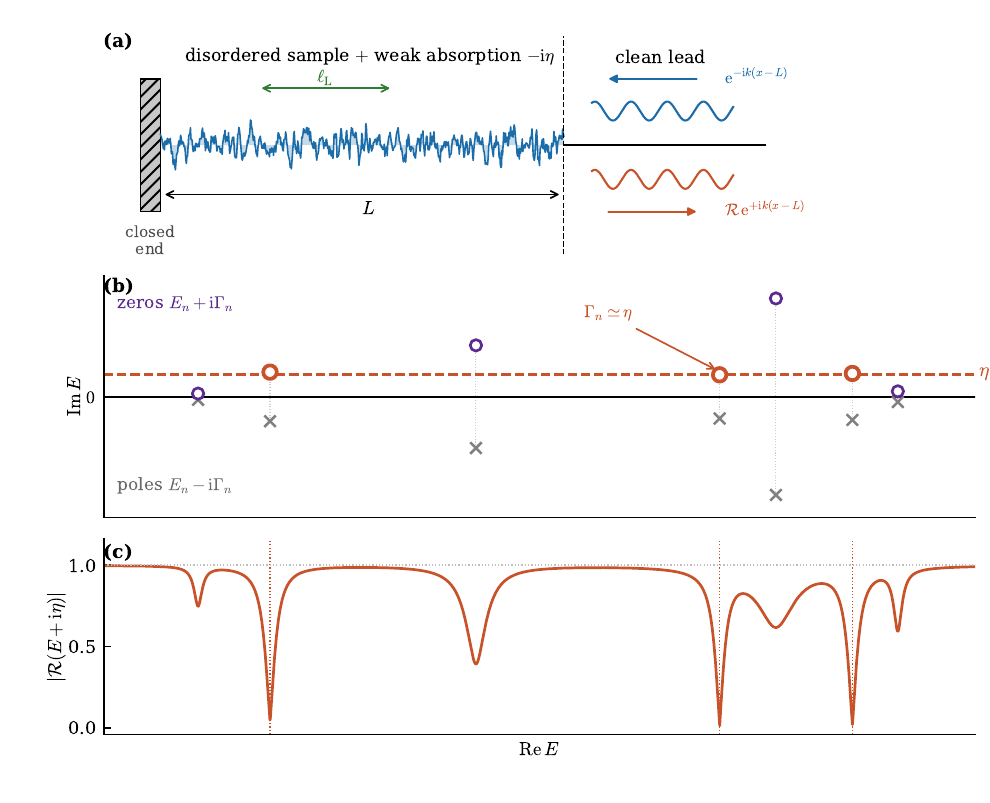}
\caption{Setup, and how a weak absorption exposes the resonances.
(a) Geometry: a disordered sample of length $L$ with a closed (Dirichlet) end
at $x=0$ and a clean lead attached at $x=L$, carrying an incident wave and a
reflected wave of amplitude $\mathcal R$.  The localization length $\ellL$ is
shown for scale; the finite-length results are controlled by $\tau=L/\ellL$.
A uniform weak absorption $-\ii\eta$ inside the sample shifts the energy
argument to $E+\ii\eta$.
(b) Resonance positions $E_n$ are drawn from a Poisson process, appropriate
to the localized regime.  Without absorption $|\mathcal R|=1$ on the real axis,
so the poles
$E_n-\ii\Gamma_n$ (crosses) are accompanied by zeros at the mirror points
$E_n+\ii\Gamma_n$ (circles).  Absorption makes the reflection amplitude
sampled along the horizontal line $\mathrm{Im}\,E=\eta$, which runs through the
\emph{zeros}.  For the narrowest states the zero and its mirror pole are
closer together than the symbol size, and appear to sit on the real axis.
(c) The resulting modulus,
$|\mathcal R(E+\ii\eta)|^2=\prod_n[(E-E_n)^2+(\eta-\Gamma_n)^2]/
[(E-E_n)^2+(\eta+\Gamma_n)^2]$.  It stays close to unity except where a zero
lies near the line: the three resonances with $\Gamma_n\simeq\eta$ (highlighted,
dotted verticals) produce near-total drops, while resonances much broader or
much narrower than $\eta$ give only shallow dips.  Since
$\frac{1}{4\pi}(\partial_E^2+\partial_\eta^2)\log|\mathcal R|^2$ is a sum of
delta functions at the zeros, sweeping $\eta$ and differentiating twice counts
the resonances without locating them; this is the content of
Eq.~\eqref{eq:main-finite-poincare-lelong}.}
\label{fig:setup}
\end{figure}

For a one-dimensional sample attached to a single lead, the scattering matrix
reduces to a complex reflection amplitude $\cR(Z)$, and the resonances are its
poles.  Resonances of
one-dimensional Schr\"odinger operators with deterministic potentials have
been studied extensively; see, for example,
Refs.~\cite{Zworski1987,Froese1997,Simon2000,Korotyaev2004}. 
 Here the potential is random and the aim is a statistical description of the
resonance poles in the Anderson-localized regime.  One possible way, both physically and mathematically,
 to approach this task is to add a small negative imaginary part (absorption) to the potential. With such an absorption 
 the reflection amplitude ceases to be unimodular for real energies, and its energy dependence 
 shows considerable fluctuations induced by counterparts of poles - the complex zeros, see  the figure \ref{fig:setup}.

The experimental relevance of scattering observables in 1D localized
systems has been demonstrated directly in microwave experiments.
Bliokh et al.~\cite{BliokhEtAl2006} showed that localized quasimodes remain
visible in reflection even when their transmission signatures are strongly
suppressed.  Spatially resolved measurements of quasimodes in random
waveguides, including the influence of absorption on their transmission and
reflection spectra, were reported in Ref.~\cite{SebbahHuKoppGenack2007}.
Earlier, Chabanov and Genack measured strongly broadened delay-time statistics
in the localized regime and found pronounced correlations between long delays
and enhanced transmission, consistent with resonant transport through
localized states~\cite{ChabanovGenack2001}.

The small-width part of the resonance distribution is governed by
localization.  In a long finite sample, progressively narrower resonances
are first associated with states localized progressively deeper inside the
sample.  A complementary effective-cavity picture identifies a resonant
segment of size of order the localization length, separated from the opening
by an exponentially opaque portion of the sample
\cite{BliokhBliokhFreilikher2004}.  This gives a more spatially resolved
interpretation of the standard $1/\Gamma$ argument: the resonance width
decreases exponentially with the distance of the localized resonant region
from the opening, while the localization centres are approximately uniformly
distributed in space.  Consequently the density is approximately uniform in
$\log\Gamma$, and hence proportional to $1/\Gamma$.

Once the localization centre reaches the vicinity of the closed end, its
distance from the opening cannot exceed $L$.  Still smaller widths require
atypically strong spatial decay across the whole available sample.  Hence
this ultranarrow part of the width distribution is driven by rare
whole-sample localization fluctuations.  We show that the resulting
finite-length cutoff takes the form of a moving front, equivalently a
first-passage law for the accumulated logarithmic decay of the wave
intensity.

Our exact starting point is a finite disordered lattice model.  The identity
relating the resonance determinant to the outgoing residual,
together with the corresponding zero-counting identities, holds sample by
sample for arbitrary real onsite potentials.  We then pass to the continuum
weak-disorder limit and specialize to centred Gaussian white noise,
$H=-\dd^2/\dd x^2+V(x)$, with
$\EE[V(x)V(x')]=D\delta(x-x')$ on $0\leq x,x'\leq L$.

The theory of one-dimensional localization has a particularly rich analytic
history.  On the physics side, the early works of Halperin
\cite{Halperin1965} and Thouless \cite{Thouless1972}, as well as the
diagrammatic approach initiated by Berezinskii \cite{Berezinskii1974},
should be especially mentioned.  Subsequent work by Abrikosov and Ryzhkin
\cite{AbrikosovRyzhkin1978} and by Mel'nikov \cite{Melnikov1980} led, in
particular, to evolution equations for transport statistics in finite
one-dimensional disordered samples.  A detailed account of this body of work
and related developments can be found in the book
\cite{LifshitsGredeskulPastur}. Later studies developed detailed statistical descriptions of local
observables, such as the distribution of the local density of states by
Altshuler and Prigodin \cite{AltshulerPrigodin1989}, 
wavefunction statistics by Kolokolov \cite{Kolokolov1993}, correlations of local
spectral densities and localized wave functions at nearby energies by Ivanov
et al.~\cite{IvanovSkvortsovOstrovskyFominov2012} and fluctuations
of finite-length Lyapunov exponents and their large-deviation statistics by
Texier and collaborators
\cite{ComtetTexierTourigny2013,Texier2020Fluctuations,Texier2020SPS}.
The invariant-embedding approach of Rammal
and Doucot \cite{Rammal1987} also deserves mention and is methodologically
especially close to the techniques used in the present paper.   A different exact approach was developed
by Schanz and Smilansky \cite{SchanzSmilansky2000}, who represented a
one-dimensional disordered chain as a quantum graph and described Anderson
localization through coherent sums over periodic-orbit families.  Their
formulation in terms of local reflection and transmission amplitudes provides
another scattering-based description of one-dimensional localization.

Our main topic here is the resonance-width distribution.  Resonances of
one-dimensional disordered systems have been studied over the years both
numerically and analytically
\cite{TitovFyodorov2000,TerraneoGuarneri2000,KunzShapiro2006,KunzShapiro2008,
Feinberg2009,Feinberg2011,GurevichShapiro2012,SorathiaEtAl2012,
HerreraGonzalezMendezBermudezIzrailev2019,GaspardSparenberg2022,GaspardSparenberg2024,FyodorovMeibohm2026}, but their
finite-length statistics remain considerably less understood than many other
aspects of one-dimensional localization.  One aim of the present work is to
develop this theory further, in particular in the ultranarrow-width regime
where finite sample length becomes essential.

On the mathematical side, a remarkably early contribution came from
Gertsenshtein and Vasil'ev \cite{GertsenshteinVasiliev1959}. Working
apparently independently of the emerging theory of Anderson localization,
they studied wave propagation in a one-dimensional randomly inhomogeneous
medium and formulated the composition of successive scattering elements as
a stochastic evolution by linear-fractional transformations.  In the
continuum limit this led to a diffusion problem on the Lobachevsky plane.
In modern language, their work may be viewed as an early probabilistic
formulation of one-dimensional random scattering and transfer-matrix
evolution, closely related in spirit to the reflection-based description
used below.  Indeed, the present paper continues this scattering-based line
of development in a rather direct way.

The corresponding mathematically rigorous spectral theory of
one-dimensional random Schr\"odinger operators began with the seminal work
of Goldsheid, Molchanov, and Pastur
\cite{GoldsheidMolchanovPastur1977} and was developed
systematically in, among other sources, the book \cite{PasturFigotin}.  Much
more recently, a series of works by Dumaz and Labb\'e has provided a detailed
rigorous description of several localization and crossover regimes
\cite{DumazLabbe2020Localization,DumazLabbe2023Delocalized,
DumazLabbe2024WhiteNoise,DumazLabbe2024Crossover}.  The Riccati diffusion
underlying many of these works is, after a change of variables, equivalent
to the reflection evolution used here.  For open systems, and in particular
for resonance statistics, the rigorous theory is much less developed than
for closed one-dimensional random operators; see, for example, a work by Klopp~\cite{Klopp2016}. 
 Another aim of the present paper is therefore to present
the resonance problem in a self-contained and well-controlled form at the
level of theoretical physics, with the hope that the resulting structure may
also provide a useful starting point for further rigorous mathematical work.

An important early methodological precursor of the present work is the
invariant-embedding analysis of weakly absorbing one-dimensional media by
Freilikher, Pustilnik, and Yurkevich
\cite{FreilikherPustilnikYurkevich1994}.  They mapped the Fokker--Planck
evolution to an imaginary-time Schr\"odinger problem and showed that weak
absorption creates a distant logarithmic wall and discretizes the spectrum.
Their main observables were transmission moments.  The same paper also
emphasized that the invariant-embedding equation for the reflection amplitude
is closed and therefore determines the reflected-intensity statistics.
In closely related work in the same year, Pradhan and Kumar
\cite{PradhanKumar1994} used invariant embedding to study the statistics of
the reflection coefficient in a one-dimensional disordered medium with
coherent amplification.  Together, these works provide important early
ingredients of the reflection-based approach developed further below.
In particular, the stationary weak-absorption density of the reflection
intensity used below was already known.  What we require in addition is its
nonstationary finite-length propagator.

This propagator also contains the classical single-channel delay-time problem.
At a real energy the reflection amplitude has unit modulus, whereas at
$E+\ii\eta$ weak absorption produces the unitarity deficit
$1-|\cR|^2$.  To first order in $\eta$ this deficit equals
$2\eta\tau_{\rm W}$, where
$\tau_{\rm W}(E)=\partial_E\arg\cR(E)$ is the Wigner delay time.
The connection between weak absorption and reflection time delay statistics was first discussed by Doron, Smilansky and Frenkel in \cite{DoronSmilanskyFrenkel1990} 
and actively used in Refs.~\cite{TitovBeenakker2000,FyodorovPNR2003}.

The finite-length distribution of $\tau_{\rm W}$ for one-dimensional random
potentials was obtained by Comtet and Texier, together with its weak-disorder
universality and finite-size log-normal tail
\cite{ComtetTexier1997,TexierComtet1999}.  Closely related studies of finite
one-dimensional disordered chains addressed the distributions of tunnelling
and delay times \cite{BoltonHeatonEtAl1999}, as well as the statistics of
scattering phases and Wigner delays in the one-channel Anderson model
\cite{OssipovKottosGeisel2000}.  The relation between localization and
reflection or transmission delay-time statistics was further explored in
Ref.~\cite{Schomerus2001}.  In the present variables the process studied in
\cite{ComtetTexier1997,TexierComtet1999} appears directly in the
weak-absorption scaling $1-|\cR|^2=O(\alpha)$ of our reflection-intensity
diffusion. Thus one and the same finite-length process controls both the delay scaling
regime near unit reflection and, at the opposite limit near zero reflection,
the ultranarrow resonance front.

There is a second connection, to eigenfunctions of the closed sample.  A
closely related precursor appeared already in the network formulation of
Chalker and Siak \cite{ChalkerSiak1990}, where in a one-dimensional version
of the model the current carried by a link was related to the energy
derivative of the total scattering phase.  A general relation between
single-channel delay statistics and local closed-system eigenfunction
intensities was derived later by Ossipov and Fyodorov
\cite{OssipovFyodorov2005}.  Their analysis also stressed that attaching the
lead at a boundary may require special care.  In the continuum model used
here the closed sample has a Dirichlet condition at the contact point, so the
boundary value of an eigenfunction vanishes identically.  The relevant local
observable is instead $|\psi_n'(L)|^2$.  We derive directly the exact
one-dimensional identity relating this derivative to the Wigner delay at a
closed-system eigenvalue and then obtain its complete finite-length
distribution.

The boundary derivative also has a simple mechanical meaning.  If the right
Dirichlet endpoint is displaced while the disorder realization is kept fixed,
then
\[
        -\partial_L E_n=|\psi_n'(L)|^2
\]
in the units used here.  Thus the distribution derived below is equivalently
the distribution of the force exerted by a normalized localized eigenmode on
the boundary. 

A complementary global constraint on resonance widths is the
Moldauer--Simonius relation~\cite{Moldauer1967,Simonius1974}.  For broken
time-reversal symmetry  it was obtained in the random-matrix analysis of quantum chaotic scattering by
Fyodorov and Sommers~\cite{FyodorovSommers1996,FyodorovSommers1997}.  Kurilov and Ostrovsky~\cite{KurilovOstrovsky2026}
 have recently given a particularly transparent and general nonlinear-$\sigma$-model derivation
and shown, within that semiclassical framework, that the average-decay-rate
relation is independent of Wigner--Dyson symmetry and sample geometry. Their formulation is however inapplicable for the present case, as  the microscopic system considered here is a strictly one-dimensional scalar Schr\"odinger operator not covered by nonlinear-$\sigma$-model formalism.  We show that the
single-channel Moldauer--Simonius sum rule nevertheless follows directly from
our reflection/Poincar\'e--Lelong formulation in the generic-bulk
weak-disorder limit.  The same derivation makes explicit how its logarithmic
divergence at perfect coupling is tied to the broad
$\rho(\Gamma)\propto\Gamma^{-2}$ tail.

Earlier studies, numerical and analytical, provided useful benchmarks for resonance-pole
statistics in one-dimensional disordered models.  Titov and Fyodorov \cite{TitovFyodorov2000} and Terraneo and Guarneri \cite{TerraneoGuarneri2000} discussed algebraic resonance-width tails in disordered 1D models.   The $1/\Gamma$  tail of resonance width density emerged in analysis by Kunz and Shapiro \cite{KunzShapiro2006,KunzShapiro2008} and was confirmed later by Feinberg \cite{Feinberg2009,Feinberg2011}, 
though general qualitative arguments in its favour were provided before, see e.g. \cite{Kottos2005} for a discussion.
Far log-normal cutoff was anticipated qualitatively by Gurevich and Shapiro
\cite{GurevichShapiro2012}, and a related rare-state mechanism appeared in the
long-time trapping problem of Muzykantskii and Khmelnitskii
\cite{MuzykantskiiKhmelnitskii1997}.  The relation between localization, transport, and coupling to the continuum
in finite one-dimensional Anderson chains has  been studied by Sorathiea et al. within the
effective non-Hermitian Hamiltonian approach
\cite{SorathiaEtAl2012}, while Herrera-Gonz\'alez, M\'endez-Berm\'udez, and Izrailev followed the pole cloud
through the perfect-coupling superradiant crossover
\cite{HerreraGonzalezMendezBermudezIzrailev2019}. Recently a more detailed reflection-based treatment by Fyodorov and Meibohm
\cite{FyodorovMeibohm2026} used a Poincar\'e--Lelong representation for the 
resonance density and, in particular, gave an explicit result for semi-infinite sample which serves as important
benchmark below.  That result however cannot describe the very narrow
resonances created by a finite sample length much longer than the localization
length.  Closing this gap is the principal aim of the present paper.  A
second methodological aim is to derive the finite-chain counting and
Poincar\'e--Lelong identities without the heuristic steps used in the earlier
presentation.  As part of this derivation we also give a self-contained
construction of the complex reflection diffusion.

\begin{figure}[t]
\centering
\includegraphics[width=0.78\textwidth]{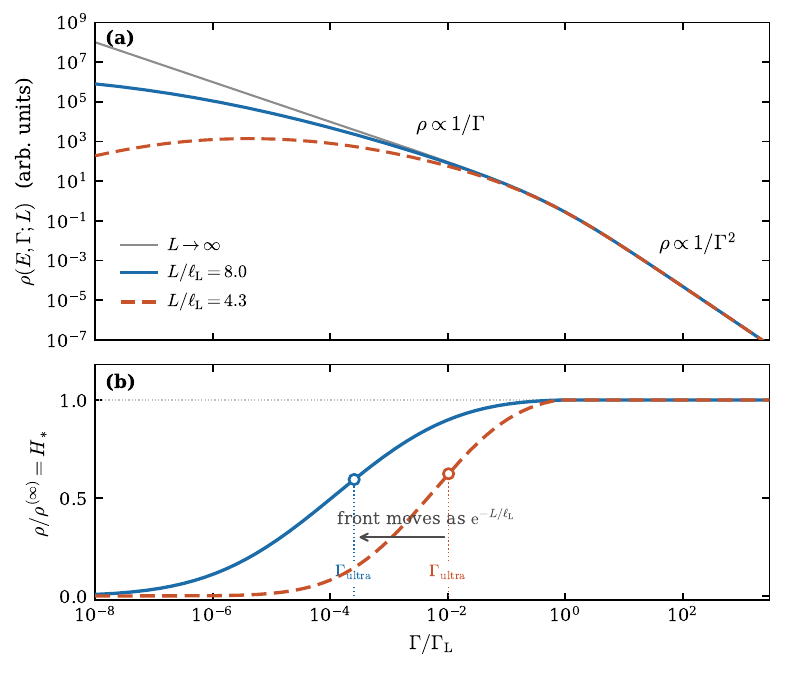}
\caption{Schematic of the finite-length width density, summarizing the results
of Sections~\ref{sec:radial} and~\ref{sec:perfect-ultranarrow}.
(a) The density itself.  Grey: the semi-infinite limit
\eqref{eq:semi_infinite_density_main}, crossing over from $\rho\propto1/\Gamma$
to $\rho\propto1/\Gamma^{2}$ at the escape width $\GamL$.  Coloured curves: two
finite lengths, which follow the same envelope over many decades and peel away
from it only at exponentially small widths.
(b) The same information with the envelope divided out.  The ratio is the
first-passage kernel \eqref{eq:two-image-first-passage-kernel}, which falls
from one to zero across the ultranarrow scale
$\Gamma_{\rm ultra}=\frac{C_*}{2}\GamL\e^{-L/\ellL}$ of
Eq.~\eqref{eq:ultranarrow_scale_main} (open circles).  Increasing $L$ translates
the front rigidly towards exponentially smaller widths, at fixed shape on the
logarithmic axis.  Note that the front is a suppression \emph{relative to} the
$1/\Gamma$ law rather than an absolute drop: at $\Gamma_{\rm ultra}$ the density
is reduced only by a factor $H_*\simeq0.6$, and $\rho$ itself continues to grow
as $\Gamma$ decreases until $\log(1/\Gamma)\simeq3L/\ellL$, which is why the
turnover in (a) lies far below $\Gamma_{\rm ultra}$.
Figure~\ref{fig:density} tests these predictions against direct
diagonalization.}
\label{fig:overview}
\end{figure}

The principal new result of this paper is the fully controlled analytical derivation 
of the long-sample ultranarrow resonance-density
profile, including the order-one front displacement
$C_*=\e^{1-\gamma_{\rm E}}$, the complete two-term transition kernel, and eventually its
log-normal-type deep tail. The figure \ref{fig:overview} visualizes the findings.

 The finite-chain determinant, projective Kac--Rice, and inverse-reflection
identities provide its exact algebraic starting point, while a supersymmetric
partner associated with the radial Fokker--Planck problem makes the
real-spectrum matching considerably simpler.  We also derive all relevant quantity for finite contact
transparency.

\begin{figure}[t]
\centering
\includegraphics[width=\textwidth]{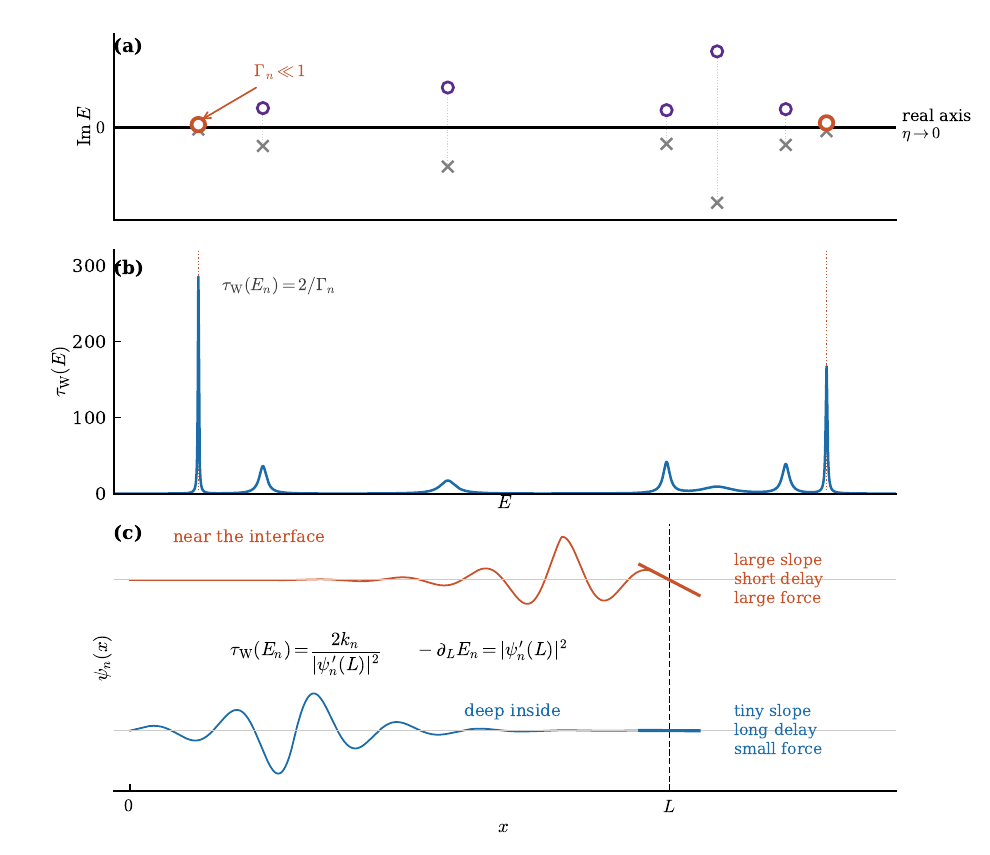}
\caption{Time delays, and their translation into boundary forces.  The
resonance set $\{E_n,\Gamma_n\}$ is the same as in Fig.~\ref{fig:setup}.\\
(a) The same zeros (circles) and poles (crosses); resonance positions
$E_n$ are drawn from a Poisson process, as expected in the localized regime,
so the spacings are strongly irregular.  The Wigner delay is a
property of the unitary problem and is read off on the real axis, i.e.\ in the
limit $\eta\to0$. 
(b) The resulting delay $\tau_{\rm W}(E)=\partial_E\arg\mathcal R
=\sum_n2\Gamma_n/[(E-E_n)^2+\Gamma_n^2]$.  Every resonance produces a peak, at
the same energies at which Fig.~\ref{fig:setup}(c) shows a dip, but the two
observables rank the resonances differently: the peak height is
$\tau_{\rm W}(E_n)=2/\Gamma_n$, so it is the \emph{narrowest} states
(highlighted, dotted verticals) that dominate here, whereas the deepest
reflection dips belonged to those with $\Gamma_n\simeq\eta$.  The two
descriptions are reconciled in the weak-absorption limit, where
$1-|\mathcal R(E+\ii\eta)|^2=2\eta\,\tau_{\rm W}(E)+O(\eta^2)$ and the dip
depth becomes proportional to the delay.
(c)  Two eigenmodes of the closed sample with the same number of nodes but different localization centres: one
sitting near the interface, one deep inside.  Both vanish at $x=L$, but their
slopes there differ by the exponential factor $\e^{-(L-x_0)/\ellL}$.  Through
the two exact identities shown, that single quantity fixes both the delay and
the force exerted on the moving boundary, so a mode localized deep inside the
sample has an exponentially long delay and an exponentially small boundary
force.  Their product is fixed, $\tau_{\rm W}(E_n)\,(-\partial_LE_n)=2k_n$.}
\label{fig:delay}
\end{figure}

The same framework yields several useful by-products.  It gives the
finite-length reflection-intensity distribution at weak absorption and
reproduces the complete Comtet--Texier delay distribution in the
scaling regime of vanishing absorption near unit reflection.  For nonideal
coupling it recovers the standard single-channel transformation of the delay
distribution.  More importantly for the closed problem, it gives the
finite-length distribution of the Dirichlet boundary force
$-\partial_L E_n=|\psi_n'(L)|^2$, a strict-1D boundary counterpart of the
Ossipov--Fyodorov time delay/eigenfunction correspondence, see figure \ref{fig:delay}.
  Finally, integrating
the finite-transparency resonance density gives the single-channel
Moldauer--Simonius sum rule directly in the same 1D weak-disorder regime.
These classical relations  all emerge from the same microscopic reflection process, while the
finite-length ultranarrow front supplies the new extreme-resonance sector.

\subsection{Scope and organization}

Denoting the localization length by $\ellL$, the sample length by $L$, and
the resonance width by $\Gamma$, we work at generic bulk energies in the
weak-disorder regime
$k\ellL\gg1$, $kL\gg1$, and $\Gamma/E\ll1$.  The ultranarrow asymptotics
additionally assume a long finite sample, $L/\ellL\gg1$, and widths near or
below the exponentially small scale $\GamL\e^{-L/\ellL}$.  The exact
finite-chain identities themselves do not require these asymptotic
assumptions.  The generic-phase reduction excludes the lattice band edges and
the usual Kappus--Wegner band-centre anomaly~\cite{KappusWegner1981}; the
sub-wavelength regime $L=O(k^{-1})$ also lies outside the present scope.

The separation between the rapidly rotating phase and the slowly evolving
reflection intensity, and the resulting phase averaging, are standard in
one-dimensional localization theory; see, for example,
Ref.~\cite{LifshitsGredeskulPastur}.  We nevertheless derive this reduction
from the complex reflection diffusion in order to keep the approximation
level and its range of validity explicit.

Section~\ref{sec:main-results} fixes the model and summarizes the principal
results.  Sections~\ref{sec:exact-kr}--\ref{sec:radial} derive the exact
counting identities, continuum reflection diffusion, and phase-averaged
intensity propagator.  Section~\ref{sec:perfect-ultranarrow} derives the
ultranarrow front from the opening real spectrum, while
Section~\ref{sec:finite-reflection} records the associated finite-length
reflection law.  Section~\ref{sec:contact} treats nonideal contacts and derives
the strict-1D Moldauer--Simonius sum rule, and Section~\ref{sec:numerics}
provides numerical checks.  The final main-text
Section~\ref{sec:reflection-delay} develops the finite-length Wigner-delay
problem and the resulting distribution of Dirichlet boundary forces.  The
appendices collect the It\^o details, the exact finite-$\alpha$ spectral
machinery, and technical details of the Wigner-delay spectral representation.

\section{Model and summary of results}
\label{sec:main-results}

This section fixes the notation and collects the principal results.  Exact
finite-chain identities are stated separately from the continuum,
generic-phase and spectral approximations used later.  The main formulas to
keep in view are the exact finite-chain Poincar\'e--Lelong identity
\eqref{eq:main-finite-poincare-lelong}, the finite-length resonance density
\eqref{eq:principal-finite-length-density}, its ultranarrow moving-front form
\eqref{eq:ultranarrow_density_main}--\eqref{eq:two_image_kernel},  and the
delay--boundary-force relations
\eqref{eq:main-boundary-force-summary}--\eqref{eq:main-boundary-delay-summary}.

\subsection{Finite chain, pole convention, and continuum scales}
\label{subsec:main-finite-model}

The disordered sample occupies sites $n=1,\ldots,N$ of a lattice with spacing
$a$.  Its isolated Hermitian Hamiltonian is defined via
\begin{equation}
\label{eq:main-Hin}
        (H_{\rm in}\psi)_n
        =
        \frac{2\psi_n-\psi_{n-1}-\psi_{n+1}}{a^2}
        +V_n\psi_n,
        \qquad
        \psi_0=\psi_{N+1}=0 .
\end{equation}
A clean semi-infinite lead with energy dispersion
\begin{equation}
\label{eq:lattice_dispersion}
        E_a(\kappa)=\frac{2}{a^2}\bigl[1-\cos(a\kappa)\bigr]
\end{equation}
is attached to site $N$ through the contact hopping parameter
$-t_{\rm c}/a^2$.  For real $0<a\kappa<\pi$, the  purely outgoing wave continuation is
\begin{equation}
\label{eq:outgoing-lead-condition}
        \psi_{N+m}=\psi_{N+1}\e^{\ii a\kappa(m-1)},
        \qquad m=1,2,\ldots .
\end{equation}
Eliminating the lead gives the effective non-Hermitian Hamiltonian
\begin{equation}
\label{eq:effective_H}
        H_{\rm eff}(\kappa)
        =H_{\rm in}
        -\frac{t_{\rm c}^2}{a^2}\e^{\ii a\kappa}|N\rangle\langle N|,
\end{equation}
with the resonance determinant
\begin{equation}
\label{eq:outgoing_determinant}
        D_N(\kappa)=\e^{-\ii a\kappa}
        \det\!\left[E_a(\kappa)\mathbf1-H_{\rm eff}(\kappa)\right].
\end{equation}
Its zeros $\kappa_j$ in
\begin{equation}
\label{eq:main-resonance-strip}
        0<\Re(a\kappa)<\pi,
        \qquad \Im\kappa<0,
\end{equation}
are the outgoing reflection resonances in the complex-wavenumber plane.  The
corresponding complex resonance energies are obtained from the analytically
continued lead dispersion,
\begin{equation}
\label{eq:main-kappa-to-resonance-energy}
        Z_j:=E_a(\kappa_j)=E_j-\ii\Gamma_j,
        \qquad \Gamma_j>0.
\end{equation}
Indeed, writing $\kappa_j=u_j-\ii v_j$, with $0<au_j<\pi$ and $v_j>0$,
gives
\begin{equation}
\begin{aligned}
        E_j
        &=
        \frac{2}{a^2}
        \left[1-\cos(au_j)\cosh(av_j)\right],
        \\
        \Gamma_j
        &=
        \frac{2}{a^2}
        \sin(au_j)\sinh(av_j)>0.
\end{aligned}
\label{eq:main-kappa-energy-components}
\end{equation}
In the continuum limit $E_a(\kappa)\to\kappa^2$, so that
$Z_j\simeq\kappa_j^2$.
The corresponding mean density of resonances in the energy--width plane is defined by 
\[
        \rho_N(E,\Gamma;L)
        =
        \EE\sum_j
        \delta(E-E_j)\delta(\Gamma-\Gamma_j),
        \qquad
        L=Na.
\]
For a generic continuous disorder distribution, the resonances are simple
with probability one. 

Let $\phi_n(\kappa)$ be the left normalized solution of the discrete Schr\"odinger recurrence
\begin{equation}
\label{eq:main-recurrence-recurrence}
        \phi_{n+1}
        =\left[2\cos(a\kappa)+a^2V_n\right]\phi_n-\phi_{n-1}, \quad  \phi_0=0,\quad \phi_1=1.
\end{equation}
With
\begin{equation}
\label{eq:main-beta-pm}
        \beta_\pm(\kappa)=t_{\rm c}^2\e^{\pm\ii a\kappa},
\end{equation}
the {\it outgoing residual} is defined by
\begin{equation}
\label{eq:outgoing_residual}
        F_N(\kappa)=\phi_{N+1}(\kappa)-\beta_+(\kappa)\phi_N(\kappa).
\end{equation}
We also use the ratios
\begin{equation}
\label{eq:main-riccati-definitions}
        \zeta_n=\frac{\phi_{n+1}}{\phi_n},
        \qquad Y_n=\partial_\kappa\zeta_n.
\end{equation}
For a unit incoming wave at the first lead site,
\begin{equation}
\label{eq:main-reflection-amplitude-definition}
        \psi_{N+1+m}=\e^{-\ii a\kappa m}
        +\cR_N(\kappa)\e^{\ii a\kappa m},
        \qquad m\geq0.
\end{equation}
Writing the sample solution as $\psi_n=C\phi_n$, contact matching between the clean lead and disordered sample gives
\begin{equation}
\label{eq:main-contact-matching}
        C\phi_{N+1}=t_{\rm c}(1+\cR_N),
        \qquad
        t_{\rm c}C\phi_N=\e^{\ii a\kappa}+\cR_N\e^{-\ii a\kappa},
\end{equation}
and hence
\begin{equation}
\label{eq:reflection_inverse_preview}
        W_N:=\cR_N^{-1}
        =-\e^{-2\ii a\kappa}
        \frac{\zeta_N-\beta_+}{\zeta_N-\beta_-}.
\end{equation}
At a resonance the denominator of $W_N$ is nonzero, so the pole condition is
exactly $W_N=0$.

The absorbing problem is evaluated at $Z_+=E+\ii\eta$, $\eta>0$, and for
real disorder is related by complex conjugation to the pole at
$E-\ii\Gamma$.  We define the reflection coefficient at complex energy as
\begin{equation} \label{r_N}
r_N(E,\eta;L):=|\cR_N(E+\ii\eta;L)|^2
\end{equation}
and finally set $\eta=\Gamma$.

For the continuum calculation the disorder is centred white noise,
\begin{equation}
\label{eq:main-white-noise}
        \EE[V(s)]=0,
        \qquad
        \EE[V(s)V(s')]=D\delta(s-s').
\end{equation}
A compatible lattice approximation has
$\EE[V_nV_m]=(D/a)\delta_{nm}$.  At the bulk energy $E=k^2$ we occasionally
write
\begin{equation}
\label{eq:main-displaced-energy}
        Z_+=E+\delta E+\ii\eta.
\end{equation}
The localization length and long-sample width scale are
\begin{equation}
\label{eq:main-continuum-scales}
        \ellL=\frac{4k^2}{D},
        \qquad
        \GamL=\frac{k}{\ellL}=\frac{D}{4k}.
\end{equation}
The scale $\GamL$ also has a direct dynamical interpretation.  Since the
continuum dispersion is $E=k^2$, the group velocity is $v_{\rm g}=2k$.
With our convention $Z=E-\ii\Gamma$, the resonance probability decays as
$\e^{-2\Gamma t}$, and hence its lifetime is $(2\Gamma)^{-1}$.  Therefore
\[
        \frac{1}{2\GamL}
        =
        \frac{\ellL}{2k}
        =
        \frac{\ellL}{v_{\rm g}},
\]
which is the propagation time through one localization length.  Thus
$\GamL$ separates broad resonances which escape before localization becomes
effective from narrow resonances whose coupling to the lead is controlled by
localization.  We measure propagation distance by $\tau=s/\ellL$ and absorption by
$\alpha=2\eta/\GamL$; for a resonance, $\alpha=2\Gamma/\GamL$.

At fixed wave energy $E=k^2$, the continuum contact is characterized
by a transparency $0<\cT\leq1$ obtained from the scaling limit of the contact hopping
\begin{equation}
\label{eq:regular-contact-scaling-model}
        t_{\rm c}^2=1-\lambda_{\rm c}a+O(a^2),
        \qquad \lambda_{\rm c}\geq0,
\end{equation}
via 
\begin{equation}
\label{eq:main-contact-definitions}
        \cT=\frac{4k^2}{4k^2+\lambda_{\rm c}^2},
        \qquad
        x_{\cT}=\frac{1-\cT}{\cT}.
\end{equation}
After generic-phase averaging we introduce the reflection coefficient
\begin{equation}
\label{eq:main-r-x-definition}
        r=|\cR|^2,
        \qquad
        x=\frac{r}{1-r}.
\end{equation}
Thus $r$ is the reflected intensity, while $x$ is its unbounded radial
coordinate.

The transition density $p_\alpha(x,\tau\mid x_0)$ is factorized as
\begin{subequations}
\label{eq:main-propagator-definitions}
\begin{align}
        \pi_\alpha(x)&=\alpha\e^{-\alpha x},
        \label{eq:main-stationary-pi}\\
        p_\alpha(x,\tau\mid x_0)&=\pi_\alpha(x)
        h_\alpha(x,\tau\mid x_0).
        \label{eq:main-p-h-factorization}
\end{align}
\end{subequations}
The perfectly reflecting closed end corresponds to $x_0=+\infty$, understood
as a limit at fixed propagation distance $\tau>0$.

\subsection{Status of the approximations}
\label{subsec:main-status}

The determinant--outgoing residual identity, the Kac--Rice formulas, the
inverse-reflection representation, and the finite-chain Poincar\'e--Lelong
identity below are exact for every fixed $N$ and $a$.  The complex reflection
equation is the controlled weak-disorder continuum limit of the lattice
M\"obius map.  The scalar intensity diffusion additionally uses generic-phase
averaging and requires
\begin{equation}
\label{eq:main-generic-bulk-regime}
        k\ellL\gg1,
        \qquad kL\gg1,
        \qquad \Gamma/E\ll1.
\end{equation}
The ultranarrow-resonance result further uses $L/\ellL\gg1$, small-absorption
spectral matching at perfect coupling, and a Poisson conversion of the
alternating discrete spectral sum in the moving-front window.  The
finite-transparency extension is uniform for fixed $0<\cT\leq1$ but not in a
singular joint limit in which the contact is closed while the front is
resolved.  The continuum phase-averaged density and the ultranarrow front are
tested numerically in Section~\ref{sec:numerics}.

\subsection{Exact counting and the reflection-intensity evolution}
\label{subsec:main-exact-results}

The outgoing residual and resonance determinant satisfy
$F_N=(-a^2)^N\e^{\ii a\kappa}D_N$, with identical zeros and multiplicities.
The exact Kac--Rice density may therefore be written as
\begin{equation}
\label{eq:main-projective-kr}
        \rho_N^{(\kappa)}(\kappa)=\EE\!\left[\delta^{(2)}(\zeta_N-\beta_+)|Y_N-\beta_+'|^2\right]
        =\EE\!\left[\delta^{(2)}(W_N)|W_N'|^2\right].
\end{equation}
For real disorder, the same zero measure gives the exact finite-chain Poincar\'e--Lelong identity
\begin{equation}
\label{eq:main-finite-poincare-lelong}
        \rho_N(E,\Gamma;L)
        =
        \left.
        \frac{1}{4\pi}
        \left(\partial_E^2+\partial_\eta^2\right)
        \EE\log r_N(E,\eta;L)
        \right|_{\eta=\Gamma}.
\end{equation}

In the weak-disorder continuum limit the complex reflection amplitude obeys
\begin{equation}
\label{eq:main-complex-reflection-diffusion}
\begin{aligned}
        \dd\cR
        =\left[
        2\ii k\cR
        +\frac{\ii\delta E-\eta}{2k}(1+\cR)^2
        -\frac{D}{4k^2}(1+\cR)^3
        \right]\dd s -\frac{\ii\sqrt D}{2k}(1+\cR)^2\dd W_s,
\end{aligned}
\end{equation}
where $\EE[(\dd W_s)^2]=\dd s$.  Generic-phase averaging gives the
following equation for the reflection coefficient defined in
Eq.~\eqref{eq:main-r-x-definition}:
\begin{equation}
\label{eq:main-radial-diffusion}
        \dd r
        =\left[(1-r)^2-\alpha r\right]\dd\tau
        +\sqrt{2r}\,(1-r)\dd B_\tau,
\end{equation}
and the transition density of $x=r/(1-r)$ satisfies the Fokker-Planck equation
\begin{equation}
\label{eq:main-reflection-intensity-forward}
        \partial_\tau p_\alpha
        =\partial_x\!\left[
        x(1+x)(\partial_xp_\alpha+\alpha p_\alpha)\right].
\end{equation}

\subsection{Finite-length density and perfect coupling}
\label{subsec:main-boundary-flux}

A nonideal contact leaves the bulk stochastic evolution of the reflection
intensity unchanged and acts only through the phase-averaged interface
observable.
\begin{equation}
\label{eq:main-interface-observable}
        f_{\cT}(x)
        =\log\max\!\left\{
        \frac{x}{1+x},
        \frac{x_{\cT}}{1+x_{\cT}}
        \right\}.
\end{equation}
Recalling Eq.~\eqref{eq:main-p-h-factorization} one defines the closed-end kernel, for fixed $\tau>0$, by
\begin{equation}\label{closed-end}
        h_\alpha(x,\tau\mid\infty):=
        \lim_{x_0\to+\infty}
        h_\alpha(x,\tau\mid x_0).
\end{equation} 
and further introduces 
\begin{equation}
\label{eq:main-F-contact}
        \mathcal F_{\alpha,\cT}(\tau)
        =\int_0^\infty
        \pi_\alpha(x)h_\alpha(x,\tau\mid\infty)
        f_{\cT}(x)\,\dd x .
\end{equation}
Within this phase-averaged weak-disorder continuum description, and
neglecting the real-energy derivatives which are relatively
$O[(\Gamma/E)^2]$, the finite-length resonance density is
\begin{equation}
\label{eq:principal-general-contact-density}
        \rho_{\cT}(E,\Gamma;L)
        \simeq
        \left.
        \frac{1}{\pi\GamL^2}
        \partial_\alpha^2
        \mathcal F_{\alpha,\cT}\!\left(\frac{L}{\ellL}\right)
        \right|_{\alpha=2\Gamma/\GamL}.
\end{equation}

At perfect coupling, $f_1(x)=\log[x/(1+x)]$ and
$\mathcal F_\alpha=\mathcal F_{\alpha,1}$.  Integration by parts gives
\begin{equation}
\label{eq:principal-endpoint-identity}
        \frac{\dd\mathcal F_\alpha}{\dd\tau}
        =\alpha\left[h_\alpha(0,\tau\mid\infty)-1\right].
\end{equation}
Since $\mathcal F_\alpha(0)=0$, the main finite-length formula takes the form
\begin{equation}
\label{eq:principal-finite-length-density}
        \rho(E,\Gamma;L)
        \simeq
        \left.
        \frac{1}{\pi\GamL^2}\partial_\alpha^2
        \left\{
        \alpha\int_0^{L/\ellL}
        [h_\alpha(0,u\mid\infty)-1]\,\dd u
        \right\}
        \right|_{\alpha=2\Gamma/\GamL}.
\end{equation}
Its stationary and small-absorption limits produce the long-sample
regimes summarized next.

\subsection{Resonance density in the semi-infinite sample and the ultranarrow-width front}
\label{subsec:main-long-sample}

Let
\begin{equation}
\label{eq:main-E1}
        E_1(\alpha)=\int_\alpha^\infty\frac{\e^{-u}}{u}\,\dd u.
\end{equation}
For a perfectly matched semi-infinite sample,
\begin{equation}
\label{eq:semi_infinite_density_main}
        \rho^{(\infty)}(E,\Gamma)
        =\frac{2k}{\pi D\Gamma}
        \left[1-\alpha\e^\alpha E_1(\alpha)\right],
        \qquad \alpha=\frac{2\Gamma}{\GamL}.
\end{equation}
This expression was derived in Ref.~\cite{FyodorovMeibohm2026}.
Its limits are
\begin{subequations}
\label{eq:main-stationary-limits}
\begin{align}
        \rho^{(\infty)}(E,\Gamma)
        &\simeq\frac{2k}{\pi D\Gamma},
        &&\Gamma\ll\GamL,\\
        \rho^{(\infty)}(E,\Gamma)
        &\simeq\frac{1}{4\pi\Gamma^2},
        &&\GamL\ll\Gamma\ll E.
\end{align}
\end{subequations}
The $1/\Gamma$ law reflects an approximately uniform distribution of
localization centres in space, hence an approximately uniform density in
$\log\Gamma$.

For a long but finite sample, define
\[
        C_*=\e^{1-\gamma_{\rm E}},
        \qquad
        T_*=\log\frac{C_*}{\alpha}
        =\log\frac{C_*\GamL}{2\Gamma}.
\]
The ultranarrow-width crossover occurs when the logarithmic distance
$T_*$ becomes comparable to the available propagation length
$\tau=L/\ellL$.  Equivalently, the crossover is centred around the
characteristic resonance width
\begin{equation}
\label{eq:ultranarrow_scale_main}
        \Gamma_{\rm ultra}(L)
        =
        \frac{C_*}{2}\GamL\e^{-L/\ellL},
\end{equation}
for which $T_*=\tau$.

In the moving-front window
\[
        \alpha\ll1,
        \qquad
        \tau=\frac{L}{\ellL}\gg1,
        \qquad
        T_*-\tau=O(\sqrt{\tau}),
\]
the resonance density is
\begin{equation}
\label{eq:ultranarrow_density_main}
        \rho_{\rm ultra}(E,\Gamma;L)
        \simeq
        \frac{2k}{\pi D\Gamma}
        H_*\!\left(T_*,\frac{L}{\ellL}\right),
\end{equation}
with the crossover kernel given by
\begin{equation}
\label{eq:two_image_kernel}
        H_*(T,\tau)
        =\frac12\erfc\!\left(\frac{T-\tau}{2\sqrt\tau}\right)
        +\frac12\e^T\erfc\!\left(\frac{T+\tau}{2\sqrt\tau}\right).
\end{equation}
Equivalently,
\[
        H_*(T,\tau)
        =\mathbb P\!\left(
        \sup_{0\leq u\leq\tau}[u+\sqrt2 B_u]\geq T
        \right),
\]
so the finite-length crossover is the cumulative first-passage probability
for the logarithmic decay of the wave intensity accumulated over the sample. 
Both terms are essential in the deep ultranarrow tail, where they contribute
at the same exponential order.

\subsection{Nonideal contacts}
\label{subsec:main-contact}
For $0<\cT<1$, applying the Fokker--Planck equation
\eqref{eq:main-reflection-intensity-forward} to
Eq.~\eqref{eq:main-F-contact} and integrating by parts gives
\begin{equation}
\begin{aligned}
\frac{\dd\mathcal F_{\alpha,\cT}}{\dd\tau}
&=
-\int_{x_{\cT}}^\infty
\bigl[\partial_x p_\alpha(x,\tau\mid\infty)
      +\alpha p_\alpha(x,\tau\mid\infty)\bigr]\,\dd x
\\
&=
p_\alpha(x_{\cT},\tau\mid\infty)
-\alpha\int_{x_{\cT}}^\infty
p_\alpha(x,\tau\mid\infty)\,\dd x .
\end{aligned}
\label{eq:main-contact-threshold-boundary-flux}
\end{equation}
The  density in the semi-infinite sample is then given by
\begin{equation}
\label{eq:contact_stationary_density_main}
        \rho_{\cT}^{(\infty)}(E,\Gamma)
        =\frac{2k}{\pi D\Gamma}
        \exp\!\left[-\alpha\frac{1-\cT}{\cT}\right]\,\,
        \left[1-\alpha\e^{\alpha/\cT}
        E_1\!\left(\frac{\alpha}{\cT}\right)\right].
\end{equation}
The small-width $1/\Gamma$ law is unchanged, whereas any fixed
$\cT<1$ exponentially suppresses sufficiently broad resonances.  In the
long-sample moving-front regime the kernel shape is unchanged and the contact
shifts the characteristic width to
\begin{equation}
\label{eq:contact_ultranarrow_scale}
        \Gamma_{{\rm ultra},\cT}(L)
        =\cT\Gamma_{\rm ultra}(L)
        =\frac{C_*}{2}\cT\GamL\e^{-L/\ellL}.
\end{equation}
At every finite length, closing the contact gives the distributional limit
\begin{equation}
\label{eq:main-closed-contact-limit}
        \rho_{\cT}(E,\Gamma;L)
        \longrightarrow
        \rho_{\rm cl}(E;L)\delta(\Gamma),
        \qquad \cT\to0.
\end{equation}
These finite-transparency results cover only the stationary and long-sample
ultranarrow regimes.

\subsection{Reflection, delay, boundary weights, and the width sum rule}
\label{subsec:main-related-results}

The finite-length reflection distribution is given by
Eq.~\eqref{eq:finite-length-reflection-density-main}.  In the weak-absorption
scaling regime $1-r=O(\alpha)$, the scaled deficit
$u=(1-r)/\alpha$ tends to
\begin{equation}
        \dd u=\dd\tau-\sqrt2\,u\,\dd B_\tau,
        \qquad
        u=\GamL\tau_{\rm W},
\label{eq:main-delay-summary}
\end{equation}
which is the finite-length Comtet--Texier time-delay process.  Its complete
finite-length distribution is given in Section~\ref{sec:reflection-delay}.

For the continuum closed sample with Dirichlet conditions, let $\psi_n$ be an
$L^2$-normalized eigenfunction.  Displacing the right endpoint gives the
boundary force
\begin{equation}
        F_n:=-\partial_L E_n=|\psi_n'(L)|^2,
\label{eq:main-boundary-force-summary}
\end{equation}
and we normalize it as
\begin{equation}
        \Upsilon_n
        =\frac{\pi\bar\rho_{\rm cl}(E_n)}{k_n}F_n.
\label{eq:main-boundary-weight-summary}
\end{equation}
At every closed eigenvalue the exact Sturm--Liouville identity gives
\begin{equation}
        \Upsilon_n=\frac{1}{\widetilde\tau_{\rm W}(E_n)},
        \qquad
        \widetilde\tau_{\rm W}
        =\frac{\Delta}{2\pi}\tau_{\rm W},
        \qquad
        \Delta=\bar\rho_{\rm cl}^{-1}.
\label{eq:main-boundary-delay-summary}
\end{equation}
After generic-phase averaging this yields an explicit Whittaker representation
for the complete finite-length density of the normalized boundary force $\Upsilon_n$;
see Section~\ref{sec:reflection-delay}.

For every fixed $0<\cT<1$, the same contact formula also gives, to leading
order in the generic-bulk weak-disorder limit,
\begin{equation}
        \int_0^\infty
        \Gamma\,\rho_{\cT}(E,\Gamma;L)\,\dd\Gamma
        =-\frac{1}{4\pi}\log(1-\cT).
\label{eq:main-MS-sum-rule}
\end{equation}
Dividing by the local closed-level density gives the single-channel
Moldauer--Simonius relation
\begin{equation}
        \langle\Gamma\rangle_E
        =-\frac{\Delta(E)}{4\pi}\log(1-\cT).
\label{eq:main-MS-average-width}
\end{equation}
The perfect-coupling divergence is nonuniform: the limit $k\ellL\to\infty$
must be taken before $\cT\uparrow1$.  In the same limit the normalized width
density has the broad tail
$\Delta\rho\sim\Delta/(4\pi\Gamma^2)$, whose first moment produces precisely
the logarithmic divergence in Eq.~\eqref{eq:main-MS-average-width}.

\section{Exact finite-chain Kac--Rice formula in reflection variables}
\label{sec:exact-kr}

We now derive the finite-chain identities summarized in
Section~\ref{sec:main-results}.  Everything in this section is exact for a
fixed lattice spacing, a fixed number of sites, and an arbitrary realization
of the real potential.  

The resonance determinant is the natural starting point, but its direct
disorder average is not adapted to the one-dimensional recursive structure.
Using invariance of the Kac--Rice measure under multiplication by a
nonvanishing analytic factor, we pass successively to the outgoing 
residual, the projective amplitude ratio, and the analytic inverse reflection
amplitude.  Each representation exposes a different part of the exact
finite-chain structure.

\subsection{Outgoing condition and determinant--outgoing residual identity}
\label{subsec:determinant-recurrence}

At the last site of the sample and the first site of the lead, the discrete
Schr\"odinger equations are
\[
        -\frac{t_{\rm c}\psi_{N+1}-2\psi_N+\psi_{N-1}}{a^2}+V_N\psi_N
        =E_a(\kappa)\psi_N
\]
and
\[
        -\frac{\psi_{N+2}-2\psi_{N+1}+t_{\rm c}\psi_N}{a^2}
        =E_a(\kappa)\psi_{N+1}.
\]
For a purely outgoing solution in the lead we have $\psi_{N+2}=\e^{\ii a\kappa}\psi_{N+1}$. 
Using $2-a^2E_a(\kappa)=2\cos(a\kappa)$ in the equation at the first
lead site gives
\[
        t_{\rm c}\psi_N=\left[2\cos(a\kappa)-\e^{\ii a\kappa}\right]\psi_{N+1}=\e^{-\ii a\kappa}\psi_{N+1},
\]
which we further rewrite as $\psi_{N+1}=t_{\rm c}\e^{\ii a\kappa}\psi_N$.

Inside the sample we use the left-normalized solution defined by
 \eqref{eq:main-recurrence-recurrence}.  At the final site, $\phi_{N+1}$ denotes the algebraic continuation of the
unit-hopping recurrence.  The interface equation at site $N$ gives
$C\phi_{N+1}=t_{\rm c}\psi_{N+1}$, where $C$ is the common normalization of
the sample solution.  The outgoing resonance condition is therefore
\begin{equation}
        \phi_{N+1}=t_{\rm c}^2\e^{\ii a\kappa}\phi_N.
\label{eq:outgoing-condition-derived}
\end{equation}
Using the definition \eqref{eq:main-beta-pm}, this condition is precisely the vanishing of the outgoing residual defined in \eqref{eq:outgoing_residual}: $F_N(\kappa)=0$.

We next verify directly that this outgoing residual is proportional to the
resonance determinant Eq.~\eqref{eq:outgoing_determinant}.  Let $H_{\rm in}^{(n)}$ be the restriction of the closed sample
Hamiltonian to the first $n$ sites and define
\begin{equation}\label{inn_det}
        \Delta_n(\kappa)
        =\det\!\left[E_a(\kappa)\mathbf 1_n-H_{\rm in}^{(n)}\right],
        \qquad
        \Delta_0(\kappa)=1.
\end{equation}
Expansion of the tridiagonal determinant gives
\[
        \Delta_n
        =-\frac{2\cos(a\kappa)+a^2V_n}{a^2}\Delta_{n-1}
        -\frac{1}{a^4}\Delta_{n-2}.
\]
After introducing $\tilde{\phi}_n(\kappa)=(-a^2)^n\Delta_n(\kappa)$,
one obtains
\[
        \tilde{\phi}_n
        =\left[2\cos(a\kappa)+a^2V_n\right]\tilde{\phi}_{n-1}
        -\tilde{\phi}_{n-2},
\]
with initial values
\[
        \tilde{\phi}_0=1,
        \qquad
        \tilde{\phi}_1=2\cos(a\kappa)+a^2V_1.
\]
These are precisely the recurrence and initial data for $\phi_{n+1}$ defined
in 
\eqref{eq:main-recurrence-recurrence}.  Therefore
\begin{equation}
        (-a^2)^n
        \det\!\left[E_a(\kappa)\mathbf 1_n-H_{\rm in}^{(n)}\right]
        =\phi_{n+1}(\kappa).
\label{eq:continuant-recurrence-identity}
\end{equation}

Eliminating the lead modifies only the last diagonal element of the sample
Hamiltonian.  Denoting
$A_N(\kappa)=E_a(\kappa)\mathbf 1-H_{\rm in}$, the definition
\eqref{eq:effective_H} gives
\[
        E_a(\kappa)\mathbf 1-H_{\rm eff}(\kappa)
        =A_N(\kappa)
        +\frac{t_{\rm c}^2}{a^2}
        \e^{\ii a\kappa}|N\rangle\langle N|.
\]
The rank-one determinant expansion, together with the cofactor identity
$\langle N|A_N^{-1}|N\rangle=\Delta_{N-1}/\Delta_N$ and
Eq.~\eqref{inn_det}, then gives
\[
\begin{aligned}
        \det\!\left[E_a(\kappa)\mathbf 1-H_{\rm eff}\right]
        &=
        \Delta_N
        +
        \frac{t_{\rm c}^2}{a^2}
        \e^{\ii a\kappa}\Delta_{N-1},
\end{aligned}
\]
which upon multiplying by $(-a^2)^N$ and using 
\eqref{eq:continuant-recurrence-identity} gives the resonance determinant defined in Eq.(\ref{eq:outgoing_determinant}):
\begin{equation}
        F_N(\kappa)
        =(-a^2)^N\det\!\left[E_a(\kappa)\mathbf 1-H_{\rm eff}\right]
        =(-a^2)^N\e^{\ii a\kappa}D_N(\kappa).
\label{eq:main-det-recurrence}
\end{equation}
Thus the resonance determinant \eqref{eq:outgoing_determinant} and the
outgoing residual \eqref{eq:outgoing_residual} have the same zeros and
multiplicities throughout the resonance strip
\eqref{eq:main-resonance-strip}.

\subsection{Complex Kac--Rice formula and its projective form}
\label{subsec:complex-kac-rice}

At a simple zero $\kappa_j$ (the generic situation for an absolutely
continuous disorder distribution)
\[
        F_N(\kappa)
        =F_N'(\kappa_j)(\kappa-\kappa_j)
        +O\!\left((\kappa-\kappa_j)^2\right).
\]
Considering this as a map from the two real coordinates of $\kappa$ to the two real
coordinates of $F_N$, its Jacobian matrix is
\[
        \begin{pmatrix}
        \Re F_N'&-\Im F_N'\\
        \Im F_N'&\Re F_N'
        \end{pmatrix},
\]
whose determinant is $|F_N'|^2$.  The ordinary two-dimensional
change-of-variables formula therefore gives
\begin{equation}
        \delta^{(2)}\!\left(F_N(\kappa)\right)|F_N'(\kappa)|^2
        =\sum_j\delta^{(2)}(\kappa-\kappa_j).
\label{eq:complex-kac-rice-sample}
\end{equation}
This is the local complex Kac--Rice formula. 

We shall repeatedly use one elementary invariance.  If
$G(\kappa)=A(\kappa)H(\kappa)$ and $A$ is analytic and nonzero on the zero set
of $H$, then, on $H=0$, one has $G'=AH'$.  Consequently,
\begin{equation}
        \delta^{(2)}(G)|G'|^2
        =\delta^{(2)}(H)|H'|^2.
\label{eq:analytic-rescaling-invariance}
\end{equation}
Multiplication of a resonance condition by a nonvanishing analytic factor
therefore changes neither its zeros nor its Kac--Rice density.  This single
identity explains the replacements $D_N\mapsto F_N$,
$F_N\mapsto\zeta_N-\beta_+$, and later $F_N\mapsto W_N$.

The amplitude ratio $\zeta_N$ at interface site of the disordered region is defined in
Eq.~\eqref{eq:main-riccati-definitions} together with its spectral derivative $Y_N$.  It is always well defined at a
resonance.  Indeed, if
$F_N(\kappa)=0$ and $\phi_N(\kappa)=0$, then also $\phi_{N+1}(\kappa)=0$.
The  recurrence applied backwards would then imply
$\phi_{N-1}=\cdots=\phi_1=0$, contradicting $\phi_1=1$.  Thus
\begin{equation}
        F_N(\kappa)=0
        \quad\Longrightarrow\quad
        \phi_N(\kappa)\neq0.
\label{eq:phiN-nonzero}
\end{equation}

The residual factorizes as
\[
        F_N=\phi_N(\zeta_N-\beta_+).
\]
with $\beta_{\pm}$ defined in Eq.~\eqref{eq:main-beta-pm}. 
Using the spectral derivative $Y_N$, see 
Eq.~\eqref{eq:main-riccati-definitions}, differentiation of the factorization
gives
\[
        F_N'
        =\phi_N'(\zeta_N-\beta_+)
        +\phi_N(Y_N-\beta_+').
\]
On the resonance condition $\zeta_N=\beta_+$, this reduces to
\begin{equation} \label{eq:on-res-cond}
    F_N'=\phi_N(Y_N-\beta_+'),
        \qquad
        \beta_+'=\ii a\beta_+.
\end{equation}

Since $\delta^{(2)}(Az)=|A|^{-2}\delta^{(2)}(z)$ for $A\neq0$, the factor
$|\phi_N|^2$ cancels exactly between the delta function and the Jacobian:
\begin{equation}
        \delta^{(2)}(F_N)|F_N'|^2
        =\delta^{(2)}(\zeta_N-\beta_+)|Y_N-\beta_+'|^2.
\label{eq:exact-projective-cancellation}
\end{equation}
Consequently, averaging over the disorder gives the resonance density as
\begin{equation}
        \rho_N^{(\kappa)}(\kappa)
        =\EE\!\left[
        \delta^{(2)}(\zeta_N-\beta_+)
        |Y_N-\ii a\beta_+|^2
        \right].
\label{eq:finite-kac-rice-riccati}
\end{equation}
 Mathematically, $\zeta_N$ is a projective
coordinate, and
\eqref{eq:exact-projective-cancellation} is the corresponding projective
invariance of the Kac--Rice density.
 For a recent development of direct Kac--Rice methods
in large non-Hermitian random matrices, see
Ref.~\cite{FyodorovKacRice2025}.

\subsection{Riccati recurrence and its spectral derivative}
\label{subsec:riccati-spectral-derivative}

Dividing the equation Eq.~\eqref{eq:main-recurrence-recurrence} by $\phi_n$ gives the exact one-site recurrence for the projective coordinate
\begin{equation}
        \zeta_n
        =2\cos(a\kappa)+a^2V_n-\frac{1}{\zeta_{n-1}}.
\label{eq:exact-riccati-recurrence}
\end{equation}
Equivalently,
\[
        \zeta_n=\Phi_{n,\kappa}(\zeta_{n-1}),
        \qquad
        \Phi_{n,\kappa}(\zeta)
        =\frac{[2\cos(a\kappa)+a^2V_n]\zeta-1}{\zeta}.
\]
Thus every one-site recurrence acts by a fractional-linear, or M\"obius,
transformation on the ratio of amplitudes.  The Dirichlet initial condition is
represented by $\zeta_0=\infty$ on the extended complex plane, and the first
finite ratio is
\[
        \zeta_1=2\cos(a\kappa)+a^2V_1.
\]
A divergence of a Riccati coordinate is not a singularity of the recurrence
solution; it only means that this particular ratio of the two components is
not a valid local coordinate.  One may use the reciprocal ratio or return to
the two-component recurrence vector.

Differentiating \eqref{eq:exact-riccati-recurrence} with respect to $\kappa$
gives
\begin{equation}
        Y_n
        =-2a\sin(a\kappa)+\frac{Y_{n-1}}{\zeta_{n-1}^2},
        \qquad
        Y_1=-2a\sin(a\kappa).
\label{eq:exact-spectral-derivative-recurrence}
\end{equation}
Once the Riccati trajectory is known, this recurrence determines the
spectral derivative required by the direct Kac--Rice Jacobian.

\subsection{Reflection amplitude and inverse-reflection variable}
\label{subsec:inverse-reflection}

The physical scattering state and its reference-plane convention were defined
in Eq.~\eqref{eq:main-reflection-amplitude-definition}.  Evaluating that
state at the first two lead sites and substituting it into the discrete
Schr\"odinger equations at the sample--lead contact gives the matching
conditions \eqref{eq:main-contact-matching}.  Eliminating the common
normalization $C$ from those equations yields
\begin{equation}
        \cR_N(\kappa)
        =-\e^{2\ii a\kappa}
        \frac{\phi_{N+1}-t_{\rm c}^2\e^{-\ii a\kappa}\phi_N}
        {\phi_{N+1}-t_{\rm c}^2\e^{\ii a\kappa}\phi_N}.
\label{eq:reflection-amplitude-recurrence}
\end{equation}

Using $\beta_-$ from Eq.~\eqref{eq:main-beta-pm}, introduce the
auxiliary incoming residual
\[
        F_N^-(\kappa)
        =\phi_{N+1}(\kappa)-\beta_-(\kappa)\phi_N(\kappa),
\]
and write $F_N^+=F_N$.  Then
\begin{equation}
        \cR_N(\kappa)
        =-\e^{2\ii a\kappa}\frac{F_N^-(\kappa)}{F_N^+(\kappa)}
        =-\e^{2\ii a\kappa}
        \frac{\zeta_N(\kappa)-\beta_-(\kappa)}
        {\zeta_N(\kappa)-\beta_+(\kappa)}.
\label{eq:reflection-residual-ratio}
\end{equation}
The signs $+$ and $-$ label the outgoing and incoming lead factors. 
They are complex conjugates for real $\kappa$, but become distinct analytic
functions upon continuation into the complex wave-number plane.

There is no pole--zero cancellation at a resonance.  Indeed,
$F_N^+=0$ implies $\zeta_N=\beta_+$, and
\[
        \zeta_N-\beta_-
        =\beta_+-\beta_-
        =2\ii t_{\rm c}^2\sin(a\kappa),
\]
which is nonzero in the resonance strip \eqref{eq:main-resonance-strip}.  In fact, for an open
contact $t_{\rm c}>0$, the incoming residual has no zeros anywhere in that
strip.  If $F_N^-(\kappa)=0$, the vector
$\phi=(\phi_1,\ldots,\phi_N)^T$ would satisfy
\[
        \left[H_{\rm in}
        -\frac{t_{\rm c}^2}{a^2}\e^{-\ii a\kappa}|N\rangle\langle N|\right]
        \phi
        =E_a(\kappa)\phi.
\]
Taking the Hermitian scalar product with $\phi$ and then the imaginary part
gives
\[
        \Im E_a(\kappa)\|\phi\|^2
        =-\frac{t_{\rm c}^2}{a^2}
        \Im\!\left(\e^{-\ii a\kappa}\right)|\phi_N|^2.
\]
For $\kappa=k-\ii q$, with $q>0$ and $0<ak<\pi$,
\[
        \Im E_a(\kappa)
        =-\frac{2}{a^2}\sin(ak)\sinh(aq)<0,
\]
whereas
\[
        -\Im\!\left(\e^{-\ii a\kappa}\right)
        =\e^{-aq}\sin(ak)>0.
\]
The two sides have opposite signs, a contradiction.  Hence
\begin{equation}
        F_N^-(\kappa)\neq0
        \qquad
        \text{throughout the resonance strip \eqref{eq:main-resonance-strip}}.
\label{eq:incoming-residual-nonzero}
\end{equation}

The inverse reflection amplitude $W_N$ defined in
Eq.~\eqref{eq:reflection_inverse_preview} is therefore analytic in the resonance strip
\eqref{eq:main-resonance-strip} and has
exactly the resonance zeros:
\begin{equation}
        W_N(\kappa)
        =-\e^{-2\ii a\kappa}\frac{F_N^+(\kappa)}{F_N^-(\kappa)}
        =-\e^{-2\ii a\kappa}
        \frac{\zeta_N(\kappa)-\beta_+(\kappa)}
        {\zeta_N(\kappa)-\beta_-(\kappa)}.
\label{eq:inverse-reflection-exact}
\end{equation}
Since $W_N$ differs from $F_N^+$ by an analytic nonvanishing factor, the
representations in terms of the outgoing residual, amplitude ratio, and
inverse reflection amplitude give precisely the equivalent finite-chain
density formulas stated in Eq.~\eqref{eq:main-projective-kr}.  They emphasize,
respectively, the outgoing boundary condition, the cancellation of the
left normalization, and the physical scattering variable.

Because $\phi_n(-\kappa)=\phi_n(\kappa)$, it implies
$F_N^-(\kappa)=F_N^+(-\kappa)$.  Together with
\eqref{eq:main-det-recurrence}, this gives
\begin{equation}
        \cR_N(\kappa)=-\frac{D_N(-\kappa)}{D_N(\kappa)}.
\label{eq:reflection-determinant-ratio}
\end{equation}
For real $\kappa$ and real disorder,
$F_N^-=\overline{F_N^+}$ and therefore $|\cR_N|=1$, as required by flux
conservation.

\subsection{Poincar\'e--Lelong and the absorbing-energy logarithmic identity}
\label{subsec:poincare-lelong}

For a holomorphic function $W(\kappa)$, the Poincar\'e--Lelong identity
(see, e.g., Ref.~\cite{BleherShiffmanZelditch2000}) for its use in the statistical theory
of complex zeros) gives
\begin{equation}
        \frac{1}{\pi}
        \partial_\kappa\partial_{\overline\kappa}
        \log|W(\kappa)|^2
        =\sum_j m_j\delta^{(2)}(\kappa-\kappa_j),
\label{eq:poincare-lelong-sample}
\end{equation}
in the sense of distributions, where $m_j$ is the multiplicity of the zero
$\kappa_j$.  For simple zeros this is equivalent to
$\delta^{(2)}(W)|W'|^2$.  In one complex variable,
Eq.~\eqref{eq:poincare-lelong-sample} follows directly from
$\Delta_\kappa\log|\kappa|^2=4\pi\delta^{(2)}(\kappa)$ and the local
factorization of a holomorphic function around its zeros.

A closely related logarithmic-determinant construction was used by Gaspard
and Sparenberg in the quantum random Lorentz gas
\cite{GaspardSparenberg2022}.  Their ``resonance potential'' expresses the
two-dimensional resonance density as the Laplacian of
$\langle\log|\det M|\rangle$; in the present terminology this is a
Poincar\'e--Lelong zero-counting relation.

A useful regularized form is
\begin{equation}
        \frac{1}{\pi}
        \partial_\kappa\partial_{\overline\kappa}
        \log\!\left(|W|^2+\epsilon^2\right)
        =
        \frac{1}{\pi}
        \frac{\epsilon^2|W'|^2}
        {\left(|W|^2+\epsilon^2\right)^2}
        \underset{\epsilon\to0^+}{\longrightarrow}
        \sum_j m_j\delta^{(2)}(\kappa-\kappa_j).
\label{eq:poincare-lelong-regularized}
\end{equation}
Thus simplicity of the resonance zeros is not required for the
zero-counting formulation.  Applying
Eq.~\eqref{eq:poincare-lelong-sample} to the analytic inverse reflection
amplitude $W_N$ gives the resonance density sample by sample.
The change from wave number to energy must be kept explicit.  In the
fundamental resonance strip
\[
        0<\Re(a\kappa)<\pi,
        \qquad
        \Im\kappa<0,
\]
one has
\[
        E_a'(\kappa)=\frac{2}{a}\sin(a\kappa)\neq0,
\]
so $E_a$ is conformal and one-to-one.  For $\eta>0$, let
$\kappa_-(E,\eta)$ denote the unique point in this strip satisfying
\[
        E_a\!\left(\kappa_-(E,\eta)\right)
        =
        E-\ii\eta.
\]
This is the branch selected by the outgoing boundary condition.  The real
two-dimensional Jacobian of the change of variables is therefore
$|E_a'(\kappa_-)|^2$, and
\begin{equation}
        \rho_N(E,\Gamma)
        =
        \frac{
        \rho_N^{(\kappa)}\!\left(\kappa_-(E,\Gamma)\right)
        }{
        \left|E_a'\!\left(\kappa_-(E,\Gamma)\right)\right|^2
        }.
\label{eq:exact-kappa-energy-jacobian}
\end{equation}
This is the exact lattice formula on the outgoing branch.  In the continuum
bulk, $E_a(\kappa)\to\kappa^2$ and
$|E_a'(\kappa_-)|^2\simeq4k^2$.

 It remains to relate the lower-half-plane resonance problem to the
upper-half-plane absorbing scattering problem. Reality of the potential implies
\[
        \overline{F_N^+(\overline\kappa)}=F_N^-(\kappa),
        \qquad
        \overline{F_N^-(\overline\kappa)}=F_N^+(\kappa).
\]
Using \eqref{eq:reflection-residual-ratio}, we obtain the exact symmetry
\begin{equation}
        W_N(\kappa)=\overline{\cR_N(\overline\kappa)}.
\label{eq:inverse-reflection-reality}
\end{equation}
Let $\kappa_-(E,\eta)$ be the point on the outgoing branch satisfying
\[
        E_a(\kappa_-)=E-\ii\eta,
        \qquad
        \eta>0,
\]
and let $\kappa_+(E,\eta)=\overline{\kappa_-(E,\eta)}$.  Define the reflection
coefficient at the absorbing energy by
\[
        r_N(E,\eta;L)
        =\left|\cR_N\!\left(\kappa_+(E,\eta)\right)\right|^2.
\]
The upper-half-plane shift $E\mapsto E+\ii\eta$ is equivalent in the
Schr\"odinger equation to adding uniform absorption.  Equation
\eqref{eq:inverse-reflection-reality} gives
\begin{equation}
        \left|W_N\!\left(\kappa_-(E,\eta)\right)\right|^2
        =r_N(E,\eta;L).
\label{eq:W-to-absorbing-r}
\end{equation}

In the energy plane the two-dimensional Laplacian $\partial_E^2+\partial_y^2$
is invariant under $y\mapsto-y$, so passing from the lower pole coordinate
$y=-\Gamma$ to the upper absorbing coordinate $\eta=\Gamma$ leaves it
unchanged.  Averaging the regularized identity, taking $\epsilon\to0^+$, changing
from wave number to energy, and using the conjugation relation proves the
exact finite-chain Poincar\'e--Lelong identity
\eqref{eq:main-finite-poincare-lelong}.
The identity is understood distributionally if necessary.  The regularized
form \eqref{eq:poincare-lelong-regularized} provides a direct way to justify
interchanging disorder averaging with the derivatives.

The direct Kac--Rice form retains the spectral derivative, whereas the
Poincar\'e--Lelong form \eqref{eq:main-finite-poincare-lelong} eliminates that
insertion and is the efficient representation for the scalar density studied
here.  Only after the discrete-to-continuum limit do we use the generic-bulk
approximation in which the real-energy derivative is subleading.

\section{From the discrete reflection map to the reflection-intensity diffusion}
\label{sec:diffusion}

This section connects the exact finite-chain construction of the previous section to the scalar
reflection diffusion used in the rest of the paper.  Three logically distinct
steps are involved.  First, the lattice Riccati recurrence is rewritten as an
exact fractional-linear map for the moving-interface reflection ratio.  Second, the
white-noise weak-disorder limit gives a complex diffusion.  Third, at generic
bulk energies, the rapidly rotating phase of the clean wave is averaged to obtain a
closed diffusion for the reflection coefficient. 

The contact parameter $t_{\rm c}$ does not occur in this bulk recurrence.  It acts only at
the final interface, after the disordered segment has been propagated.  We
therefore derive the complex-amplitude and reflection-intensity evolutions once and later combine the
resulting universal bulk propagator with the interface observable appropriate
to the chosen transparency $\cT$.

\subsection{Moving-interface reflection ratio and exact M\"obius map}
\label{subsec:exact-reflection-map}

Fix a real  wave number $k$ in the interior of the lattice band and put
\begin{equation}\label{def-Delta}
        z=\e^{\ii ak},
        \qquad
        \Delta=z-z^{-1}=2\ii\sin(ak),
        \qquad
        E=E_a(k).
\end{equation}

At finite lattice spacing, we consider the complex energy $Z=E_a(k)+\delta E+\ii\eta,\quad\eta>0$,
where $\delta E$ is a real detuning from the wave energy $E_a(k)$.
This is the lattice counterpart of the continuum parametrization
\eqref{eq:main-displaced-energy}.  The eigenvalue equation
$H_{\rm in}\phi=Z\phi$, with $H_{\rm in}$ defined in
Eq.~\eqref{eq:main-Hin}, can then be written exactly as
\[
        \phi_{n+1}=\bigl[z+z^{-1}+h_n\bigr]\phi_n-\phi_{n-1},
        \qquad
        h_n=a^2\bigl(V_n-\delta E-\ii\eta\bigr).
\]
Thus the ratio $\zeta_n=\phi_{n+1}/\phi_n$ obeys the Riccati recurrence
\begin{equation}\label{Riccati-recur}
        \zeta_n=z+z^{-1}+h_n-\frac{1}{\zeta_{n-1}}.
\end{equation}

Consider first the truncated disordered segment consisting of sites
$1,\ldots,n$.  At the auxiliary interface immediately to the right of
site $n$, decompose the two-site boundary data into waves
propagating towards increasing and decreasing site number:  
\[
        \phi_n=A_n^++A_n^-,
        \qquad
        \phi_{n+1}=zA_n^++z^{-1}A_n^-.
\]
Here $A_n^+$ and $A_n^-$ are respectively the local outgoing and incoming
amplitudes, with the wave phases measured from site $n$.  We define the
moving-interface reflection ratio as
\[
        R_n=\frac{A_n^+}{A_n^-}.
\]
It follows that
\[
        \zeta_n
        =
        \frac{\phi_{n+1}}{\phi_n}
        =
        \frac{zR_n+z^{-1}}{1+R_n}
        =
        \frac{z^2R_n+1}{z(1+R_n)}.
\]
Solving for $R_n$ gives
\begin{equation}
        R_n
        =
        -\frac{\zeta_n-z^{-1}}{\zeta_n-z},
        \qquad
        \zeta_n
        =
        \frac{z^2R_n+1}{z(1+R_n)}.
\label{eq:interface-reflection-coordinate}
\end{equation}
The left Dirichlet condition $\phi_0=0$ implies
$A_0^++A_0^-=0$, and therefore $R_0=-1$.

  Substitution into the Riccati
recurrence Eq.~\eqref{Riccati-recur} yields the exact one-site map
\begin{equation}
        R_n
        =
        \frac{
        \Delta z^2R_{n-1}
        +h_n\bigl(1+z^2R_{n-1}\bigr)
        }{
        \Delta-h_n\bigl(1+z^2R_{n-1}\bigr)
        },
\label{eq:exact-interface-mobius}
\end{equation}
where $\Delta$ was defined in Eq.~\eqref{def-Delta}. For $h_n=0$, one has $R_n=z^2R_{n-1}$ and hence
$R_n=-z^{2n}$, which fixes the reference-plane phase convention.
The variable $R_n$ is the reflection amplitude seen at the moving interface,
whereas $\cR_N$ in Section~\ref{sec:exact-kr} is the physical reflection
amplitude measured in the external lead.

\subsection{Weak-disorder expansion and conditional moments}
\label{subsec:weak-disorder-moments}

Discretize the white-noise potential defined in
Eq.~\eqref{eq:main-white-noise} by averaging it over each lattice interval,
\[
        V_n
        =
        \frac{1}{a}\int_{(n-1)a}^{na}V(s)\,\dd s,
        \qquad
        \xi_n=aV_n
        =
        \int_{(n-1)a}^{na}V(s)\,\dd s.
\]
The variables $\xi_n$ are independent centered real Gaussians of the order $O_{\mathbb P}(\sqrt a)$  satisfying
\[
        \EE\xi_n=0,
        \qquad
        \EE(\xi_n\xi_m)=Da\,\delta_{nm}.
\]
Here $O_{\mathbb P}(\sqrt a)$ means a random correction whose typical size is
of order $\sqrt a$: more precisely, after division by $\sqrt a$ it remains
bounded with probability tending to one as $a\to0$.

With $\Delta=z-z^{-1}$ as defined in Eq.~\eqref{def-Delta}, set
\[
        R=R_{n-1},
        \qquad
        A=1+z^2R,
        \qquad
        \delta_n=\frac{h_n}{\Delta}.
\]
Then the exact map \eqref{eq:exact-interface-mobius} takes the form
\begin{equation}
\label{eq:exact-map-rescaled}
        R_n
        =
        \frac{z^2R+\delta_nA}{1-\delta_nA}.
\end{equation}

In the continuum scaling $a\to0$ at fixed wave number $k$, so that
$ka\ll1$, the lattice factors have the expansions
\[
        z^2-1
        =
        2\ii ka+O(a^2),
        \qquad
        \Delta
        =
        2\ii ka-\frac{\ii}{3}k^3a^3+O(a^5).
\]
and hence
\[
        \frac{1}{\Delta}
        =
        \frac{1}{2\ii ka}
        \left[
        1+\frac{k^2a^2}{6}+O(a^4)
        \right].
\]
Since
\[
        h_n
        =
        a\xi_n-a^2(\delta E+\ii\eta),
\]
we obtain
\begin{equation}
        \delta_n=\frac{\xi_n}{2\ii k}-a\frac{\delta E+\ii\eta}{2\ii k}+o_{\mathbb P}(a)\quad \mbox{and}\quad 
        \delta_n^2=-\frac{\xi_n^2}{4k^2}+o_{\mathbb P}(a).
\label{eq:delta-expansion-main}
\end{equation}
Thus, in view of $\xi_n=O_{\mathbb P}(\sqrt a)$ we have $\delta_n=O_{\mathbb P}(\sqrt a)$ and $\delta_n^2=O_{\mathbb P}(a)$.
The term proportional to $\delta_n^2$ must therefore be retained in order
to recover the full continuum drift. Using the exact algebraic identity
\[
        \frac{1}{1-y}
        =
        1+y+y^2+\frac{y^3}{1-y},
\]
Eq.~\eqref{eq:exact-map-rescaled} becomes
\[
        R_n
        =
        z^2R+\delta_nA^2+\delta_n^2A^3
        +\frac{\delta_n^3A^4}{1-\delta_nA}.
\]
For bounded $R$, one has
$\delta_nA=O_{\mathbb P}(\sqrt a)$ in the weak-disorder scaling, so the
denominator in Eq.~\eqref{eq:exact-map-rescaled} differs from unity by a
term of order $\sqrt a$.  The remainder in the exact expansion is therefore
$O_{\mathbb P}(a^{3/2})$, and
\begin{equation}
        R_n
        =
        z^2R+\delta_nA^2+\delta_n^2A^3
        +O_{\mathbb P}(a^{3/2}).
\label{eq:mobius-second-order-main}
\end{equation}

Using
\[
        A^2=(1+R)^2+O(a),
        \qquad
        A^3=(1+R)^3+O(a),
\]
together with Eq.~\eqref{eq:delta-expansion-main}, gives the one-step
increment (not to be confused with the finite-chain determinant
$\Delta_n(\kappa)$ of Eq.~\eqref{inn_det})
\begin{equation}
\begin{aligned}
        \Delta_nR
        &:=
        R_n-R_{n-1}
        \\
        &=
        \left[
        2\ii kR
        -\frac{\delta E+\ii\eta}{2\ii k}(1+R)^2
        \right]a
        +\frac{(1+R)^2}{2\ii k}\,\xi_n
        -\frac{(1+R)^3}{4k^2}\,\xi_n^2
        +o_{\mathbb P}(a).
\end{aligned}
\label{eq:one-step-reflection-increment}
\end{equation}

Separating the mean and centered parts of the term quadratic in the lattice disorder,
\[
        \xi_n^2
        =
        Da+\bigl(\xi_n^2-Da\bigr),
\]
we may equivalently write
\[
\begin{aligned}
        \Delta_nR
        &=
        \left[
        2\ii kR
        -\frac{\delta E+\ii\eta}{2\ii k}(1+R)^2
        -\frac{D}{4k^2}(1+R)^3
        \right]a
        \\
        &\quad
        +\frac{(1+R)^2}{2\ii k}\,\xi_n
        -\frac{(1+R)^3}{4k^2}
        \bigl(\xi_n^2-Da\bigr)
        +o_{\mathbb P}(a).
\end{aligned}
\]
For the Gaussian increments,
\[
        \EE\!\left[
        \xi_n\bigl(\xi_n^2-Da\bigr)
        \right]
        =0,
        \qquad
        \EE\!\left[
        \bigl(\xi_n^2-Da\bigr)^2
        \right]
        =
        2D^2a^2.
\]
Thus the centered quadratic term does not contribute to the limiting
one-step covariance.  Over a fixed macroscopic length containing
$O(a^{-1})$ lattice spacings its accumulated variance is $O(a)$ and vanishes
as $a\to0$.  Its nonzero mean $Da$, however, survives and produces the cubic
contribution to the drift.  This is the discrete origin of the cubic
It\^o drift.  A brief primer on It\^o calculus is given in
Appendix~A.

Let  $\EE[\,\cdot\mid R]$ denotes the average over the disorder in one
lattice step, with the current value of the reflection intensity $R$ held
fixed.  The limiting first two moments of the corresponding change
$\Delta_n R$ are therefore
\begin{equation}
\begin{aligned}
        \lim_{a\to0}
        \frac{\EE[\Delta_nR\mid R]}{a}
        &=
        2\ii kR
        -\frac{\delta E+\ii\eta}{2\ii k}(1+R)^2
        -\frac{D}{4k^2}(1+R)^3, \\
        \lim_{a\to 0}
        \frac{\EE[(\Delta_nR)^2\mid R]}{a}= &
        -\frac{D}{4k^2}(1+R)^4, \quad \lim_{a\to0}
        \frac{\EE[|\Delta_nR|^2\mid R]}{a}=
        \frac{D}{4k^2}|1+R|^4.
\end{aligned}
\label{eq:three-reflection-moments}
\end{equation}

These limiting conditional moments determine the drift and diffusion
covariance of the continuum process.  In the notation of
Appendix~\ref{app:complex-Ito-diffusions}, the second and third lines of
Eq.~\eqref{eq:three-reflection-moments} satisfy $|Q(R)|=C(R)$.  The real
$2\times2$ covariance matrix therefore has rank one and may be represented
by a single real Brownian motion.  A convenient noise coefficient is
\[
        \sigma(R)
        =
        -\frac{\ii\sqrt D}{2k}(1+R)^2,
\]
for which $\sigma(R)^2=Q(R)$ and $|\sigma(R)|^2=C(R)$.  Together with the
first conditional moment this yields the It\^o  differential
equation \eqref{eq:main-complex-reflection-diffusion} for the continuum
reflection amplitude $\cR(s)$.

\subsection{Complex reflection diffusion and exact intensity equation}
\label{subsec:complex-reflection-diffusion}

We now derive the evolution of the reflected intensity directly from the
complex diffusion.  Write Eq.~\eqref{eq:main-complex-reflection-diffusion} as
\[
        \dd\cR=b(\cR)\,\dd s+\sigma(\cR)\,\dd W_s,
\]
where
\[
\begin{aligned}
        b(\cR)
        =
        2\ii k\cR
        +\frac{\ii\delta E-\eta}{2k}(1+\cR)^2
        -\frac{D}{4k^2}(1+\cR)^3,
        \qquad
        \sigma(\cR)=-\frac{\ii\sqrt D}{2k}(1+\cR)^2,
\end{aligned}
\]
and $W_s$ is a standard real Brownian motion.  Introduce the intensity and
phase by
\[
        \cR=\sqrt r\,\e^{\ii\varphi},
        \qquad
        r=|\cR|^2.
\]
It\^o's formula gives
\[
        \dd r
        =
        \overline{\cR}\,\dd\cR
        +\cR\,\dd\overline{\cR}
        +\dd\cR\,\dd\overline{\cR}.
\]
Substituting the drift and noise coefficients above and collecting the
Fourier harmonics of $\varphi$ yields the exact intensity equation associated
with the continuum complex diffusion:
\begin{equation}
\begin{aligned}
        \dd r
        &=
        A_r(r,\varphi)\,\dd s
        +B_r(r,\varphi)\,\dd W_s,
        \\
        A_r(r,\varphi)
        &=
        -\frac{\eta}{k}
        \left[2r+\sqrt r(1+r)\cos\varphi\right]
        +\frac{\delta E}{k}\sqrt r(1-r)\sin\varphi
        \\
        &\quad
        +\frac{D}{4k^2}(1-r)
        \left[
        1-r+2\sqrt r\cos\varphi+2r\cos(2\varphi)
        \right],
        \\
        B_r(r,\varphi)
        &=
        -\frac{\sqrt D}{k}\sqrt r(1-r)\sin\varphi.
\end{aligned}
\label{eq:full-intensity-equation-before-average}
\end{equation}
No phase averaging has yet been used in
Eq.~\eqref{eq:full-intensity-equation-before-average}.

Several physical checks are immediate.  If $D=\eta=\delta E=0$, then
$\dd r=0$ and the clean solution satisfying the closed-end condition is
$\cR(s)=-\e^{2\ii ks}$. 
At real energy, $\eta=0$, there is no absorption, and flux conservation in
the one-lead geometry with a closed far end requires
$|\cR|=r=1$.  Correspondingly,
\[
        A_r(1,\varphi)=0,
        \qquad
        B_r(1,\varphi)=0.
\]
Thus both the radial drift and the radial noise vanish on the unit circle:
a reflection amplitude initially satisfying $|\cR|=1$ remains on that circle,
while its phase $\varphi$ continues to evolve. For $\eta>0$,
\[
        A_r(1,\varphi)
        =
        -\frac{4\eta}{k}\cos^2\frac{\varphi}{2}\leq0,
        \qquad
        B_r(1,\varphi)=0,
\]
so absorption drives the process into the unit disk. Although the polar phase is undefined at $r=0$, the radial coefficients have
regular, phase-independent limits,
\[
        A_r(0,\varphi)=\frac{D}{4k^2},
        \qquad
        B_r(0,\varphi)=0.
\]
Thus $R=0$ is not an absorbing boundary of the diffusion: although the
noise amplitude vanishes there, the disorder-induced drift remains positive
and drives the reflection intensity away from the origin.

It remains to identify the separation between the phase and intensity
scales. For $r>0$, applying It\^o's formula locally to $\log\cR$ gives
\[
        \dd\varphi
        =
        \Im\!\left[
        \frac{b(\cR)}{\cR}
        -\frac{\sigma(\cR)^2}{2\cR^2}
        \right]\dd s
        +
        \Im\!\left[
        \frac{\sigma(\cR)}{\cR}
        \right]\dd W_s.
\]
The explicit coefficients are given in
Appendix~\ref{app:reflection-phase-evolution}. 
For $r$ bounded away from the phase-indeterminate point $r=0$, they have the scale
\[
        \dd\varphi
        =
        \left[
        2k
        +O\!\left(\frac{|\delta E|}{k}\right)
        +O\!\left(\frac{D}{k^2}\right)
        +O\!\left(\frac{\eta}{k}\right)
        \right]\dd s
        +O\!\left(\frac{\sqrt D}{k}\right)\dd W_s.
\]
Thus the phase rotates on the microscopic scale $k^{-1}$, whereas the
intensity changes on the parametrically longer localization and absorption
scales $\ellL$, defined in Eq.~\eqref{eq:main-continuum-scales}, and
$\ell_{\mathrm A}=k/\eta$, respectively.

\subsection{Generic-phase averaging}
\label{subsec:generic-phase-averaging}

We now assume that the wave number $k$ lies in the generic bulk regime and satisfies
\begin{equation}
        k\ellL\gg1,
        \qquad
        kL\gg1,
        \qquad
        \frac{\eta}{E}\ll1,
        \qquad
        \frac{|\delta E|}{E}\ll1,
\label{eq:generic-phase-conditions}
\end{equation}
away from band edges and the the band centre
anomalies. 

Under the conditions \eqref{eq:generic-phase-conditions}, the exact
intensity equation
\eqref{eq:full-intensity-equation-before-average} contains two well
separated longitudinal scales.  The phase $\varphi$ rotates with leading
angular velocity $2k$, whereas the reflection intensity $r$ changes only
on the much longer localization and absorption scales.  Thus, over an
intermediate interval satisfying
\[
        k^{-1}\ll\Delta s\ll\ellL,\ell_{\mathrm A},
\]
the intensity remains approximately fixed while the phase performs many
rotations.

These rapid rotations sample the phase uniformly to leading order.  We therefore define the phase average at fixed $r$ by
\[
        \langle f\rangle_{\varphi}
        =
        \frac{1}{2\pi}
        \int_0^{2\pi}f(r,\varphi)\,\dd\varphi.
\]

The coarse-grained drift is obtained by averaging $A_r$ over the rapid
phase, whereas the effective noise strength is determined by the phase
average of the quadratic variation $B_r^2$.

Using
\[
        \langle\cos\varphi\rangle_\varphi
        =
        \langle\sin\varphi\rangle_\varphi
        =
        \langle\cos(2\varphi)\rangle_\varphi
        =0,
        \qquad
        \langle\sin^2\varphi\rangle_\varphi
        =
        \frac12,
\]
we obtain directly from
Eq.~\eqref{eq:full-intensity-equation-before-average}
\begin{equation}
\begin{aligned}
        \langle A_r\rangle_\varphi
        =
        -\frac{2\eta}{k}r
        +\frac{D}{4k^2}(1-r)^2,
        \qquad 
        \langle B_r^2\rangle_\varphi=\frac{D}{2k^2}r(1-r)^2.
\end{aligned}
\label{eq:phase-averaged-intensity-coefficients}
\end{equation}

In particular, one must average $B_r^2$, rather than the noise amplitude
$B_r$ itself; the latter has zero phase average and would incorrectly
suggest the absence of diffusion.  A nonzero detuning $\delta E$
satisfying Eq.~\eqref{eq:generic-phase-conditions} changes the rapid phase
rotation but does not contribute to the leading phase-averaged evolution
of the reflection intensity.

Replacing in Eq.~\eqref{eq:full-intensity-equation-before-average} the
drift by $\langle A_r\rangle_\varphi$ and the quadratic variation by
$\langle B_r^2\rangle_\varphi$, as given in
Eq.~\eqref{eq:phase-averaged-intensity-coefficients}, yields the
phase-averaged intensity process.  Using
\[
        \frac{D}{4k^2}=\frac{1}{\ellL},
        \qquad
        \frac{\eta}{k}=\frac{1}{\ell_{\mathrm A}},
        \qquad
        \ell_{\mathrm A}=\frac{k}{\eta},
\]
we obtain
\begin{equation}
        \dd r
        =
        \left[
        -\frac{2}{\ell_{\mathrm A}}r
        +\frac{1}{\ellL}(1-r)^2
        \right]\dd s
        +\sqrt{\frac{2}{\ellL}}\,
        \sqrt r(1-r)\,\dd B_s,
\label{eq:radial-sde-dimensional}
\end{equation}
where $B_s$ is an effective standard real Brownian motion. 
In the dimensionless variables
\[
        \tau=\frac{s}{\ellL},
        \qquad
        \alpha=\frac{2\ellL}{\ell_{\mathrm A}}
        =\frac{2\eta}{\GamL},
\]
this gives the phase-averaged reflection-intensity diffusion
\eqref{eq:main-radial-diffusion}.  In
Section~\ref{subsec:x-fokker-planck} we transform this process to the
variable $x=r/(1-r)$ and obtain the corresponding Fokker--Planck, or
forward Kolmogorov, equation
\eqref{eq:main-reflection-intensity-forward}.

The exact closed-end condition is
\[
        \cR(0)=-1,
        \qquad
        r(0)=1,
        \qquad
        \varphi(0)=\pi.
\]
The clean phase rotation moves the process away from this special phase over
a microscopic distance $O(k^{-1})$. During this initial stage the relative change of $r$ is only of order
$1/(k\ellL)$ or $\eta/E$.   The phase-averaged
process therefore starts from
\[
        r(0)=1,
        \qquad
        x(0)=+\infty,
        \qquad
        x=\frac{r}{1-r}.
\]
For $\alpha>0$, the propagator corresponding to the closed-end initial condition is understood as the limit
$x_0\to+\infty$, see Eq.~\eqref{closed-end}. The physical forward solution has vanishing probability current at
$x=0$ and as $x\to\infty$.  At $\alpha=0$, a trajectory starting at $r=1$ (equivalently, $x=\infty$)
remains there under the stochastic evolution, as required by flux
conservation at real energy. 

Corrections to the uniform-phase approximation are controlled by
\[
        \frac{1}{k\ellL},
        \qquad
        \frac{\eta}{E},
        \qquad
        \frac{|\delta E|}{E}.
\]
The additional condition $kL\gg1$ ensures that the sample contains many wave
oscillations, so that the influence of the initial boundary condition is
confined to a microscopically short interval. At this stage, no stationary, large-$L$, or small-$\alpha$
asymptotic limit has yet been taken.

\section{Reflection-intensity evolution and interface logarithmic observables}
\label{sec:radial}

The exact finite-chain Poincar\'e--Lelong identity
\eqref{eq:main-finite-poincare-lelong} contains both
$\partial_E^2$ and $\partial_\eta^2$. The relative size of the terms generated by these derivatives must be checked
because the dimensionless continuum variables depend on $E=k^2$.  At fixed
$\eta$, $L$, and $D$,
\[
        \alpha=\frac{8k\eta}{D},
        \qquad
        \tau=\frac{LD}{4k^2},
\]
and hence, for any reduced observable $F(\alpha,\tau)$ after introducing $\mathcal D_E
        :=\frac{\alpha}{2}\partial_\alpha-\tau\partial_\tau$ we have 
\begin{equation}
        \partial_E F
        =\frac1E\mathcal D_EF,
        \quad
        \partial_E^2F
        =\frac1{E^2}(\mathcal D_E^2-\mathcal D_E)F.
\label{eq:energy-derivative-chain-rule}
\end{equation}
By contrast, at fixed $E$,
\begin{equation}
        \partial_\eta^2F
        =\left(\frac{\alpha}{\eta}\right)^2
        \partial_\alpha^2F.
\label{eq:absorption-derivative-chain-rule}
\end{equation}
Thus the real-energy term is suppressed by $(\eta/E)^2$, multiplied by a
dimensionless ratio involving both $\alpha$- and $\tau$-derivatives.  In the
stationary regime the $\tau$ derivatives vanish.  In the moving-front window
they generate only polynomial powers of $\tau$, while
$\Gamma/E=(k\ellL)^{-1}\exp[-\tau+O(\sqrt\tau)]$ is exponentially small.
The $\partial_E^2$ contribution is therefore parametrically subleading in the
stationary and ultranarrow resonance regimes retained below.  The full Laplacian remains
the exact identity; retaining only the absorption derivative is an additional
generic-bulk approximation justified by the scale separation
$\eta\sim\Gamma\ll E$.

The phase-averaged diffusion of Section~\ref{sec:diffusion}  supplies the
bulk propagator independently of the contact transparency at the open end
of the sample. The opening is encoded separately by
the interface function \eqref{eq:main-interface-observable}. 
Thus the same bulk propagator $h_\alpha(x,\tau\mid\infty)$ is combined with the
transparency-dependent observable $f_{\cT}(x)$. 
We first establish this separation between the bulk evolution and the dependence
on the contact transparency, and then analyse perfect coupling, for which the
contact value is $x=0$ and the boundary-flux identity takes its simplest form.
The transparency-dependent interface relation and Jensen-formula calculation are derived
in Subsection~\ref{subsec:regular-contact-main}.  The exact finite-$\alpha$
endpoint resolvent is constructed in Appendix~\ref{app:heun}, and its
small-$\alpha$ matching is performed in
Section~\ref{sec:perfect-ultranarrow}.

\subsection{Fokker--Planck evolution, closed-end limit, and stationary law}
\label{subsec:x-fokker-planck}

We use the reflection-intensity variable $x=r/(1-r)$ defined in
Eq.~\eqref{eq:main-r-x-definition}.  The closed end $r=1$ corresponds to
$x=+\infty$, while $x=0$ describes zero reflection.  Applying It\^o's
formula to Eq.~\eqref{eq:main-radial-diffusion}, as detailed in
Appendix~\ref{app:complex-Ito-diffusions}, gives
\begin{equation}
        \dd x
        =
        \bigl[1+2x-\alpha x(1+x)\bigr]\dd\tau
        +\sqrt{2x(1+x)}\,\dd B_\tau.
\label{eq:x-radial-sde-main}
\end{equation}
The transition density $p_\alpha(x,\tau\mid x_0)$ therefore obeys the
Fokker--Planck equation
\eqref{eq:main-reflection-intensity-forward}.  Equivalently,
\begin{equation}\label{probcurrent}
        \partial_\tau p_\alpha=-\partial_xJ_\alpha, \qquad J_\alpha
        =-x(1+x)\bigl(\partial_xp_\alpha+\alpha p_\alpha\bigr),
\end{equation}
and the physical solution has vanishing probability current at both
endpoints.

The closed-end condition $x_0=+\infty$ has a direct local meaning.  It is
convenient to introduce the reflection unitarity deficit
\[
        y=1-r=\frac{1}{1+x}.
\]
We derive its evolution directly from the reflection-intensity diffusion
\eqref{eq:main-radial-diffusion}.  Since $y=1-r$ is a linear function of
$r$, It\^o's formula (see Appendix \ref{Ito}) contains no second-derivative correction and simply
gives $\dd y=-\dd r$. Using
\[
        r=1-y, \qquad (1-r)^2-\alpha r = y^2-\alpha(1-y),
\]
and
\[
        \sqrt{2r}\,(1-r) = \sqrt{2(1-y)}\,y,
\]
we obtain
\begin{equation}
        \dd y
        =
        \bigl[\alpha(1-y)-y^2\bigr]\dd\tau
        -\sqrt{2(1-y)}\,y\,\dd B_\tau.
\label{eq:closed-end-y-main}
\end{equation}

For $\alpha>0$, the drift at $y=0$ equals $\alpha$ while the noise
coefficient vanishes.  The process therefore enters the interval $y>0$
immediately, and to leading order
\[
        y(\tau)=\alpha\tau+o_{\mathbb P}(\tau),
        \qquad
        x(\tau)\sim\frac{1}{\alpha\tau},
        \qquad
        \tau\to0^+.
\]
This is the physical meaning of the limit $x_0\to+\infty$.  At $\alpha=0$, both the drift and noise coefficients in
Eq.~\eqref{eq:closed-end-y-main} vanish at $y=0$, so the real-energy process
remains at unit reflection.

A stationary density must have zero current and hence according to Eq.~\eqref{probcurrent} satisfies
\[
        \partial_x\pi_\alpha+\alpha\pi_\alpha=0.
\]
Normalization on $x\geq0$ gives
\[
        \pi_\alpha(x)=\alpha\e^{-\alpha x},
\]
in agreement with Eq.~\eqref{eq:main-stationary-pi}.  In terms of the
reflection coefficient $r=x/(1+x)$, this becomes
\begin{equation}
        P_\alpha^{\rm st}(r)
        =
        \frac{\alpha}{(1-r)^2}
        \exp\!\left[-\frac{\alpha r}{1-r}\right],
        \qquad 0\leq r<1.
\label{eq:stationary-r-main}
\end{equation}
reproducing the solution first obtained in \cite{FreilikherPustilnikYurkevich1994} and \cite{PradhanKumar1994}.

For finite propagation length the same change of variables gives the full
reflection-intensity density, within the phase-averaged continuum theory,
\begin{equation}
        P_\alpha(r,\tau\mid\infty)
        =\frac{\alpha}{(1-r)^2}
        \exp\!\left[-\frac{\alpha r}{1-r}\right]
        h_\alpha\!\left(
        \frac{r}{1-r},\tau\mid\infty
        \right),
        \qquad 0\leq r<1.
\label{eq:finite-length-reflection-density-main}
\end{equation}
Equation~\eqref{eq:finite-length-reflection-density-main} is exact once the
phase-averaged diffusion has been reached.  Its stationary limit is
Eq.~\eqref{eq:stationary-r-main}; later we use the ultranarrow spectral
analysis to make its small-$\alpha$, long-sample outer tail explicit.

We factor the transition density as
$p_\alpha=\pi_\alpha h_\alpha$, as in
Eq.~\eqref{eq:main-p-h-factorization}.  With this convention
$h_\alpha\to1$ as $\tau\to\infty$, and the physical closed-end propagator
$h_\alpha(x,\tau\mid\infty)$ is defined by the limit
$x_0\to+\infty$ at fixed $\tau>0$.

\subsection{Perfect coupling: regularized endpoint boundary-flux identity}
\label{subsec:boundary-flux-endpoint-main}

For $\cT=1$, write
\[
        \mathcal F_\alpha(\tau)
        =\mathcal F_{\alpha,1}(\tau)
        =\EE_\infty f_1(x(\tau)),
        \qquad
        f_1(x)=\log\frac{x}{1+x}.
\]
The closed-end condition gives $\mathcal F_\alpha(0)=0$.  Since $f_1$ is
logarithmically singular at $x=0$, we regularize it before integrating by
parts:
\[
        f_{1,\epsilon}(x)
        =
        \log\frac{x+\epsilon}{1+x+\epsilon},
        \qquad
        \epsilon>0,
\]
and define
\[
        \mathcal F_{\alpha,\epsilon}(\tau)
        =
        \int_0^\infty
        p_\alpha(x,\tau\mid\infty)f_{1,\epsilon}(x)\,\dd x.
\]
Using the conservative form of the Fokker--Planck equation in  Eq.~\eqref{probcurrent}  and the
vanishing current at the endpoints gives
\[
\begin{aligned}
        \frac{\dd\mathcal F_{\alpha,\epsilon}}{\dd\tau}
        &=-\int_0^\infty
        f_{1,\epsilon}'(x)x(1+x)
        \bigl(\partial_xp_\alpha+\alpha p_\alpha\bigr)\,\dd x,
\end{aligned}
\]
where
\[
        f_{1,\epsilon}'(x)x(1+x)
        =
        \frac{x(1+x)}{(x+\epsilon)(1+x+\epsilon)}.
\]

This factor lies between zero and one and tends to unity for every $x>0$
as $\epsilon\to0^+$.  For every fixed $\tau>0$, the physical transition density
$p_\alpha(x,\tau\mid\infty)$ is regular at $x=0$ and decays sufficiently
rapidly as $x\to\infty$ for the integrations by parts below to be justified,
with no boundary contribution from $\partial_x p_\alpha+\alpha p_\alpha$
at infinity.
  The endpoint behaviour of the propagator
therefore allows the limit to be taken under the integral, yielding
\[
        \frac{\dd\mathcal F_\alpha}{\dd\tau} =-\int_0^\infty
        \bigl(\partial_xp_\alpha+\alpha p_\alpha\bigr)\,\dd x =p_\alpha(0,\tau\mid\infty)-\alpha.
\]
Since
$p_\alpha(0,\tau\mid\infty)=\pi_\alpha(0)
 h_\alpha(0,\tau\mid\infty)=\alpha h_\alpha(0,\tau\mid\infty)$,
we recover the endpoint boundary-flux identity
\eqref{eq:principal-endpoint-identity}.  Integrating it from $0$ to
$\tau$ gives
\begin{equation}
        \mathcal F_\alpha(\tau)
        =
        \alpha\int_0^\tau
        \bigl[h_\alpha(0,u\mid\infty)-1\bigr] \,\dd u.
\label{eq:logarithmic-endpoint-integrated-main}
\end{equation}

\subsection{Perfect coupling: finite-length and stationary densities}
\label{subsec:finite-stationary-density-main}

Combining Eq.~\eqref{eq:principal-general-contact-density} at $\cT=1$
with Eq.~\eqref{eq:logarithmic-endpoint-integrated-main} gives the
finite-length density formula
\eqref{eq:principal-finite-length-density}.  Expanding the two derivatives
with respect to $\alpha$ gives the equivalent form
\begin{equation}
\begin{aligned}
        \rho(E,\Gamma;L)
        &\simeq
        \left.
        \frac{1}{\pi\GamL^2}
        \int_0^{L/\ellL}
        \left[
        2\,\partial_\alpha h_\alpha(0,u\mid\infty)
        +\alpha\,\partial_\alpha^2h_\alpha(0,u\mid\infty)
        \right]\dd u
        \right|_{\alpha=2\Gamma/\GamL}.
\end{aligned}
\label{eq:finite-length-density-expanded-main}
\end{equation}
This form is well suited for the spectral analysis of
Section~\ref{subsec:radial-spectral-main}, where the endpoint value
$h_\alpha(0,u\mid\infty)$ is obtained from the mode expansion of the
reduced propagator.

For an infinitely long sample, the logarithmic observable is averaged with
respect to the stationary density given in Eq.~\eqref{eq:main-stationary-pi}:
\[
\begin{aligned}
        \mathcal F_\alpha^{(\infty)}
        &=
        \alpha\int_0^\infty
        \e^{-\alpha x}\log\frac{x}{1+x}\,\dd x
        \\
        &=
        \alpha\int_0^\infty\e^{-\alpha x}\log x\,\dd x
        -
        \alpha\int_0^\infty\e^{-\alpha x}\log(1+x)\,\dd x.
\end{aligned}
\]
The first integral is
\[
        \alpha\int_0^\infty\e^{-\alpha x}\log x\,\dd x
        =-\gamma_{\rm E}-\log\alpha,
\]
while integration by parts in the second gives
\[
        \alpha\int_0^\infty
        \e^{-\alpha x}\log(1+x)\,\dd x
        =
        \int_0^\infty\frac{\e^{-\alpha x}}{1+x}\,\dd x
        =\e^\alpha E_1(\alpha),
\]
where $E_1$ is defined in Eq.~\eqref{eq:main-E1}.  Hence
\begin{equation}
        \mathcal F_\alpha^{(\infty)}
        =
        -\gamma_{\rm E}-\log\alpha-\e^\alpha E_1(\alpha).
\label{eq:stationary-logarithmic-average-main}
\end{equation}
Using $\frac{\dd E_1(\alpha)}{\dd\alpha}=-\frac{\e^{-\alpha}}{\alpha}$
we obtain
\[
        \frac{\dd\mathcal F_\alpha^{(\infty)}}{\dd\alpha}
        =-\e^\alpha E_1(\alpha),
        \qquad
        \frac{\dd^2\mathcal F_\alpha^{(\infty)}}{\dd\alpha^2}
        =\frac{1-\alpha\e^\alpha E_1(\alpha)}{\alpha}.
\]
Substitution into Eq.~\eqref{eq:principal-general-contact-density}, with
$\alpha=2\Gamma/\GamL$ and $\GamL=D/(4k)$, gives the resonance density in the semi-infinite sample, 
\eqref{eq:semi_infinite_density_main}. The factor $1-\alpha\e^\alpha E_1(\alpha)$
is the stationary mean reflected intensity, $1-\alpha\e^\alpha E_1(\alpha) =
        \langle r\rangle_{\rm st}$.
Equivalently,
\[
        \left\langle\frac{1}{1+x}\right\rangle_{\rm st}
        =
        \alpha\e^\alpha E_1(\alpha).
\]

For $\alpha\to0$,
\[
        1-\alpha\e^\alpha E_1(\alpha)
        =
        1-\alpha\left(
        \log\frac{1}{\alpha}-\gamma_{\rm E}
        \right)
        +O\!\left(\alpha^2\log\frac{1}{\alpha}\right),
\]
whereas for $\alpha\to\infty$,
\[
        1-\alpha\e^\alpha E_1(\alpha)
        =
        \frac{1}{\alpha}-\frac{2}{\alpha^2}
        +\frac{6}{\alpha^3}+O(\alpha^{-4}).
\]
These expansions reproduce
\[
        \rho^{(\infty)}(E,\Gamma)
        \sim\frac{2k}{\pi D\Gamma},
        \qquad \Gamma\ll\GamL,
\]
and
\[
        \rho^{(\infty)}(E,\Gamma)
        \sim\frac{1}{4\pi\Gamma^2},
        \qquad \GamL\ll\Gamma\ll E.
\]
The two power laws have different physical origins.  The
$\Gamma^{-1}$ narrow-resonance tail is a standard consequence of exponential
localization.  Indeed, a state localized at a distance $x$ from the opening
has, parametrically, $\Gamma(x)\propto \e^{-2x/\ellL}$. 
Since the localization centres are approximately uniformly distributed along
a long sample, $\dd x\propto\dd\Gamma/\Gamma$, which directly produces the
inverse-width law.   The limit
$\Gamma\ll\GamL$ in the first formula is taken after the long-sample limit and
is therefore not uniform in the finite-length ultranarrow-width regime.

The $\Gamma^{-2}$ broad-resonance tail was obtained explicitly for the
strictly one-dimensional pure-reflection problem in
Ref.~\cite{FyodorovMeibohm2026}.  Its origin is, however, more general and admits a simple general physical
interpretation.  For an ideally coupled opening, escape on sufficiently short
time scales is not suppressed by an interface barrier.  If the corresponding
escape-time density is regular, $P_{\rm esc}(t)\to{\rm const}$, then the
inverse relation $\Gamma\sim\hbar/t$ gives
\[
        P(\Gamma)
        =
        P_{\rm esc}(\hbar/\Gamma)\frac{\hbar}{\Gamma^2}
        \propto \Gamma^{-2}.
\]
This argument was used in the random-matrix theory of chaotic scattering to
explain the same inverse-square law \cite{FyodorovSommers1997}.  In spatially disordered systems the
broad resonances are associated with states concentrated close to the
opening, so that their decay probes predominantly this local escape physics
rather than transport through the bulk.  Perfect coupling is essential for
the asymptotic power law: incomplete coupling introduces an additional
suppression of sufficiently broad resonances,  see
Refs.~\cite{FyodorovSkvortsovTikhonov2024,FyodorovMeibohm2026} for further
discussion.

\subsection{Bulk spectral problem and supersymmetric partner}
\label{subsec:radial-spectral-main}

The finite-length density requires the reduced propagator rather than only
its stationary limit.  Substituting $p_\alpha=\pi_\alpha h_\alpha$ into
Eq.~\eqref{eq:main-reflection-intensity-forward} gives
\begin{equation}
        \partial_\tau h_\alpha=\cL_\alpha h_\alpha,
        \qquad
        \cL_\alpha
        =\frac{1}{\pi_\alpha(x)}
        \partial_x\!\left[
        x(1+x)\pi_\alpha(x)\partial_x
        \right].
\label{eq:h-evolution-main}
\end{equation}
For finite $x_0$ the normalization and initial condition are
\[
        \int_0^\infty
        h_\alpha(x,\tau\mid x_0)\pi_\alpha(x)\,\dd x=1,
        \qquad
        h_\alpha(x,0\mid x_0)
        =\frac{\delta(x-x_0)}{\pi_\alpha(x_0)}.
\]
The physical closed-end propagator is obtained, at every fixed $\tau>0$, by
\[
        h_\alpha(x,\tau\mid\infty)
        =\lim_{x_0\to\infty}h_\alpha(x,\tau\mid x_0).
\]

Under the zero-current boundary conditions, $\cL_\alpha$ is symmetric with
respect to the stationary measure $\pi_\alpha(x)\,\dd x$.  Indeed,
\[
\begin{aligned}
        \int_0^\infty\overline f\,\cL_\alpha g\,\pi_\alpha\,\dd x
        =-\int_0^\infty
        x(1+x)\pi_\alpha(x)\overline{f'(x)}g'(x)\,\dd x
        =\int_0^\infty
        \overline{\cL_\alpha f}\,g\,\pi_\alpha\,\dd x,
\end{aligned}
\]
where the boundary terms vanish.  Separating the stationary mode
$\phi_0(x)=1$ and choosing the remaining eigenmodes orthonormal with respect
to $\pi_\alpha(x)\,\dd x=\alpha e^{-\alpha x}\,\dd x$, one has
\begin{equation}
        h_\alpha(x,\tau\mid x_0)
        =1+\sum_{n\geq1}
        e^{-\lambda_n\tau}\phi_n(x)\phi_n(x_0),
        \qquad
        -\cL_\alpha\phi_n=\lambda_n\phi_n.
\label{eq:bulk-spectral-expansion-main}
\end{equation}
The bulk spectrum depends on the disorder and on absorption, but not on the
contact transparency.

For later use it is helpful to display the finite-$\alpha$ eigenvalue
equation explicitly.  From Eq.~\eqref{eq:h-evolution-main},
\begin{equation}
        x(1+x)\phi_\lambda''(x)
        +\left[1+2x-\alpha x(1+x)\right]\phi_\lambda'(x)
        +\lambda\phi_\lambda(x)=0.
\label{eq:ultranarrow-finite-alpha-eigenvalue}
\end{equation}
This equation is the exact starting point of the spectral analysis.  The
confluent-Heun form and the precise self-adjoint boundary conditions are
collected in Appendix~\ref{app:heun}.

For the small-absorption analysis, introduce the hyperbolic coordinate
\begin{equation}
        x=\sinh^2\frac{\vartheta}{2},
        \qquad 0\leq\vartheta<\infty.
\label{eq:hyperbolic-coordinate-main}
\end{equation}
Then
\[
        \cL_\alpha
        =\partial_\vartheta^2
        +b_\alpha(\vartheta)\partial_\vartheta,
        \qquad
        b_\alpha(\vartheta)
        =\coth\vartheta-\frac{\alpha}{2}\sinh\vartheta.
\]
The stationary measure becomes
\begin{equation}
        \pi_\alpha(x)\,\dd x
        =m_\alpha(\vartheta)\,\dd\vartheta,
        \qquad
        m_\alpha(\vartheta)
        =\frac{\alpha}{2}\sinh\vartheta
        \exp\!\left[-\frac{\alpha}{2}(\cosh\vartheta-1)\right],
\label{eq:hyperbolic-invariant-weight}
\end{equation}
and
\[
        \frac{\dd}{\dd\vartheta}\log m_\alpha(\vartheta)
        =b_\alpha(\vartheta),
        \qquad
        \cL_\alpha
        =\frac1{m_\alpha}\partial_\vartheta
        \left(m_\alpha\partial_\vartheta\right).
\]
For an eigenmode, define
\[
        \Psi_\lambda(\vartheta)
        =\sqrt{m_\alpha(\vartheta)}\,\phi_\lambda(\vartheta).
\]
A direct differentiation converts the eigenvalue equation to
\begin{equation}
        \mathcal H_-\Psi_\lambda
        :=\left[-\partial_\vartheta^2+U_{-,\alpha}(\vartheta)\right]
        \Psi_\lambda
        =\lambda\Psi_\lambda,
\label{eq:radial-schrodinger-main}
\end{equation}
with
\begin{equation}
        U_{-,\alpha}(\vartheta)
        =\frac{\alpha^2}{16}\sinh^2\vartheta
        -\frac{\alpha}{2}\cosh\vartheta
        +\frac14-\frac{1}{4\sinh^2\vartheta}.
\label{eq:radial-potential-main}
\end{equation}
The subscript ``$-$'' anticipates a factorization that will be useful for the
finite-length front.

Indeed, define
\begin{equation}
\begin{aligned}
        Q_\alpha
        &=\partial_\vartheta-\frac12 b_\alpha(\vartheta),
        &
        Q_\alpha^\dagger
        &=-\partial_\vartheta-\frac12 b_\alpha(\vartheta).
\end{aligned}
\label{eq:susy-Q-main}
\end{equation}
Since
$U_{-,\alpha}=\frac12 b_\alpha'+\frac14b_\alpha^2$, the radial Hamiltonian
factorizes exactly as
\begin{equation}
        \mathcal H_-=Q_\alpha^\dagger Q_\alpha.
\label{eq:susy-factorization-main}
\end{equation}
This is the standard supersymmetric, or Darboux, factorization associated
with the ground-state transformation of a reversible diffusion \cite{Risken1996,CooperKhareSukhatme1995}.  Its partner
is
\begin{equation}
\begin{aligned}
        \mathcal H_+
        :=Q_\alpha Q_\alpha^\dagger
        =-\partial_\vartheta^2+U_{+,\alpha}(\vartheta),
        \qquad
        U_{+,\alpha}(\vartheta)
        =\frac14
        +\frac{3}{4\sinh^2\vartheta}
        +\frac{\alpha^2}{16}\sinh^2\vartheta.
\end{aligned}
\label{eq:partner-hamiltonian-main}
\end{equation}
The useful simplification is visible directly in
Eq.~\eqref{eq:partner-hamiltonian-main}: the term linear in
$\alpha\cosh\vartheta$ has cancelled.

The factorization also has a transparent meaning for the diffusion modes.
Since
$\partial_\vartheta\log\sqrt{m_\alpha}=b_\alpha/2$,
\begin{equation}
        Q_\alpha\!\left[\sqrt{m_\alpha}\,\phi\right]
        =\sqrt{m_\alpha}\,\partial_\vartheta\phi.
\label{eq:susy-derivative-identity-main}
\end{equation}
Thus the partner problem describes the derivative, or current, sector of the
radial diffusion.  In particular,
\[
        \Psi_0=\sqrt{m_\alpha},
        \qquad
        Q_\alpha\Psi_0=0,
\]
so the stationary mode is removed, whereas $\mathcal H_-$ and $\mathcal H_+$ share the same positive eigenvalues.

The boundary structure is also simpler in the partner problem.  The physical
solution of $\mathcal H_-$ behaves as $\vartheta^{1/2}$ at the origin; the logarithmic
companion is excluded by the zero-current condition of the underlying
$x$-space diffusion.  By contrast,
\[
        U_{+,\alpha}(\vartheta)
        =\frac{3}{4\vartheta^2}+O(1),
        \qquad \vartheta\to0,
\]
so the two possible solutions behave as $\vartheta^{3/2}$ and
$\vartheta^{-1/2}$, and only the first is square integrable.  At large
$\vartheta$ the positive $\alpha^2\sinh^2\vartheta/16$ term confines both
problems.  Consequently, for every fixed $\alpha>0$ the spectrum is discrete
and simple.  Moreover, since
\[
        \mathcal H_+-\frac14
        =-\partial_\vartheta^2
        +\frac{3}{4\sinh^2\vartheta}
        +\frac{\alpha^2}{16}\sinh^2\vartheta
\]
has a strictly positive quadratic form on every nonzero admissible state,
\begin{equation}
        0=\lambda_0(\alpha)
        <\frac14
        <\lambda_1(\alpha)<\lambda_2(\alpha)<\cdots,
        \qquad \alpha>0.
\label{eq:heun-discrete-spectrum}
\end{equation}
The domain statement and the strict inequality are proved in
Appendix~\ref{app:heun-self-adjoint}.

Finally, the continuum onto which the excited levels condense is already
visible by setting $\alpha=0$ before the distant wall is reached.  Equation
\eqref{eq:ultranarrow-finite-alpha-eigenvalue} then becomes
\begin{equation}
        x(1+x)\phi_\lambda''(x)
        +(1+2x)\phi_\lambda'(x)
        +\lambda\phi_\lambda(x)=0,
\label{eq:ultranarrow-wall-free-eigenvalue}
\end{equation}
or, with $1+2x=\cosh\vartheta$,
\begin{equation}
        \left[
        \partial_\vartheta^2
        +\coth\vartheta\,\partial_\vartheta
        +\lambda
        \right]\phi(\vartheta)=0.
\label{eq:Legendre-radial-equation}
\end{equation}
Writing $\lambda=\frac14+q^2,\, q\geq0$, the solution regular at $\vartheta=0$ and normalized by $\phi_q(0)=1$ is
\begin{equation}
        \phi_q^{\rm in}(\vartheta)
        =P_{-\frac12+\ii q}(\cosh\vartheta).
\label{eq:Legendre-regular-mode}
\end{equation}
Hence the wall-free continuum begins at $\lambda=1/4$.  As
$\alpha\to0$, the excited levels of Eq.~\eqref{eq:heun-discrete-spectrum}
condense onto $[1/4,\infty)$, while the stationary state at $\lambda=0$
remains isolated.  The next section determines this condensation by matching
the real continuum modes to the distant absorption wall.  The matching is
performed in the partner problem, where the wall equation is particularly
simple.

\section{Ultranarrow-resonance front for a perfectly matched contact}
\label{sec:perfect-ultranarrow}

The ultranarrow resonance front is a finite-length effect and therefore
cannot be obtained from the stationary small-width density alone.  At small
positive $\alpha$, the excited spectrum is discretized by an absorption wall
at logarithmic distance $\vartheta\sim\log(1/\alpha)$.  The real level
positions and their spectral weights retain the information about that
finite distance, and their Poisson reconstruction produces the moving front.

The relevant boundary value of the reduced propagator is
\begin{equation}
        H_\alpha(\tau)
        :=h_\alpha(0,\tau\mid\infty).
\label{eq:ultranarrow-endpoint-propagator}
\end{equation}
 The factorization of the preceding subsection identifies a convenient
representation for the matching. Since $Q_\alpha$ annihilates the stationary
state while preserving all positive eigenvalues, we first reconstruct
\begin{equation}
        K_\alpha(\tau):=\partial_\tau H_\alpha(\tau)
\label{eq:partner-K-definition}
\end{equation}
from the positive spectrum of $\mathcal H_+$. After the real-$q$ matching and Poisson
summation are completed, one integration in $\tau$ recovers $H_\alpha$.
The advantage of this representation is that the matching near the absorbing
boundary reduces to a modified-Bessel equation.

\subsection{Real-$q$ matching to the absorption wall}
\label{subsec:ultranarrow-spectrum-opening}
\label{subsec:ultranarrow-partner-matching}

We begin with the wall-free mode
Eq.~\eqref{eq:Legendre-regular-mode}.  For $1\ll\vartheta$, its two-wave
expansion is
\begin{equation}
\begin{aligned}
        P_{-\frac12+\ii q}(\cosh\vartheta)
        &=B_+(q)e^{(-\frac12+\ii q)\vartheta}
        +B_-(q)e^{(-\frac12-\ii q)\vartheta}
        +O(e^{-3\vartheta/2}),
\end{aligned}
\label{eq:Legendre-large-theta}
\end{equation}
where
\begin{equation}
        B_+(q)=\frac{\Gamma(\ii q)}
        {\sqrt\pi\,\Gamma(\frac12+\ii q)},
        \qquad
        B_-(q)=\frac{\Gamma(-\ii q)}
        {\sqrt\pi\,\Gamma(\frac12-\ii q)}.
\label{eq:Legendre-scattering-coefficients}
\end{equation}
These coefficients follow from the hypergeometric representation of the
Legendre function and its large-argument connection formula; see
Ref.~\cite{NISTHandbook}, Secs.~14.3 and 15.8.  For real $q$,
$B_-(q)=\overline{B_+(q)}$.  At $q=0$ the two powers coalesce, so the
threshold limit must be taken only after the two terms are combined.  The
quantized levels below have $q_n>0$, and the continuation used for the front
is performed only after the real-spectrum matching has been completed.

For a positive eigenvalue $\lambda=1/4+q^2$, normalize the partner
eigenfunction by
\begin{equation}
        \widetilde\Psi_q
        :=\frac1\lambda Q_\alpha\Psi_q.
\label{eq:partner-eigenfunction-normalization}
\end{equation}
The factor $1/\lambda$ is convenient because it makes the Wronskian relation
used below especially simple.  In the overlap region, where absorption is
still weak, Eq.~\eqref{eq:susy-derivative-identity-main} and
Eq.~\eqref{eq:Legendre-large-theta} give
\begin{equation}
\begin{aligned}
        \widetilde\Psi_q^{\rm in}
        &\simeq
        e^{\alpha/4-T_0/2}
        \left[
        C_+(q)e^{\ii qT_0}\xi^{\ii q}
        +C_-(q)e^{-\ii qT_0}\xi^{-\ii q}
        \right],
\end{aligned}
\label{eq:Legendre-inner-overlap}
\end{equation}
where
\begin{equation}
        C_+(q)
        =-\frac{B_+(q)}{\frac12+\ii q}
        =-\frac{\Gamma(\ii q)}
        {\sqrt\pi\,\Gamma(\frac32+\ii q)},
        \qquad
        C_-(q)
        =-\frac{B_-(q)}{\frac12-\ii q}
        =-\frac{\Gamma(-\ii q)}
        {\sqrt\pi\,\Gamma(\frac32-\ii q)}.
\label{eq:partner-inner-coefficients}
\end{equation}
Here we have introduced the wall coordinate
\begin{equation}
        \xi=e^{\vartheta-T_0},
        \qquad
        T_0=\log\frac4\alpha.
\label{eq:partner-wall-coordinate}
\end{equation}
Indeed, $\alpha e^\vartheta=O(1)$ when $\vartheta=T_0+O(1)$, so
$\xi=O(1)$ describes the vicinity of the absorption wall, while
$1\ll\vartheta\ll T_0$ is the common matching region.

The reason for using the partner Hamiltonian $\mathcal H_+$ becomes especially clear
in the wall region. From Eq.~\eqref{eq:partner-hamiltonian-main},
\begin{equation}
        U_{+,\alpha}(\vartheta)
        =\frac14+\frac{\xi^2}{4}
        +O(e^{-2\vartheta})+O(\alpha^2)
\label{eq:partner-wall-potential}
\end{equation}
in that region.  Therefore, at $\lambda=1/4+q^2$,
\begin{equation}
        \frac{\dd^2\widetilde\Psi}{\dd\vartheta^2}
        +\left[q^2-\frac{\xi^2}{4}\right]\widetilde\Psi=0,
\label{eq:partner-wall-equation-theta}
\end{equation}
or, since $\partial_\vartheta=\xi\partial_\xi$,
\begin{equation}
        \xi^2\frac{\dd^2\widetilde\Psi}{\dd\xi^2}
        +\xi\frac{\dd\widetilde\Psi}{\dd\xi}
        +\left[q^2-\frac{\xi^2}{4}\right]\widetilde\Psi=0.
\label{eq:partner-wall-equation-xi}
\end{equation}
This is the modified-Bessel equation. The solution decaying in the strongly absorbing region beyond the wall
is normalized as
\begin{equation}
        \widetilde\Psi_q^{\rm out}(\vartheta)
        =\frac{e^{\alpha/4}}{\sqrt\pi}
        K_{\ii q}\!\left(\frac\xi2\right).
\label{eq:partner-Bessel-decaying-solution}
\end{equation}
For $\xi\ll1$,
\begin{equation}
        \widetilde\Psi_q^{\rm out}
        \simeq e^{\alpha/4}
        \left[D_+(q)\xi^{\ii q}+D_-(q)\xi^{-\ii q}\right],
\label{eq:partner-Bessel-small-xi}
\end{equation}
with
\begin{equation}
        D_+(q)=\frac{2^{-1-2\ii q}\Gamma(-\ii q)}{\sqrt\pi},
        \qquad
        D_-(q)=\frac{2^{-1+2\ii q}\Gamma(\ii q)}{\sqrt\pi}.
\label{eq:partner-Bessel-coefficients}
\end{equation}
For real $q$, both $C_-=\overline{C_+}$ and
$D_-=\overline{D_+}$.

The inner and wall solutions represent the same eigenmode precisely when the
ratios of their two overlap coefficients agree:
\begin{equation}
        \frac{D_+(q)}{D_-(q)}
        =\frac{C_+(q)}{C_-(q)}e^{2\ii qT_0}.
\label{eq:partner-quantization-ratio}
\end{equation}
Define
\begin{equation}
        \widetilde{\mathcal M}(q):=C_+(q)D_-(q)
        =|\widetilde{\mathcal M}(q)|e^{\ii\delta(q)},
        \qquad \delta(0)=0,
\label{eq:partner-matching-product}
\end{equation}
and
\begin{equation}
        \Phi_\alpha(q):=qT_0+\delta(q).
\label{eq:ultranarrow-total-phase}
\end{equation}
Then the leading real-$q$ quantization condition is
\begin{equation}
        \Phi_\alpha(q_n)=\pi n,
        \qquad n=1,2,\ldots .
\label{eq:ultranarrow-phase-quantization-leading}
\end{equation}
Thus the distant wall converts the continuum parameter $q$ into a closely
spaced sequence of positive real values $q_n$.  These are exactly the
positive eigenvalues of the finite-$\alpha$ radial problem, written as
$\lambda_n=1/4+q_n^2$.

\subsection{Quantized levels and partner spectral residues}
\label{subsec:ultranarrow-quantized-levels}

Setting $x=0$ and $x_0=\infty$ in the exact spectral expansion gives
\begin{equation}
        H_\alpha(\tau)
        =1+\sum_{n\geq1}
        \frac{e^{-\lambda_n\tau}}
        {\partial_\lambda\mathscr W_\alpha(\lambda_n)}.
\label{eq:ultranarrow-exact-spectral-sum}
\end{equation}
Here $\mathscr W_\alpha$ is the exact weighted Wronskian of the regular and
decaying diffusion-sector solutions; the above identity is
proved in Appendix~\ref{app:heun-wronskian}.
Differentiating Eq.~\eqref{eq:ultranarrow-exact-spectral-sum} gives
\begin{equation}
        K_\alpha(\tau)
        =-\sum_{n\geq1}
        \frac{\lambda_n e^{-\lambda_n\tau}}
        {\partial_\lambda\mathscr W_\alpha(\lambda_n)}.
\label{eq:K-diffusion-spectral-sum}
\end{equation}
The stationary mode contribution has disappeared identically.

To express the remaining weights directly in the partner problem, let
$\Psi_u$ and $\Psi_v$ be the Schr\"odinger representatives of the regular
and decaying solutions and put
\[
        \widetilde\Psi_u=\frac1\lambda Q_\alpha\Psi_u,
        \qquad
        \widetilde\Psi_v=\frac1\lambda Q_\alpha\Psi_v.
\]
For any two solutions $f,g$ of $\mathcal H_-f=\lambda f$ and $\mathcal H_-g=\lambda g$,
straightforward differentiation gives
\[
        W[Q_\alpha f,Q_\alpha g]=\lambda W[f,g].
\]
Therefore
\begin{equation}
        \widetilde{\mathscr W}_\alpha(\lambda)
        :=W[\widetilde\Psi_u,\widetilde\Psi_v]
        =\frac{\mathscr W_\alpha(\lambda)}{\lambda},
        \qquad \lambda\neq0.
\label{eq:partner-Wronskian-relation}
\end{equation}
At a positive eigenvalue,
\begin{equation}
        \frac1{\partial_\lambda
        \widetilde{\mathscr W}_\alpha(\lambda_n)}
        =\frac{\lambda_n}
        {\partial_\lambda\mathscr W_\alpha(\lambda_n)}.
\label{eq:partner-residue-relation}
\end{equation}
Hence
\begin{equation}
        K_\alpha(\tau)
        =-\sum_{n\geq1}
        \frac{e^{-\lambda_n\tau}}
        {\partial_\lambda\widetilde{\mathscr W}_\alpha(\lambda_n)}.
\label{eq:K-partner-spectral-sum}
\end{equation}

Using Eqs.~\eqref{eq:Legendre-inner-overlap} and
\eqref{eq:partner-Bessel-small-xi}, the partner Wronskian in the overlap
region is
\begin{equation}
\begin{aligned}
        \widetilde{\mathscr W}_\alpha\!\left(\frac14+q^2\right)
        &=2\ii q\,e^{\alpha/2-T_0/2}
        \left[
        C_-(q)D_+(q)e^{-\ii qT_0}
        -C_+(q)D_-(q)e^{\ii qT_0}
        \right]
        +\Delta\widetilde{\mathscr W}_\alpha(q)
        \\
        &=4q\,e^{\alpha/2-T_0/2}
        |\widetilde{\mathcal M}(q)|
        \left[\sin\Phi_\alpha(q)+\mathcal R_\alpha(q)\right],
\end{aligned}
\label{eq:ultranarrow-real-q-Wronskian}
\end{equation}
where $\Phi_\alpha(q)$ is the matching phase and
$\widetilde{\mathcal M}(q)$ is the corresponding smooth amplitude.
The terms $\Delta\widetilde{\mathscr W}_\alpha(q)$ and
$\mathcal R_\alpha(q)$ collect the corrections due to the finite overlap
between the inner and wall approximations.  These corrections are
exponentially small in $T_0$ throughout the range relevant to the first
front, including the analytic continuation used below.
The partner potential makes the overlap particularly benign. In the inner region, the first correction produced by the absorption term is
of order $\xi^2$, whereas the term neglected in the wall equation
\eqref{eq:partner-wall-equation-xi} is $O(e^{-2\vartheta})$.  At the symmetric matching
point $\vartheta_{\rm m}=T_0/2$, both are $O(e^{-T_0})$. 
For the front analysis it is sufficient that the matching error, together
with the few derivatives needed below, remains exponentially small in $T_0$
throughout the range of $q$ relevant to the first front and its analytic
continuation.

The remainder may be absorbed into a corrected phase,
\[
        \widetilde\Phi_\alpha(q)
        =\Phi_\alpha(q)+O(e^{-c_0T_0}),
\]
where the correction remains exponentially small throughout the range relevant
to the first front and under the analytic continuation used below.  The exact
excited levels in the front window therefore satisfy
\begin{equation}
        \widetilde\Phi_\alpha(q_n)=\pi n.
\label{eq:ultranarrow-phase-quantization}
\end{equation}
Their local density is
\begin{equation}
        \frac{\dd n}{\dd q}
        =\frac{\widetilde\Phi_\alpha'(q)}{\pi}
        =\frac{T_0+\delta'(q)}{\pi}+O(e^{-c_0T_0}).
\label{eq:ultranarrow-q-density}
\end{equation}
To keep the notation light, we suppress the tilde below; every use of
$\Phi_\alpha$ in the discrete-to-continuous conversion is understood in this
corrected sense.

At a root $q=q_n$, differentiation of
Eq.~\eqref{eq:ultranarrow-real-q-Wronskian}, together with
$\partial_\lambda=(2q)^{-1}\partial_q$, gives
\begin{equation}
        \frac1{\partial_\lambda
        \widetilde{\mathscr W}_\alpha(\lambda_n)}
        =\frac{(-1)^n e^{T_0/2-\alpha/2}}
        {2|\widetilde{\mathcal M}(q_n)|\Phi_\alpha'(q_n)}
        \left[1+O(e^{-c_0T_0})\right].
\label{eq:ultranarrow-boundary-residue}
\end{equation}
The same derivative $\Phi_\alpha'(q_n)$ that controls the local density of
levels therefore appears in the residue denominator.  The residues also
alternate in sign.  Substitution into Eq.~\eqref{eq:K-partner-spectral-sum}
yields the real discrete sum
\begin{equation}
        K_\alpha(\tau)
        =-\frac{e^{T_0/2-\alpha/2}}{2}
        \sum_{n\geq1}
        \frac{(-1)^n e^{-(q_n^2+1/4)\tau}}
        {|\widetilde{\mathcal M}(q_n)|\Phi_\alpha'(q_n)}
        +\text{subleading terms}.
\label{eq:ultranarrow-quantized-spectral-sum}
\end{equation}

\subsection{Poisson summation and the front contour}
\label{subsec:ultranarrow-poisson}

The corrected phase is odd,
$\Phi_\alpha(-q)=-\Phi_\alpha(q)$, with the branch chosen continuously from
the origin.  Extend the roots by $q_{-n}=-q_n$.  Since
$|\widetilde{\mathcal M}(q)|^{-1}=O(q^2)$ as $q\to0$, the formal $n=0$
term vanishes, and Eq.~\eqref{eq:ultranarrow-quantized-spectral-sum} becomes
\begin{equation}
        K_\alpha(\tau)
        \simeq-\frac{e^{T_0/2-\alpha/2}}{4}
        \sum_{n\in\mathbb Z}
        \frac{(-1)^n e^{-(q_n^2+1/4)\tau}}
        {|\widetilde{\mathcal M}(q_n)|\Phi_\alpha'(q_n)}.
\label{eq:ultranarrow-symmetric-spectral-sum}
\end{equation}
The quantization condition implies
\[
        \delta\!\left(\Phi_\alpha(q)-\pi n\right)
        =\frac{\delta(q-q_n)}{\Phi_\alpha'(q_n)}.
\]
Using the alternating Dirac comb
\[
        \sum_{n\in\mathbb Z}(-1)^n\delta(x-\pi n)
        =\frac1\pi\sum_{m\in\mathbb Z}e^{\ii(2m+1)x},
\]
one obtains, for any sufficiently regular $F$,
\begin{equation}
        \sum_{n\in\mathbb Z}
        \frac{(-1)^nF(q_n)}{\Phi_\alpha'(q_n)}
        =\frac1\pi\sum_{m\in\mathbb Z}
        \int_{-\infty}^{\infty}
        F(q)e^{\ii(2m+1)\Phi_\alpha(q)}\,\dd q.
\label{eq:ultranarrow-Poisson-formula}
\end{equation}
The cancellation of $\Phi_\alpha'$ between the level density and the
spectral residue is now explicit.  Hence
\begin{equation}
        K_\alpha(\tau)
        \simeq-\frac{e^{T_0/2-\alpha/2}}{4\pi}
        \sum_{m\in\mathbb Z}
        \int_{-\infty}^{\infty}
        \frac{e^{-(q^2+1/4)\tau}}
        {|\widetilde{\mathcal M}(q)|}
        e^{\ii(2m+1)\Phi_\alpha(q)}\,\dd q.
\label{eq:ultranarrow-Poisson-expanded}
\end{equation}

\paragraph{The first Poisson pair.}
The first front is generated by the harmonics $p:=2m+1=\pm1$.  For the positive
harmonic,
\[
        \frac{e^{\ii\delta(q)}}{|\widetilde{\mathcal M}(q)|}
        =\frac1{C_-(q)D_+(q)}.
\]
Using Eqs.~\eqref{eq:partner-inner-coefficients} and
\eqref{eq:partner-Bessel-coefficients},
\begin{equation}
        -\frac1{C_-(q)D_+(q)}
        =2^{1+2\ii q}\pi
        \frac{\Gamma(\frac32-\ii q)}{\Gamma(-\ii q)^2}.
\label{eq:partner-first-harmonic-weight}
\end{equation}
The two first harmonics are equal after $q\mapsto-q$.  Their sum gives
\begin{equation}
        K_\alpha(\tau)
        \simeq e^{T_0/2-\alpha/2}
        \int_{-\infty}^{\infty}
        e^{-(q^2+1/4)\tau+\ii q(T_0+2\log2)}
        \frac{\Gamma(\frac32-\ii q)}{\Gamma(-\ii q)^2}\,\dd q.
\label{eq:ultranarrow-leading-spectral-integral}
\end{equation}
This is the central discrete-to-continuous representation: every ingredient
up to this point came from the real quantized spectrum and its real-$q$
matching data.

The contributions of the remaining odd harmonics are parametrically smaller,
as will be demonstrated below.  

\paragraph{Continuation to the front contour.}
The first-pair spectral factor is regular at $q=\ii/2$:
\[
        \left.
        \frac{\Gamma(\frac32-\ii q)}{\Gamma(-\ii q)^2}
        \right|_{q=\ii/2}
        =\frac1\pi.
\]
Consequently, the regularity at $q=\ii/2$ allows us to deform the real contour
through this point.  On the deformed contour the higher odd harmonics acquire
additional factors $e^{-(p-1)T_0/2}$.  Thus, in the moving-front regime
$T_*-\tau=O(\sqrt{\tau})$, the next pair, $p=3$, is already
$O(e^{-T_0})$, while all subsequent pairs are smaller still.

Set
\begin{equation}
        q=\ii\left(\frac12+\sigma\right),
        \qquad
        \lambda=\frac14+q^2=-\sigma(1+\sigma),
\label{eq:front-contour-sigma}
\end{equation}
with $0<c_\sigma:=\Re\sigma<\sigma_0$.   Direct
substitution gives
\[
        2^{-1-2\sigma}
        \frac{\Gamma(2+\sigma)}
        {\Gamma(\frac12+\sigma)^2}
        =\frac{1+2\sigma}{2\pi}\,\mathcal A_*(\sigma),
\]
where
\begin{equation}
        \mathcal A_*(\sigma)
        :=\frac{\pi(1+\sigma)\Gamma(1+\sigma)}
        {2^{2\sigma}(1+2\sigma)
        \Gamma(\frac12+\sigma)^2}.
\label{eq:stationary-amplitude-exact}
\end{equation}
After accounting for the orientation of the transformed contour,
Eq.~\eqref{eq:ultranarrow-leading-spectral-integral} becomes
\begin{equation}
        K_\alpha(\tau)
        \simeq\frac{e^{-\alpha/2}}{2\pi\ii}
        \int_{c_\sigma-\ii\infty}^{c_\sigma+\ii\infty}
        (1+2\sigma)\mathcal A_*(\sigma)
        e^{\tau\sigma(1+\sigma)-T_0\sigma}\,\dd\sigma.
\label{eq:ultranarrow-spectral-contour}
\end{equation}

\subsection{Front displacement and the first-passage kernel}
\label{subsec:ultranarrow-spectral-front}

The explicit expression above gives $\mathcal A_*(0)=1$. Moreover,
\[
        \partial_\sigma\log\mathcal A_*(\sigma)
        =\frac1{1+\sigma}+\psi(1+\sigma)-2\log2
        -\frac2{1+2\sigma}
        -2\psi\!\left(\frac12+\sigma\right).
\]
Using $\psi(1)=-\gamma_{\rm E}$ and
$\psi(1/2)=-\gamma_{\rm E}-2\log2$ gives
\begin{equation}
        \log\mathcal A_*(\sigma)
        =(\gamma_{\rm E}+2\log2-1)\sigma+O(\sigma^2).
\label{eq:stationary-amplitude-linear}
\end{equation}
Hence
\[
        \mathcal A_*(\sigma)e^{-\sigma T_0}
        =e^{-\sigma T_*}[1+O(\sigma^2)],
\]
where
\begin{equation}
        T_*=T_0-\gamma_{\rm E}-2\log2+1
        =\log\frac{C_*}{\alpha},
        \qquad
        C_*=e^{1-\gamma_{\rm E}}.
\label{eq:stationary-shifted-coordinate}
\end{equation}
Thus the matching amplitude fixes the order-one displacement of the front,
$T_0\to T_*$, and hence the factor $C_*=e^{1-\gamma_{\rm E}}$.

In the moving-front regime $T_*-\tau=O(\sqrt{\tau})$, the saddle point and
the width of the dominant integration region are both of order
$\tau^{-1/2}$.  Hence only $\sigma=O(\tau^{-1/2})$ contributes to leading
order.  Keeping the linear
term in Eq.~\eqref{eq:stationary-amplitude-linear} and using
$e^{-\alpha/2}=1+O(\alpha)$ gives
\begin{equation}
        K_\alpha(\tau)\simeq K_*(T_*,\tau),
        \qquad
        K_*(T,\tau)
        :=\frac1{2\pi\ii}
        \int_{c-\ii\infty}^{c+\ii\infty}
        (1+2\sigma)e^{\tau\sigma(1+\sigma)-T\sigma}\,\dd\sigma.
\label{eq:ultranarrow-universal-spectral-contour}
\end{equation}
This integral is Gaussian.  Since
\[
        \tau\sigma(1+\sigma)-T\sigma
        =\tau\sigma^2+(\tau-T)\sigma,
\]
the saddle lies at $\sigma_{\rm s}=\frac{T-\tau}{2\tau}$.
The integrand is entire, so the vertical contour may be shifted to the
saddle.  With $\sigma=\sigma_{\rm s}+\ii y$,
\[
        \tau\sigma(1+\sigma)-T\sigma
        =-\frac{(T-\tau)^2}{4\tau}-\tau y^2,
        \qquad
        1+2\sigma=\frac{T}{\tau}+2\ii y.
\]
The odd part integrates to zero, and the remaining Gaussian gives
\begin{equation}
        K_*(T,\tau)
        =\frac{T}{2\sqrt\pi\,\tau^{3/2}}
        \exp\!\left[-\frac{(T-\tau)^2}{4\tau}\right].
\label{eq:two-image-first-passage-kernel}
\end{equation}
This is the first-passage density associated with the cumulative distribution
$H_*(T,\tau)$, see the discussion below Eq.~\eqref{eq:two_image_kernel} and further in this section,
see Eq.~(\ref{eq:two-image-drift-equation})

For $T>0$, the closed-end-to-zero-reflection propagator vanishes at
$\tau=0$, and therefore
\begin{equation}
        H_*(T,\tau)=\int_0^\tau K_*(T,u)\,\dd u.
\label{eq:H-from-K-integral}
\end{equation}
Evaluating this elementary integral gives
\begin{equation}
        H_*(T,\tau)
        =\frac12\erfc\!\left(\frac{T-\tau}{2\sqrt\tau}\right)
        +\frac12e^T
        \erfc\!\left(\frac{T+\tau}{2\sqrt\tau}\right).
\label{eq:two-image-kernel-derived}
\end{equation}
This is the two-term crossover kernel stated in the Main Results,
Eq.~\eqref{eq:two_image_kernel}.  The two error-function terms are most
naturally viewed here as the closed form of the single cumulative integral
\eqref{eq:H-from-K-integral}.  Equivalently, $H_*$ satisfies
\begin{equation}
        \partial_\tau H_*=(\partial_T^2-\partial_T)H_*,
\label{eq:two-image-drift-equation}
\end{equation}
with
\[
        H_*(T,0)=0\quad(T>0),
        \qquad H_*(0,\tau)=1,
        \qquad H_*(T,\tau)\to0\quad(T\to\infty),
\]
and $H_*(T,\tau)\to1$ as $\tau\to\infty$ at fixed $T$.  Thus
$K_*(T,\tau)$ is the first-passage-time density and $H_*(T,\tau)$ its
cumulative distribution for a diffusion with unit drift to cover the
logarithmic distance $T$.

\subsection{Ultranarrow resonance density and regime of validity}
\label{subsec:ultranarrow-density-validity}

To leading front order,
\[
        H_\alpha(\tau)=h_\alpha(0,\tau\mid\infty)
        \simeq H_*(T_*,\tau).
\]
The boundary-flux identity
Eq.~\eqref{eq:logarithmic-endpoint-integrated-main} gives
\[
        \mathcal F_\alpha(\tau)
        =\alpha\int_0^\tau[H_*(T_*,u)-1]\,\dd u.
\]
Define
\[
        G(T,\tau)=\int_0^\tau[H_*(T,u)-1]\,\dd u.
\]
Since $T_*=\log(C_*/\alpha)$ and
$\partial_\alpha T_*=-1/\alpha$,
\[
        \partial_\alpha[\alpha G(T_*,\tau)]=G-G_T,
        \qquad
        \partial_\alpha^2[\alpha G(T_*,\tau)]
        =\frac1\alpha(G_{TT}-G_T).
\]
Using Eq.~\eqref{eq:two-image-drift-equation},
\[
        G_{TT}-G_T
        =\int_0^\tau\partial_uH_*(T,u)\,\dd u
        =H_*(T,\tau)-H_*(T,0).
\]
For $T>0$, $H_*(T,0)=0$, and hence
\begin{equation}
        \partial_\alpha^2\mathcal F_\alpha(\tau)
        \simeq\frac1\alpha H_*(T_*,\tau).
\label{eq:two-image-alpha-derivative}
\end{equation}
Using the finite-length density relation
\eqref{eq:principal-finite-length-density}, Eq.~\eqref{eq:two-image-alpha-derivative}
then gives
\begin{equation}
\begin{aligned}
        \rho_{\rm ultra}(E,\Gamma;L)
        \simeq\frac{1}{\pi\GamL^2\alpha}H_*(T_*,\tau)
        =\frac{2k}{\pi D\Gamma}
        H_*\!\left(
        \log\frac{C_*\GamL}{2\Gamma},
        \frac{L}{\ellL}
        \right).
\end{aligned}
\label{eq:ultranarrow-density-derived-main}
\end{equation}
This is Eq.~\eqref{eq:ultranarrow_density_main}.  Here
$\alpha=2\Gamma/\GamL$ and $\tau=L/\ellL$.  The centre of the front is fixed
by $T_*=\tau$, giving
\begin{equation}
        \Gamma_{\rm ultra}(L)
        =\frac{C_*}{2}\GamL e^{-L/\ellL}
        =\frac{e^{1-\gamma_{\rm E}}}{2}\GamL e^{-L/\ellL}.
\label{eq:ultranarrow-scale-derived-main}
\end{equation}
The crossover width is
\[
        |T_*-\tau|=O(\sqrt\tau),
        \qquad\text{equivalently}\qquad
        \log\frac{\Gamma_{\rm ultra}(L)}{\Gamma}
        =O\!\left(\sqrt{\frac{L}{\ellL}}\right).
\]

\paragraph{Deep tail of ultranarrow resonances.}
For
\[
        T>\tau,
        \qquad T-\tau\gg\sqrt\tau,
\]
the two terms in Eq.~\eqref{eq:two_image_kernel} have the same leading
exponential.  Their sum gives
\begin{equation}
        H_*(T,\tau)
        \sim\frac{2T\sqrt\tau}{\sqrt\pi\,(T^2-\tau^2)}
        \exp\!\left[-\frac{(T-\tau)^2}{4\tau}\right].
\label{eq:two-image-deep-tail}
\end{equation}
Thus the density has a log-normal-type ultranarrow tail: the leading
exponential is Gaussian in
$\log[\Gamma_{\rm ultra}(L)/\Gamma]$, while the two-term kernel fixes the
additional algebraic prefactor in the logarithmic width variable.

\paragraph{Regime of validity.}
The front asymptotics requires weak absorption, $\alpha\ll1$, so that the
inner region and the absorption wall are well separated.  It applies to long
samples,
\[
        \tau=\frac{L}{\ellL}\gg1,
        \qquad
        T_*-\tau=O(\sqrt{\tau}),
\]
where the front is resolved on its natural $O(\sqrt{\tau})$ scale.  In this
regime only the linear part of the matching amplitude affects the leading
front position and fixes the factor $C_*$; higher-order terms contribute only
to subleading corrections.

As throughout the continuum phase-averaged analysis, we also assume
$k\ellL\gg1$, $kL\gg1$, and $\Gamma\ll E$.  The derivation above is for
perfect coupling.  Finite contact transparency modifies the interface
condition but leaves the bulk front mechanism unchanged; this extension is
discussed in Section~\ref{sec:contact}.
\section{Finite-length reflection statistics}
\label{sec:finite-reflection}

The same propagator that produces the resonance front also determines the
reflection statistics of a finite absorbing sample.  Equation
\eqref{eq:finite-length-reflection-density-main} is the exact relation between
the transition density of $x$ and the reflected intensity $r=|\cR|^2$.  The
spectral analysis of Section~\ref{sec:perfect-ultranarrow} makes one part of
that distribution explicit without solving a new stochastic problem.

Consider first the outer weak-absorption regime in which $x$ stays fixed as
$\alpha\to0$.  After the continuation
$q=\ii(\frac12+\sigma)$ used in the front calculation, the regular inner mode
at a finite observation point is
\[
        P_{-1-\sigma}(1+2x)=P_\sigma(1+2x).
\]
For small $\sigma$,
\begin{equation}
        P_\sigma(1+2x)
        ={}_2F_1(-\sigma,1+\sigma;1;-x)
        =1+\sigma\log(1+x)+O(\sigma^2).
\label{eq:reflection-legendre-small-sigma}
\end{equation}
Thus the finite observation point changes the leading front factor
$\exp(-\sigma T_*)$ only by a displacement,
\[
        T_*\longmapsto T_x
        =T_*-\log(1+x).
\]
Since $1+x=(1-r)^{-1}$, it is convenient to write
\begin{equation}
        T_r
        :=\log\frac{C_*(1-r)}{\alpha}.
\label{eq:reflection-front-coordinate}
\end{equation}
In the same moving-front scaling as before,
$T_r-\tau=O(\sqrt\tau)$ with $1-r\gg\alpha$, we therefore obtain
\begin{equation}
        h_\alpha\!\left(
        \frac{r}{1-r},\tau\mid\infty
        \right)
        \simeq H_*(T_r,\tau),
\label{eq:finite-reflection-h-front}
\end{equation}
and hence
\begin{equation}
        P_\alpha(r,\tau\mid\infty)
        \simeq
        \frac{\alpha}{(1-r)^2}
        \exp\!\left[-\frac{\alpha r}{1-r}\right]
        H_*\!\left(
        \log\frac{C_*(1-r)}{\alpha},\tau
        \right).
\label{eq:finite-reflection-front-density}
\end{equation}
This is an outer-tail formula rather than a separately normalized
approximation to the whole reflection distribution: for weak absorption most
of the probability remains concentrated near $r=1$.

The complementary scaling $1-r=O(\alpha)$ near unit reflection produces the
finite-length Wigner-delay problem.  We return to this limit in
Section~\ref{sec:reflection-delay}, where it is developed together with the
closed-system boundary-force distribution.

\section{Nonideal contacts}
\label{sec:contact}

We now return to the interface observable $f_{\cT}$ defined in
Eq.~\eqref{eq:main-interface-observable}.  No new bulk stochastic or
spectral problem is required.  Finite transparency moves the contact threshold
from the endpoint $x=0$ to the interior point
$x_{\cT}=(1-\cT)/\cT$ and changes the projections of the unchanged bulk modes.

\subsection{Continuum contact, interface map, and phase average}
\label{subsec:regular-contact-main}

Recall from Eq.~\eqref{eq:main-riccati-definitions} that the interface
amplitude ratio is
\[
        \zeta_N=\frac{\phi_{N+1}}{\phi_N}.
\]
In the continuum limit, the interface lattice amplitudes are identified as
\[
        \phi_N=\psi(L),
        \qquad
        \phi_{N+1}=\psi(L+a).
\]
Introduce the continuum logarithmic derivative
\[
        y(L)\coloneqq\frac{\psi'(L)}{\psi(L)}.
\]
Taylor expansion of $\psi(L+a)$ gives
\begin{equation}
\begin{aligned}
        \zeta_N
        =\frac{\psi(L+a)}{\psi(L)}
        =1+a\,\frac{\psi'(L)}{\psi(L)}+O(a^2)
        =1+a\,y(L)+O(a^2)
\end{aligned}
\label{eq:interface-ratio-continuum-expansion}
\end{equation}

Substituting this expansion, the regular contact scaling
\eqref{eq:regular-contact-scaling-model}, and
$\e^{\pm\ii ak}=1\pm\ii ak+O(a^2)$ into the exact finite-chain reflection
formula \eqref{eq:reflection-residual-ratio} gives
\[
\begin{aligned}
        \zeta_N-t_{\rm c}^2\e^{-\ii ak}=a\bigl[y(L)+\lambda_{\rm c}+\ii k\bigr]+O(a^2),
        \qquad
        \zeta_N-t_{\rm c}^2\e^{\ii ak}
        =a\bigl[y(L)+\lambda_{\rm c}-\ii k\bigr]+O(a^2).
\end{aligned}
\]
Since $\e^{2\ii ak}=1+O(a)$, the continuum reflection amplitude is
\begin{equation}
        \cR_{\cT}
        =-
        \frac{y(L)+\lambda_{\rm c}+\ii k}
             {y(L)+\lambda_{\rm c}-\ii k},
\label{eq:continuum-contact-reflection-main}
\end{equation}
where $y(L)=\psi'(L)/\psi(L)$.  After analytic continuation to a complex
energy $Z$, a resonance pole is a value of $Z$ for which the denominator
of Eq.~\eqref{eq:continuum-contact-reflection-main} vanishes.  Equivalently,
the solution $\psi(x;Z)$ satisfying the left boundary condition must also
satisfy at $x=L$ the outgoing Robin condition
\begin{equation}
        \psi'(L;Z)
        =\bigl[\ii k(Z)-\lambda_{\rm c}\bigr]\psi(L;Z),
\label{eq:contact-robin-main}
\end{equation}
with $k(Z)=\sqrt{Z}$ on the outgoing branch.
The inverse reflection amplitude remains analytic in the resonance strip:
the incoming residual is nonzero there by
Eq.~\eqref{eq:incoming-residual-nonzero}.  Thus the regular contact introduces
no additional pole--zero cancellation.

Let $\cR=-\frac{y+\ii k}{y-\ii k}$ denote the perfectly matched-contact reflection amplitude and introduce the parameter
$c:=\frac{\lambda_{\rm c}}{2k}$.
Solving the matched relation for $y$ and substituting into
Eq.~\eqref{eq:continuum-contact-reflection-main} gives the leading interface
fractional-linear map at the real wave energy.  The exact absorbing-energy
map is obtained by replacing $k$ by $k_+=\sqrt{E+\ii\eta}$ throughout. 
Since $k_+=\sqrt{E+\ii\eta}=k\left[1+O\!\left(\frac{\eta}{E}\right)\right]$,
the phase-averaged interface observable may be evaluated with the real
wavenumber $k$ to leading order in the generic-bulk regime, so that:
\begin{equation}
        \cR_{\cT}
        =
        \frac{(1-\ii c)\cR-\ii c}
             {\ii c\cR+(1+\ii c)}.
\label{eq:interface-contact-map-main}
\end{equation}
For a clean sample, $\cR=0$, and hence
\[
        |\cR_{\cT}|^2=\frac{c^2}{1+c^2}.
\]
The contact transparency is the transmitted flux fraction
$\cT=1-|\cR_{\cT}|^2$ of this clean interface.  Therefore
\[
        \cT
        =\frac{1}{1+c^2}
        =\frac{4k^2}{4k^2+\lambda_{\rm c}^2},
\]
in agreement with Eq.~\eqref{eq:main-contact-definitions}.  Holding a fixed
lattice value $t_{\rm c}<1$ as $a\to0$ instead sends
$\lambda_{\rm c}\to\infty$.  Thus lattice closure $t_{\rm c}\to0$ and
continuum closure $\lambda_{\rm c}\to\infty$ represent the same physical
limit through technically different scalings.

It remains to identify the phase-averaged interface observable.  Write
$\cR=\sqrt r\,\e^{\ii\varphi}$.  For complex numbers $A$ and $B$, Jensen's
formula gives
\[
        \frac{1}{2\pi}\int_0^{2\pi}
        \log|A\e^{\ii\varphi}+B|^2\,\dd\varphi
        =2\log\max\{|A|,|B|\}.
\]
For the numerator of Eq.~\eqref{eq:interface-contact-map-main},
\[
        |A_{\rm N}|^2=(1+c^2)r,
        \qquad
        |B_{\rm N}|^2=c^2,
\]
whereas for the denominator,
\[
        |A_{\rm D}|^2=c^2r
        <1+c^2
        =|B_{\rm D}|^2.
\]
Subtracting the two Jensen averages yields
\begin{equation}
        \frac{1}{2\pi}\int_0^{2\pi}
        \log|\cR_{\cT}|^2\,\dd\phi
        =\log\max\{r,1-\cT\}
\label{eq:contact-Jensen-main}
\end{equation}
With $x=r/(1-r)$ and $x_{\cT}=(1-\cT)/\cT$, this reproduces the interface
function $f_{\cT}$ in Eq.~\eqref{eq:main-interface-observable}.  In particular,
\begin{equation}
        f_{\cT}'(x)
        =\frac{\Theta(x-x_{\cT})}{x(1+x)},
        \qquad
        \partial_x\!\left[x(1+x)f_{\cT}'(x)\right]
        =\delta(x-x_{\cT})
\label{eq:contact-interface-derivatives-main}
\end{equation}
The Jensen identity is exact for a uniform phase.  Its use here is the same
generic-phase approximation employed in deriving the phase-averaged bulk
diffusion.

\subsection{Threshold boundary-flux identity and spectral projections}
\label{subsec:contact-threshold-main}
Starting from Eq.~\eqref{eq:main-F-contact}, the forward equation
\eqref{eq:main-reflection-intensity-forward} and
$x(1+x)f_{\cT}'(x)=\Theta(x-x_{\cT})$ give
\[
\begin{aligned}
        \frac{\dd\mathcal F_{\alpha,\cT}}{\dd\tau}=-\int_{x_{\cT}}^\infty
        \bigl[\partial_xp_\alpha+\alpha p_\alpha\bigr] \,\dd x
        =p_\alpha(x_{\cT},\tau\mid\infty)
        -\alpha\int_{x_{\cT}}^\infty
        p_\alpha(x,\tau\mid\infty)\,\dd x.
\end{aligned}
\]
This is the threshold boundary-flux identity
\eqref{eq:main-contact-threshold-boundary-flux}.  No endpoint regularization
is needed for $x_{\cT}>0$, because the interface observable is bounded at
$x=0$.  For $\cT=1$, normalization reduces the formula to the perfect-contact
endpoint identity \eqref{eq:principal-endpoint-identity}.  At stationarity the
two terms cancel because
\[
        \pi_\alpha(x_{\cT})
        =\alpha\int_{x_{\cT}}^\infty\pi_\alpha(x)\,\dd x
        =\alpha\e^{-\alpha x_{\cT}}.
\]

Expanding the unchanged bulk propagator gives
\[
        \mathcal F_{\alpha,\cT}(\tau)
        =\mathcal F_{\alpha,\cT}^{(\infty)}
        +\sum_{\nu\geq1}
        \e^{-\lambda_\nu\tau}
        \phi_\nu(\infty)f_{\cT,\nu},
\]
where a first integration by parts yields
\begin{equation}
        f_{\cT,\nu}
        =
        \frac{
        \alpha\displaystyle\int_{x_{\cT}}^\infty
        \pi_\alpha(x)\phi_\nu(x)\,\dd x
        -\pi_\alpha(x_{\cT})\phi_\nu(x_{\cT})
        }{\lambda_\nu}
\label{eq:contact-mode-coefficient-main}
\end{equation}
Using the divergence form of the eigenvalue equation,
\[
-\partial_x\!\left[
x(1+x)\pi_\alpha(x)\phi_\nu'(x)
\right]
=
\lambda_\nu\pi_\alpha(x)\phi_\nu(x),
\]
and integrating from $x_{\cT}$ to infinity gives
\[
\begin{aligned}
\lambda_\nu
\int_{x_{\cT}}^\infty
\pi_\alpha(x)\phi_\nu(x)\,\dd x
={}
x_{\cT}(1+x_{\cT})
\pi_\alpha(x_{\cT})\phi_\nu'(x_{\cT})
-\lim_{x\to\infty}
x(1+x)\pi_\alpha(x)\phi_\nu'(x).
\end{aligned}
\]
The last term vanishes by the self-adjoint boundary condition at infinity.
Therefore
\[
\int_{x_{\cT}}^\infty
\pi_\alpha(x)\phi_\nu(x)\,\dd x
=
\frac{x_{\cT}(1+x_{\cT})
\pi_\alpha(x_{\cT})\phi_\nu'(x_{\cT})}
{\lambda_\nu}.
\]
Consequently, the coefficient has the entirely local form
\begin{equation}
        f_{\cT,\nu}
        =
        \frac{\alpha x_{\cT}(1+x_{\cT})
        \pi_\alpha(x_{\cT})\phi_\nu'(x_{\cT})}
        {\lambda_\nu^2}
        -
        \frac{\pi_\alpha(x_{\cT})\phi_\nu(x_{\cT})}
        {\lambda_\nu}
\label{eq:contact-coefficient-local-main}
\end{equation}
At perfect coupling, $x_{\cT}=0$ and $\pi_\alpha(0)=\alpha$, so this reduces
to $f_{1,\nu}=-\alpha\phi_\nu(0)/\lambda_\nu$.  Thus the contact changes
only the interface projection of the unchanged bulk modes.

\subsection{Stationary density, positivity, and limiting regimes}
\label{subsec:stationary-contact-density-main}

For $0<\cT<1$, splitting the stationary integral at $x_{\cT}$ and integrating
by parts gives
\begin{equation}
\begin{aligned}
        \mathcal F_{\alpha,\cT}^{(\infty)}
        &=\log\frac{x_{\cT}}{1+x_{\cT}}
        +\int_{x_{\cT}}^\infty
        \frac{\e^{-\alpha x}}{x(1+x)}\,\dd x
        \\
        &=\log\frac{x_{\cT}}{1+x_{\cT}}
        +E_1(\alpha x_{\cT})
        -\e^\alpha E_1\!\bigl(\alpha(1+x_{\cT})\bigr)
\end{aligned}
\label{eq:stationary-contact-observable-main}
\end{equation}
The perfect-contact result follows by taking $x_{\cT}\rightarrow 0_+$.  The first
two derivatives are
\[
        \partial_\alpha
        \mathcal F_{\alpha,\cT}^{(\infty)}
        =-\e^\alpha E_1\!\bigl(\alpha(1+x_{\cT})\bigr)
\]
and
\[
\begin{aligned}
        \partial_\alpha^2
        \mathcal F_{\alpha,\cT}^{(\infty)}
        =\frac{\e^{-\alpha x_{\cT}}}{\alpha}
        -\e^\alpha E_1\!\bigl(\alpha(1+x_{\cT})\bigr)
        =\frac{\e^{-\alpha x_{\cT}}}{\alpha}
        \left[
        1-\alpha\e^{\alpha(1+x_{\cT})}
        E_1\!\bigl(\alpha(1+x_{\cT})\bigr)
        \right].
\end{aligned}
\]
Taking two derivatives with respect to the absorption parameter therefore
gives the stationary contact density stated in
Eq.~\eqref{eq:contact_stationary_density_main}.  Positivity follows from
\[
        \e^zE_1(z)<\frac1z,
        \qquad z>0.
\]

For every fixed $0<\cT\leq1$, the leading narrow-width law remains
\[
        \rho_{\cT}^{(\infty)}(E,\Gamma)
        \sim\frac{2k}{\pi D\Gamma}.
\]
For small contact transparency, the validity of this regime additionally requires
\[
        \frac{\alpha}{\cT}\ll1,
        \qquad\text{equivalently}\qquad
        \Gamma\ll\cT\GamL.
\]
For fixed $0<\cT<1$ and $\alpha\to\infty$,
\[
        1-\alpha\e^{\alpha/\cT}E_1(\alpha/\cT)
        =(1-\cT)+\frac{\cT^2}{\alpha}+O(\alpha^{-2}).
\]
Hence, within $\GamL\ll\Gamma\ll E$,
\begin{equation}
        \rho_{\cT}^{(\infty)}(E,\Gamma)
        \sim
        \frac{2k(1-\cT)}{\pi D\Gamma}
        \exp\!\left[
        -\frac{2\Gamma}{\GamL}
        \frac{1-\cT}{\cT}
        \right]
\label{eq:contact-broad-tail-main}
\end{equation}
Thus every fixed nonideal contact replaces the perfect-contact algebraic
broad-width tail by an exponential cutoff.  The limit $\cT\to1$ is nonuniform
because the leading coefficient $1-\cT$ vanishes; the crossover is controlled
by
\[
        \alpha x_{\cT}
        =\frac{2\Gamma}{\GamL}\frac{1-\cT}{\cT}.
\]

\subsection{Moldauer--Simonius sum rule from the 1D reflection formulation}
\label{subsec:MS-main}

The finite-transparency formula contains a useful global sum rule.  Define
the first width moment per unit energy at fixed $L$ by
\begin{equation}
        M_1(E;L)
        :=\int_0^\infty
        \Gamma\,\rho_{\cT}(E,\Gamma;L)\,\dd\Gamma .
\label{eq:MS-first-moment-main}
\end{equation}
Within the same generic-bulk approximation in which the absorption derivative
dominates the Poincar\'e--Lelong Laplacian,
Eq.~\eqref{eq:principal-general-contact-density} and
$\Gamma=(\GamL/2)\alpha$ give
\begin{equation}
        M_1(E;L)
        =\frac{1}{4\pi}
        \int_0^\infty
        \alpha\,\partial_\alpha^2
        \mathcal F_{\alpha,\cT}(\tau)\,\dd\alpha .
\label{eq:MS-alpha-integral-main}
\end{equation}
An integration by parts reduces the whole width integral to the two absorption
limits,
\begin{equation}
        M_1(E;L)
        =\frac{1}{4\pi}
        \left[
        \alpha\partial_\alpha\mathcal F_{\alpha,\cT}
        -\mathcal F_{\alpha,\cT}
        \right]_{\alpha=0}^{\alpha=\infty}.
\label{eq:MS-boundary-main}
\end{equation}
For zero absorption the closed-end sample is lossless, $r=1$, and therefore
$f_{\cT}(\infty)=0$ gives
$\mathcal F_{0,\cT}=0$.  At strong absorption a wave returning from the
disordered bulk is suppressed before it reaches the interface.  The remaining
reflection is that of the contact itself, with intensity $1-\cT$; equivalently
the radial distribution is concentrated at $x=0$ and
\begin{equation}
        \mathcal F_{\infty,\cT}=f_{\cT}(0)=\log(1-\cT).
\label{eq:MS-F-limits-main}
\end{equation}
For every fixed $0<\cT<1$ the approach to both limits is fast enough that
$\alpha\partial_\alpha\mathcal F_{\alpha,\cT}$ vanishes at the corresponding
boundary.  Hence
\begin{equation}
        M_1(E;L)=-\frac{1}{4\pi}\log(1-\cT).
\label{eq:MS-first-moment-result-main}
\end{equation}
The result is independent of $L$ and of the disorder strength: localization
redistributes the resonance widths but does not change this leading
semiclassical first-moment density.

Let $\bar\rho_{\rm cl}(E;L)=\Delta(E;L)^{-1}$ be the local mean density of
closed-system levels.  Dividing Eq.~\eqref{eq:MS-first-moment-result-main} by
$\bar\rho_{\rm cl}$ gives
\begin{equation}
        \langle\Gamma\rangle_E
        =-\frac{\Delta(E;L)}{4\pi}\log(1-\cT).
\label{eq:MS-average-main}
\end{equation}
This is the single-channel Moldauer--Simonius relation in the pole convention
$Z=E-\ii\Gamma$.  The relation was introduced in nuclear scattering
\cite{Moldauer1967,Simonius1974}; for broken time-reversal symmetry it follows
from the random-matrix treatment of Ref.~\cite{FyodorovSommers1997}.
Kurilov and Ostrovsky~\cite{KurilovOstrovsky2026}
recently gave a general nonlinear-$\sigma$-model
derivation valid, within its semiclassical domain, independently of
Wigner--Dyson symmetry and sample geometry.  Equation~\eqref{eq:MS-average-main} provides an
independent derivation for the strict one-dimensional Schr\"odinger problem
considered here, using only the reflection formulation and the same
phase-averaged weak-disorder limit used elsewhere in the paper.

There is one important order-of-limits qualification.  At fixed
$0<\cT<1$, the contact cuts the width integral off around
$\alpha(1-\cT)/\cT=O(1)$.  This lies inside $\Gamma\ll E$ when
$k\ellL\to\infty$.  The perfect-coupling limit should therefore be taken only
after the semiclassical weak-disorder limit.  As $\cT\uparrow1$,
Eq.~\eqref{eq:MS-average-main} diverges logarithmically.  At perfect coupling
the broad part of our 1D density is
\begin{equation}
        \rho(E,\Gamma)
        \simeq\frac{1}{4\pi\Gamma^2},
        \qquad
        \GamL\ll\Gamma\ll E,
\label{eq:MS-broad-tail-main}
\end{equation}
so that the normalized width density
$\Delta\rho$ has the universal form
$\Delta/(4\pi\Gamma^2)$.  Its first moment is logarithmic.  Thus in the
present microscopic 1D model the broad $1/\Gamma^2$ law and the
Moldauer--Simonius divergence originate from the same
large-width range controlled by the contact transparency, in agreement with the general discussion of
Ref.~\cite{KurilovOstrovsky2026}.

At fixed $\Gamma>0$, Eq.~\eqref{eq:contact_stationary_density_main} gives
\[
        \lim_{\cT\to0}\rho_{\cT}^{(\infty)}(E,\Gamma)=0.
\]
For a finite sample, however, closing the contact does not remove the levels:
the resonance poles approach the real eigenvalues of the closed system, so that
\[
        \rho_{\cT}(E,\Gamma;L)
        \longrightarrow
        \rho_{\rm cl}(E;L)\delta(\Gamma),
        \qquad \cT\to0.
\]
Thus the limits $L\to\infty$ and $\cT\to0$ should not be interchanged.

\subsection{Contact shift of the ultranarrow front}
\label{subsec:contact-ultranarrow-main}

Recall that $x=r/(1-r)=\sinh^2(\vartheta/2)$.  Perfect coupling probes
$x=0$, or $\vartheta=0$, whereas finite transparency moves the interface
threshold to
\[
        x_{\cT}=\frac{1-\cT}{\cT},
        \qquad
        \vartheta_{\cT}
        =2\,\operatorname{arsinh}\sqrt{x_{\cT}}.
\]
The bulk absorption wall remains at
$\vartheta\simeq T_0=\log(4/\alpha)$, so finite transparency changes the
effective distance from the interface threshold to the wall, but not the
position of the wall itself.
Dividing Eq.~\eqref{eq:contact-coefficient-local-main} by the perfect-contact
coefficient gives the exact ratio
\begin{equation}
        \frac{f_{\cT,\nu}}{f_{1,\nu}}
        =\e^{-\alpha x_{\cT}}
        \frac{
        \phi_\nu(x_{\cT})
        -\alpha x_{\cT}(1+x_{\cT})
        \phi_\nu'(x_{\cT})/\lambda_\nu
        }{\phi_\nu(0)}
\label{eq:contact-mode-ratio-main}
\end{equation}
 In the moving-front regime, $\alpha\to0$ at fixed $x_{\cT}$, the derivative
term and $\e^{-\alpha x_{\cT}}-1$ are subleading.  Continuing the regular
wall-free mode \eqref{eq:Legendre-regular-mode} to
$q=\ii(\frac12+\sigma)$ gives
\[
        \phi_\sigma(x)\sim P_\sigma(1+2x),
        \qquad
        P_\sigma(1)=1.
\]
Therefore
\[
        \frac{f_{\cT,\sigma}}{f_{1,\sigma}}
        \sim P_\sigma(1+2x_{\cT}).
\]
Using the hypergeometric representation
\[
        P_\sigma(1+2x_{\cT})
        ={}_2F_1(-\sigma,1+\sigma;1;-x_{\cT}),
\]
we find
\begin{equation}
        P_\sigma(1+2x_{\cT})
        =1+\sigma\log(1+x_{\cT})+O(\sigma^2)
\label{eq:Legendre-contact-expansion-main}
\end{equation}
Indeed, $(-\sigma)_n=-\sigma(n-1)!+O(\sigma^2)$, and hence the linear term is
\[
        -\sigma\sum_{n\geq1}\frac{(-x_{\cT})^n}{n}
        =\sigma\log(1+x_{\cT}).
\]
Multiplying the perfect-contact factor $\exp(-\sigma T_*)$ by this interface
factor replaces
\begin{equation}
        T_*=\log\frac{C_*}{\alpha}
        \quad\longmapsto\quad
        T_{*,\cT}
        =T_*-\log(1+x_{\cT})
        =\log\frac{C_*\cT}{\alpha}
\label{eq:contact-front-coordinate-main}
\end{equation}
because $1+x_{\cT}=1/\cT$.  To leading moving-front order, the two-term
kernel is unchanged in shape:
\begin{equation}
        \rho_{{\rm ultra},\cT}(E,\Gamma;L)
        \simeq
        \frac{2k}{\pi D\Gamma}
        H_*\!\left(
        \log\frac{C_*\cT\GamL}{2\Gamma},
        \frac{L}{\ellL}
        \right)
\label{eq:contact-ultranarrow-density-main}
\end{equation}
The contact shift gives the characteristic width
\eqref{eq:contact_ultranarrow_scale}.  The factor $\e^{-L/\ellL}$ describes
the probability for a localized state to reach the open end, while $\cT$ is
the probability of passing through the barrier once that end is reached.
Terms of order $\sigma^2$ affect subleading front-shape and prefactor
corrections.  

The finite-transparency results above apply for fixed $0<\cT\leq 1$ in the
generic-bulk, long-sample regime; the joint limit $\cT\to 0$ together with the
ultranarrow scaling is not considered here.

\section{Finite-length Wigner time delays and boundary forces of localized modes}
\label{sec:reflection-delay}

The weak-absorption region near unit reflection contains a second finite-length
observable of the same reflection process.  It is the single-channel Wigner time delay.
Here we make this connection explicit and then use it to obtain the
finite-length statistics of a closed-system boundary force.

\subsection{Wigner time-delay scaling near unit reflection}

Put
\begin{equation}
        u_\alpha=\frac{1-r}{\alpha}.
\label{eq:delay-scaled-deficit-main}
\end{equation}
Equation~\eqref{eq:closed-end-y-main} gives exactly
\begin{equation}
        \dd u_\alpha
        =\left(1-\alpha u_\alpha-\alpha u_\alpha^2\right)\dd\tau
        -\sqrt{2(1-\alpha u_\alpha)}\,u_\alpha\dd B_\tau.
\label{eq:delay-scaled-sde-main}
\end{equation}
At fixed $u$ and $\tau$, the weak-absorption limit gives
\begin{equation}
        \dd u=\dd\tau-\sqrt2\,u\,\dd B_\tau,
        \qquad u(0)=0.
\label{eq:wigner-delay-sde-main}
\end{equation}
On the real axis write $\cR(E)=e^{\ii\varphi(E)}$.  Analytic continuation to
$E+\ii\eta$ gives
\[
        1-|\cR(E+\ii\eta)|^2
        =2\eta\,\partial_E\varphi(E)+O(\eta^2).
\]
With $\tau_{\rm W}=\partial_E\varphi$ and
$\alpha=2\eta/\GamL$, Eq.~\eqref{eq:delay-scaled-deficit-main} becomes
\begin{equation}
        u=\GamL\tau_{\rm W}.
\label{eq:wigner-delay-scaling-main}
\end{equation}
Thus Eq.~\eqref{eq:wigner-delay-sde-main} is precisely the finite-length
one-dimensional Wigner time delay process in our units.  The time delay scaling near
unit reflection and the outer reflection front
\eqref{eq:finite-reflection-front-density} are complementary limits of the
same finite-$\alpha$ reflection propagator.

\subsection{Complete finite-length time delay distribution}

Changing $B\mapsto-B$ does not change the distribution of the process.  The
solution of Eq.~\eqref{eq:wigner-delay-sde-main} can therefore be written as
the exponential Brownian functional
\begin{equation}
        u_\tau\overset{\rm d}{=}
        \int_0^\tau
        \exp\!\left(\sqrt2 B_s-s\right)\dd s.
\label{eq:delay-exponential-functional-main}
\end{equation}
Its density $P(u,\tau)$ obeys
\begin{equation}
        \partial_\tau P
        =-\partial_uP+\partial_u^2(u^2P),
        \qquad P(u,0)=\delta(u),
        \qquad u>0,
\label{eq:delay-fokker-planck-main}
\end{equation}
and has stationary limit
\begin{equation}
        P_\infty(u)=\frac1{u^2}e^{-1/u}.
\label{eq:delay-stationary-main}
\end{equation}
The complete finite-length solution is the Comtet--Texier density
\cite{ComtetTexier1997,TexierComtet1999},
\begin{equation}
\begin{aligned}
        P(u,\tau)
        =\frac1{u^2}e^{-1/u}
        +\frac{2}{\pi u}e^{-1/(2u)}
        \int_0^\infty
        \frac{q\sinh(\pi q)}{q^2+\frac14}
        W_{1,\ii q}\!\left(\frac1u\right)
        e^{-(q^2+\frac14)\tau}\,\dd q.
\end{aligned}
\label{eq:delay-exact-density-main}
\end{equation}
The first term is the semi-infinite result, including the familiar
$\tau_{\rm W}^{-2}$ tail, while the integral contains the complete finite-$L$
correction.

It is worth being precise about how this spectral problem is related to the one
solved in Section~\ref{sec:perfect-ultranarrow}.  It is not the same problem,
but its degeneration in which only the absorption wall survives.  Taking the
radial generator $\cL_\alpha=\partial_\vartheta^2+b_\alpha(\vartheta)
\partial_\vartheta$ into the wall coordinate $\xi=\e^{\vartheta-T_0}$ of
Eq.~\eqref{eq:partner-wall-coordinate}, where $\coth\vartheta\to1$ and
$\frac{\alpha}{2}\sinh\vartheta\to\xi$, gives
\[
        \cL_\alpha\longrightarrow
        \xi^2\partial_\xi^2+(2\xi-\xi^2)\partial_\xi,
\]
which is exactly the generator of Eq.~\eqref{eq:wigner-delay-sde-main} written
in the variable $1/u$.  The two arguments agree as they should: $\xi\simeq\alpha
x$ for $\vartheta\gg1$, while $1/u=\alpha(1+x)=\alpha+\xi$.  Hence the common
Whittaker index $\mu=\ii q$ and eigenvalue $q^2+\frac14$.

The front calculation uses the full problem instead, since it needs the
Legendre region $\vartheta=O(1)$, the Whittaker wall, and the matching between
them.  The order-one displacement $C_*=\e^{1-\gamma_{\rm E}}$ is assembled in
$\mathcal A_*(\sigma)$ from the connection coefficients of both regions, and so
has no counterpart in the time-delay distribution, which never resolves the
inner region.  Appendix~\ref{app:wigner-delay} records the short spectral route
from Eq.~\eqref{eq:delay-fokker-planck-main} to
Eq.~\eqref{eq:delay-exact-density-main} and its relation to the original
Comtet--Texier variables.

\subsection{Finite contact transparency}
\label{subsec:contact-delay-main}

The fractional-linear interface relation of Section~\ref{sec:contact} also
gives the delay distribution for a nonideal contact.  Since the matrix in
Eq.~\eqref{eq:interface-contact-map-main} is an $SU(1,1)$ transformation, one
has the exact algebraic identity
\begin{equation}
        1-|\cR_{\cT}|^2
        =\frac{1-|\cR|^2}
        {|1+\ii c+\ii c\cR|^2}.
\label{eq:contact-unitarity-deficit-main}
\end{equation}
On the real axis put $\cR=e^{\ii\varphi}$.  Combining this identity with the
weak-absorption relation gives, for the disorder-induced part of the delay,
\begin{equation}
        \tau_{\rm W,\cT}
        =J_{\cT}(\varphi)\tau_{\rm W,1}
        +O(E^{-1}),
        \qquad
        J_{\cT}(\varphi)
        =\frac{1}{|1+\ii c+\ii c e^{\ii\varphi}|^2}.
\label{eq:contact-delay-factor-main}
\end{equation}
The $O(E^{-1})$ term is the direct microscopic delay of the physical
$\delta$-contact and is smaller by $O[(k\ellL)^{-1}]$ than the localization
scale retained here.  After a shift of the uniformly distributed phase,
\begin{equation}
        g=\frac{2}{\cT}-1,
        \qquad
        A_{\cT}(\theta)
        =g-\sqrt{g^2-1}\cos\theta,
        \qquad
        J_{\cT}=A_{\cT}^{-1}.
\label{eq:contact-delay-A-main}
\end{equation}
If $P_1(u,\tau)$ denotes Eq.~\eqref{eq:delay-exact-density-main}, the
finite-transparency density is
\begin{equation}
        P_{\cT}(v,\tau)
        =\frac{1}{2\pi}\int_0^{2\pi}
        A_{\cT}(\theta)
        P_1\!\left(A_{\cT}(\theta)v,\tau\right)\dd\theta,
        \qquad
        v=\GamL\tau_{\rm W,\cT}.
\label{eq:contact-delay-transform-main}
\end{equation}
Thus all finite-length dependence remains in the perfect-contact density,
while the contact enters through a one-dimensional phase average.  Closely
related arbitrary-coupling relations are well known in the literature on random matrix approach to quantum chaotic scattering, see \cite{OssipovFyodorov2005,SavinFyodorovSommers2001} and references within.

\subsection{Dirichlet eigenfunctions and force on the boundary}
\label{subsec:boundary-force-main}

We now show how the finite-length Wigner-delay distribution determines the
statistics of the boundary derivative of closed-system eigenfunctions and,
through it, the force exerted on a Dirichlet boundary.  For each energy $E$,
let $\phi(x,E)$ denote the solution of
\begin{equation}
        -\phi''(x,E)+V(x)\phi(x,E)=E\phi(x,E),
        \qquad
        \phi(0,E)=0,
        \qquad
        \phi'(0,E)=1.
\label{eq:boundary-fundamental-solution-main}
\end{equation}
Thus $\phi(x,E)$ is the left-normalized fundamental solution.  The
eigenvalues $E_n$ of the closed system with Dirichlet conditions at both
ends are precisely the energies satisfying
\[
        \phi(L,E_n)=0.
\]

If a perfectly matched lead is attached at $x=L$, introduce the endpoint
logarithmic derivative
\[
        m(E)=\frac{\phi'(L,E)}{\phi(L,E)}.
\]
The reflection amplitude is then
\begin{equation}
        \cR(E)=-\frac{m(E)+\ii k}{m(E)-\ii k},
        \qquad E=k^2.
\label{eq:boundary-reflection-m-main}
\end{equation}

Differentiating the Schr\"odinger equation with respect to $E$ and combining
the result with the original equation gives the standard Sturm--Liouville
identity
\[
        \frac{\dd}{\dd x}
        \left[
        (\partial_E\phi)\phi'
        -(\partial_E\phi')\phi
        \right]
        =\phi^2.
\]
Integrating from $0$ to $L$ gives no contribution from the left endpoint,
since $\phi(0,E)=0$ and $\phi'(0,E)=1$ are independent of $E$.
Evaluating the result at a closed-system eigenvalue $E_n$, where
$\phi(L,E_n)=0$, therefore gives
\[
        \partial_E\phi(L,E_n)\phi'(L,E_n)
        =\int_0^L\phi(x,E_n)^2\,\dd x.
\]

Near $E_n$,
\[
        \phi(L,E)
        =(E-E_n)\partial_E\phi(L,E_n)
        +O\!\left((E-E_n)^2\right),
        \qquad
        \phi'(L,E)=\phi'(L,E_n)+O(E-E_n).
\]
Hence
\[
\begin{aligned}
        m(E)=\frac{\phi'(L,E)}{\phi(L,E)}=
        \frac{\phi'(L,E_n)^2}
        {(E-E_n)\displaystyle\int_0^L\phi(x,E_n)^2\,\dd x}
        +O(1),
\end{aligned}
\]
where in the second line we used the Sturm--Liouville identity above.
For the normalized closed eigenfunction
\[
        \psi_n(x)
        =
        \frac{\phi(x,E_n)}
        {\left[\displaystyle\int_0^L\phi(y,E_n)^2\,\dd y\right]^{1/2}},
\]
this becomes
\begin{equation}
        m(E)=\frac{|\psi_n'(L)|^2}{E-E_n}+O(1),
        \qquad E\to E_n.
\label{eq:boundary-m-residue-main}
\end{equation}
To extract the Wigner delay, write
\[
        A_n:=|\psi_n'(L)|^2,
        \qquad
        \delta E:=E-E_n.
\]
Equation~\eqref{eq:boundary-m-residue-main} can then be written more explicitly as
\[
        m(E)=\frac{A_n}{\delta E}+B_n+O(\delta E),
\]
with a finite regular part $B_n$.  Substitution into
Eq.~\eqref{eq:boundary-reflection-m-main} gives
\[
\begin{aligned}
        \cR(E)=
        -\frac{
        A_n+(B_n+\ii k_n)\delta E+O(\delta E^2)
        }{
        A_n+(B_n-\ii k_n)\delta E+O(\delta E^2)
        }=
        -\left[
        1+\frac{2\ii k_n}{A_n}\delta E
        +O(\delta E^2)
        \right]=
        -\exp\!\left[
        \frac{2\ii k_n}{A_n}\delta E
        +O(\delta E^2)
        \right].
\end{aligned}
\]
Thus, choosing the phase continuously through $\cR(E_n)=-1$,
\[
        \arg\cR(E)
        =
        \pi+\frac{2k_n}{A_n}(E-E_n)
        +O\!\left((E-E_n)^2\right).
\]
Since $\tau_{\rm W}(E)=\partial_E\arg\cR(E)$, we obtain the exact identity
\begin{equation}
        \tau_{\rm W}(E_n)
        =
        \frac{2k_n}{|\psi_n'(L)|^2}.
\label{eq:boundary-delay-derivative-main}
\end{equation}
No disorder averaging or phase approximation enters this step.

The same boundary derivative also has an independent mechanical interpretation.
Let $E_n(L)$ and $\psi_n(x;L)$ denote a normalized eigenpair when the right
Dirichlet endpoint at $x=L$ is allowed to move.  
The moving Dirichlet condition is $\psi_n(L;L)=0$.  Differentiating it with
respect to $L$ and using the chain rule gives
\[
        0
        =
        \frac{\dd}{\dd L}\psi_n(L;L)
        =
        \psi_n'(L;L)+\partial_L\psi_n(L;L),
\]
and hence
\[
        \partial_L\psi_n(L;L)=-\psi_n'(L;L).
\]
Let
\[
        \dot\psi_n(x;L):=\partial_L\psi_n(x;L),
        \qquad
        \dot E_n:=\partial_L E_n .
\]
Differentiating the eigenvalue equation
\[
        \bigl[-\partial_x^2+V(x)\bigr]\psi_n
        =E_n\psi_n
\]
with respect to $L$ gives, for $0<x<L$,
\[
        \bigl[-\partial_x^2+V(x)-E_n\bigr]\dot\psi_n
        =\dot E_n\,\psi_n .
\]
Multiplying by $\overline{\psi_n}$ and integrating over $0<x<L$, with
$\int_0^L|\psi_n|^2\,\dd x=1$, gives
\[
        \dot E_n
        =
        \int_0^L
        \overline{\psi_n}
        \bigl[-\partial_x^2+V-E_n\bigr]\dot\psi_n\,\dd x .
\]
Using the eigenvalue equation for $\psi_n$ and integrating by parts,
\[
\begin{aligned}
        \dot E_n
        &=
        \left[
        \overline{\psi_n'}\,\dot\psi_n
        -
        \overline{\psi_n}\,\dot\psi_n'
        \right]_{0}^{L}.
\end{aligned}
\]
At the fixed left endpoint,
$\psi_n(0;L)=\dot\psi_n(0;L)=0$, while at the moving right endpoint
$\psi_n(L;L)=0$ and, as shown above, $\dot\psi_n(L;L)=-\psi_n'(L;L)$. 
Consequently,
\[
        \frac{\partial E_n}{\partial L}
        =
        \overline{\psi_n'(L)}\,\dot\psi_n(L)
        =
        -|\psi_n'(L)|^2.
\]
Thus the force exerted by the normalized eigenmode on the moving Dirichlet
boundary is
\begin{equation}
        F_n:=-\frac{\partial E_n}{\partial L}
        =|\psi_n'(L)|^2.
\label{eq:boundary-force-hadamard-main}
\end{equation}

Thus Eq.~\eqref{eq:boundary-delay-derivative-main} can be read as an inverse
relation between the Wigner time delay and the force exerted by the corresponding
closed eigenmode on the boundary,
\begin{equation}
        \tau_{\rm W}(E_n)=\frac{2k_n}{F_n}.
\label{eq:delay-force-reciprocity-main}
\end{equation}
In a higher-dimensional waveguide the analogous force divided by the boundary
area is a pressure.  Here we keep the strictly one-dimensional terminology
``boundary force.''

\subsection{Finite-length distribution of the boundary force}

Introduce the dimensionless force
\begin{equation}
        \Upsilon_n
        =\frac{\pi\bar\rho_{\rm cl}(E_n)}{k_n}F_n,
        \qquad
        \widetilde\tau_{\rm W}
        =\frac{\Delta}{2\pi}\tau_{\rm W},
        \qquad
        \Delta=\bar\rho_{\rm cl}^{-1}.
\label{eq:boundary-scaled-force-main}
\end{equation}
Equation~\eqref{eq:delay-force-reciprocity-main} becomes
\begin{equation}
\Upsilon_n=\frac{1}{\widetilde\tau_{\rm W}(E_n)}.
\label{eq:boundary-force-delay-reciprocity-main}
\end{equation}
For the high-energy white-noise regime,
$\bar\rho_{\rm cl}(E)\simeq L/(2\pi k)$, so
$\Upsilon_n\simeq L|\psi_n'(L)|^2/(2k_n^2)$.  A clean Dirichlet interval has
$\Upsilon_n=1$, fixing the normalization.

The closed eigenvalues are the phase crossings
\[
        \cR(E)=-1,
        \qquad
        \varphi(E_n)=(2n+1)\pi,
\]
where $\cR(E)=e^{\ii\varphi(E)}$.  Since
$\tau_{\rm W}(E)=\partial_E\varphi(E)$, the density of such crossings in
energy can be written locally as
\[
        \sum_n\delta(E-E_n)
        =
        \tau_{\rm W}(E)
        \sum_m
        \delta\!\left(\varphi(E)-(2m+1)\pi\right).
\]
Thus sampling the delay at the closed-system eigenvalues differs from
sampling it at a generic energy by one factor of $\tau_{\rm W}$.

Let $P_{\widetilde\tau}(t)$ denote the density of
\[
        t=\widetilde\tau_{\rm W}
        =\frac{\Delta}{2\pi}\tau_{\rm W}
\]
at a generic energy, and let $P_{\rm cr}(t)$ denote the corresponding
density sampled at the phase crossings.  With the same generic-phase
averaging used in the continuum theory, the phase is uniformly distributed,
and therefore
\[
        P_{\rm cr}(t)=t\,P_{\widetilde\tau}(t).
\]
Here the normalization follows from
$\langle\widetilde\tau_{\rm W}\rangle=1$, equivalently
$\langle\tau_{\rm W}\rangle=2\pi/\Delta$.

At a crossing, Eq.~\eqref{eq:boundary-force-delay-reciprocity-main} gives
$\Upsilon=1/t$.  The ordinary change of variables between the crossing-sampled
densities therefore gives
\[
\begin{aligned}
        P_\Upsilon(\Upsilon)=
        P_{\rm cr}\!\left(\frac1\Upsilon\right)
        \left|\frac{\dd}{\dd \Upsilon}\frac1\Upsilon\right|=
        \frac{1}{\Upsilon^3}
        P_{\widetilde\tau}\!\left(\frac1\Upsilon\right).
\end{aligned}
\]
Equivalently,
\begin{equation}
        P_{\widetilde\tau}(t)
        =\frac{1}{t^3}P_\Upsilon\!\left(\frac1t\right).
\label{eq:boundary-force-delay-density-main}
\end{equation}
Since $\widetilde\tau_{\rm W}=u/\tau$, combining
Eqs.~\eqref{eq:delay-exact-density-main} and
\eqref{eq:boundary-force-delay-density-main} gives
\begin{equation}
        P_\Upsilon(\Upsilon;\tau)
        =\frac{\tau}{\Upsilon^3}
        P\!\left(\frac{\tau}{\Upsilon},\tau\right).
\label{eq:boundary-force-transform-main}
\end{equation}
Explicitly,
\begin{equation}
\begin{aligned}
        P_\Upsilon(\Upsilon;\tau)
        ={}&\frac{1}{\tau \Upsilon}e^{-\Upsilon/\tau}
        +\frac{2}{\pi \Upsilon^2}e^{-\Upsilon/(2\tau)}
        \int_0^\infty
        \frac{q\sinh(\pi q)}{q^2+\frac14}
        W_{1,\ii q}\!\left(\frac{\Upsilon}{\tau}\right)
        e^{-(q^2+\frac14)\tau}\,\dd q.
\end{aligned}
\label{eq:boundary-force-density-main}
\end{equation}
This is the complete finite-length distribution of the normalized Dirichlet
boundary force within the generic-phase weak-disorder regime.

The evolution with sample length is most transparent on a logarithmic force
scale.  Writing $s=\log \Upsilon$,
\begin{equation}
        \mathcal P_\tau(s)=e^sP_\Upsilon(e^s;\tau),
\label{eq:log-boundary-force-density-main}
\end{equation}
we have $\mathcal P_\tau(s)\to\delta(s)$ for $\tau\to0$, since a short nearly
clean interval has $\Upsilon\simeq1$.  For $\tau\gg1$, away from the extreme
small-force cutoff, the stationary part of the delay distribution gives
\begin{equation}
        P_\Upsilon(\Upsilon;\tau)\simeq\frac{1}{\tau \Upsilon}e^{-\Upsilon/\tau},
        \qquad
        \mathcal P_\tau(s)\simeq\frac1\tau e^{-e^s/\tau}.
\label{eq:boundary-force-bulk-asymptotic-main}
\end{equation}
Thus localization produces a broad distribution on the logarithmic force
scale.  The finite-length term in Eq.~\eqref{eq:boundary-force-density-main}
is essential at very small $\Upsilon$, where it cuts off the apparent $1/\Upsilon$
behavior; by Eq.~\eqref{eq:delay-force-reciprocity-main}, this is precisely the
sector of exceptionally long Wigner delays.

\begin{figure}[t]
\centering
\includegraphics[width=0.78\textwidth]{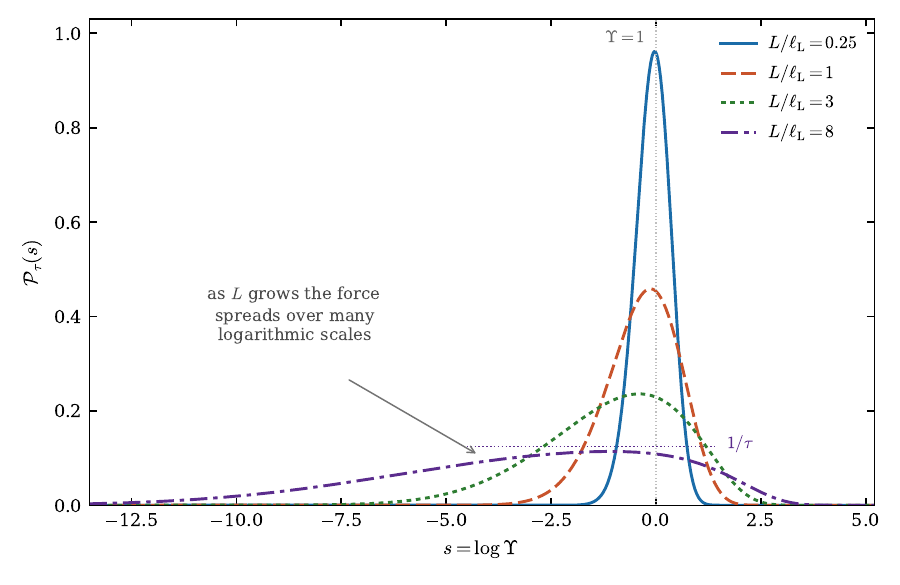}
\caption{Finite-length distribution of the normalized boundary force on a
logarithmic scale: $\mathcal P_\tau(s)$ with $s=\log\Upsilon$, from
Eq.~\eqref{eq:log-boundary-force-density-main}, for several $\tau=L/\ellL$.
The curves are obtained by solving the Fokker--Planck equation
\eqref{eq:delay-fokker-planck-main} for the delay variable and applying the
crossing-sampled transformation
\eqref{eq:boundary-force-transform-main}; they agree with the closed form
\eqref{eq:boundary-force-density-main} to five decimal places, and each is
normalized to unity over the plotted range.  A short, nearly clean sample is
concentrated at the clean value $\Upsilon=1$ (dotted vertical).  As $L/\ellL$
grows the distribution flattens towards the plateau $\mathcal P_\tau\simeq1/\tau$
(dotted horizontal), which is the crossing-sampled image of the
$\tau_{\rm W}^{-2}$ tail of the delay distribution: at $\tau=8$ the force is
spread almost uniformly over some $3.7$ decades, cut off at small force by the
finite length of the sample.}
\label{fig:boundary-force-distribution}
\end{figure}

Parametric derivatives of energy levels in disordered and chaotic systems have
a long history.  Examples include their interpretation in terms of persistent
currents under variation of an Aharonov--Bohm flux
\cite{KravtsovZirnbauer1992}, the universal statistics of level velocities
\cite{SimonsAltshuler1993,FyodorovLevelVelocity1994}, and their relation to
eigenfunction fluctuations in disordered metals
\cite{FyodorovMirlin1995,ZharekeshevKramer1999}.  The observable considered
here is the response to a local geometrical displacement of a Dirichlet
endpoint: in this case the spectral derivative is the force exerted on the
boundary and, in strict one dimension, is tied exactly to the Wigner delay.

\section{Numerical checks}
\label{sec:numerics}

 The numerical tests address the
approximations entering:  the weak-disorder
continuum limit, generic-phase averaging, and the matched spectral asymptotics
of the ultranarrow widths front.  Direct diagonalisation of the lattice model tests
the first two together with the final density formulas, while a numerical solution of the finite-$\alpha$ spectral problem provides
an independent determination of the front constant $C_*$, without using
disorder-averaged resonance data.

\subsection{Extraction of the poles}

For a chain of $N$ sites with $a=1$ and $t_{\rm c}=1$ the resonance condition
$\phi_{N+1}=w\,\phi_N$, with $w=\e^{\ii\kappa}$ and $2\cos\kappa=w+w^{-1}$, is
quadratic in $w$.  In the reciprocal variable $\mu=w^{-1}$ it acquires an
invertible leading coefficient,
\begin{equation}
\mu^{2}x=\mu A_1x+A_2x,\qquad
A_1=\text{hopping}-\mathrm{diag}(V_n),\qquad
A_2=-\mathrm{diag}(1,\dots,1,0),
\label{eq:numerics-qep}
\end{equation}
and therefore linearises to a \emph{standard} $2N\times2N$ eigenvalue problem
with companion matrix $\bigl(\begin{smallmatrix}A_1&A_2\\ \mathbf 1&0\end{smallmatrix}\bigr)$.
All poles of a given realisation are obtained in one diagonalisation, with no
root search; retaining $|w|>1$ and $0<\arg w<\pi$ selects the fundamental
strip, and $E-\ii\Gamma=2-w-w^{-1}$. Poles were validated against the Riccati condition
$\zeta_N=\beta_+$ propagated independently.  The relative residual is below
$10^{-10}$ for $\Gamma\gtrsim10^{-5}$, but the numerical accuracy begins to
deteriorate near $\Gamma\sim10^{-7}$.  We therefore restrict the analysis
below to $\Gamma>3\times10^{-7}$.  For the binned counts used here,
realisation-labelled runs give
$\mathrm{Var}/\mathrm{mean}=1.00\pm0.06$ across well-populated bins, so the
error bars shown are the Poisson counting errors $\sqrt{n}$.

Two parameter sets are used, both with Gaussian on-site disorder:
$k=1$, $D=0.06$ ($\ellL=47.2$, $\GamL=1.78\times10^{-2}$, $k\ellL=47$) and
$k=2$, $D=0.03$ ($\ellL=110.2$, $\GamL=8.25\times10^{-3}$, $k\ellL=220$). On the lattice, the continuum factor $2\ii k$ is replaced by
$\Delta=2\ii\sin k$ in the exact M\"obius map, so
\eqref{eq:main-continuum-scales} becomes $\ellL=4\sin^2\!k/D$ and
$\GamL=D/(4\sin k)$; both energies lie well away from the band centre and the
band edges.

\subsection{The width density}

\begin{figure}[t]
\centering
\includegraphics[width=\textwidth]{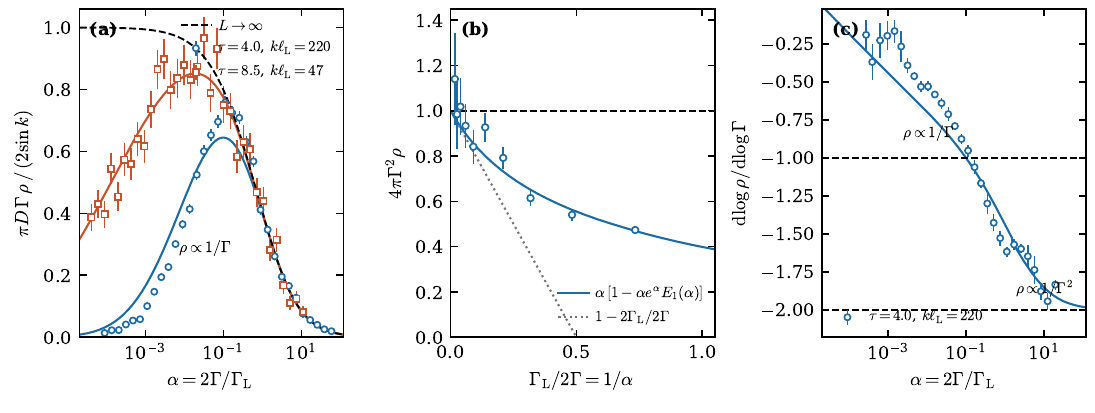}
\caption{Width density on a linear ordinate.  Circles: $k=2$, $D=0.03$,
$N=441$, $410$ realisations, $|E-E_0|<0.25$ ($k\ellL=220$, $\tau=4.0$).
Squares: $k=1$, $D=0.06$, $N=400$, $200$ realisations, $|E-E_0|<0.15$
($k\ellL=47$, $\tau=8.5$).  (a) $\pi D\Gamma\rho/(2\sin k)$ against
$\alpha=2\Gamma/\GamL$; solid curves are
$[1-\alpha\e^{\alpha}E_1(\alpha)]H_*(T_*,\tau)$, dashed the semi-infinite limit
\eqref{eq:semi_infinite_density_main}.  The abscissa must remain logarithmic:
the two features are separated by $\Gamma_{\rm ultra}/\GamL=\e^{-L/\ellL}$.
(b) The broad-width regime on fully linear axes, testing
\eqref{eq:numerics-broad}; dotted, the leading correction $1-2/\alpha$.
(c) Local exponent $\dd\log\rho/\dd\log\Gamma$ from sliding five-point weighted
regressions.  The excess over the predicted curve at $\alpha\lesssim10^{-3}$ is
the finite-$\tau$ effect of Fig.~\ref{fig:front}(b).}
\label{fig:density}
\end{figure}

Figure~\ref{fig:density} tests \eqref{eq:semi_infinite_density_main} and
\eqref{eq:ultranarrow_density_main} together.  Panel (a) plots the
dimensionless combination $\pi D\Gamma\rho/(2\sin k)$ on a linear ordinate,
where the theory predicts $[1-\alpha\e^{\alpha}E_1(\alpha)]H_*(T_*,\tau)$; the
three regimes then appear as a rounded maximum flanked by two distinct
falls --- the ultranarrow cut-off on the left, migrating as
$\Gamma_{\rm ultra}\propto\e^{-L/\ellL}$, and the crossover to
$\rho\propto1/\Gamma^{2}$ on the right.

The sharpest statement is panel (b).  In the broad-width regime
\eqref{eq:semi_infinite_density_main} predicts
\begin{equation}
4\pi\Gamma^{2}\rho=\alpha\bigl[1-\alpha\e^{\alpha}E_1(\alpha)\bigr]
=1-\frac{2}{\alpha}+\frac{6}{\alpha^{2}}-\dots,
\label{eq:numerics-broad}
\end{equation}
a statement with no free parameter and no fitted amplitude: on axes linear in
$\Gamma_{\rm L}/2\Gamma$ the data must reach exactly unity at the origin with
slope $-2$.  The three highest points give $4\pi\Gamma^{2}\rho=1.02\pm0.13$,
$0.98\pm0.15$ and $1.14\pm0.20$ at $\Gamma/E=0.037,\,0.057,\,0.086$.  Panel (c)
shows the same crossover through the local exponent, which runs from
$-0.17\pm0.10$ inside the cut-off, through the $-1$ plateau, to $-1.94\pm0.05$.
Reaching the $1/\Gamma^{2}$ regime requires $\GamL\ll\Gamma\ll E$, that is
$k\ellL\gg1$; at $k\ellL=47$ the accessible $\alpha$ is exhausted while
$\Gamma/E$ has already grown to $0.13$, and the measured exponent levells off near
$-1.8$.  This is a useful practical remark: the broad-width law is a statement
about a window that closes rapidly as $k\ellL$ decreases.

\subsection{The ultranarrow front}

\begin{figure}[t]
\centering
\includegraphics[width=\textwidth]{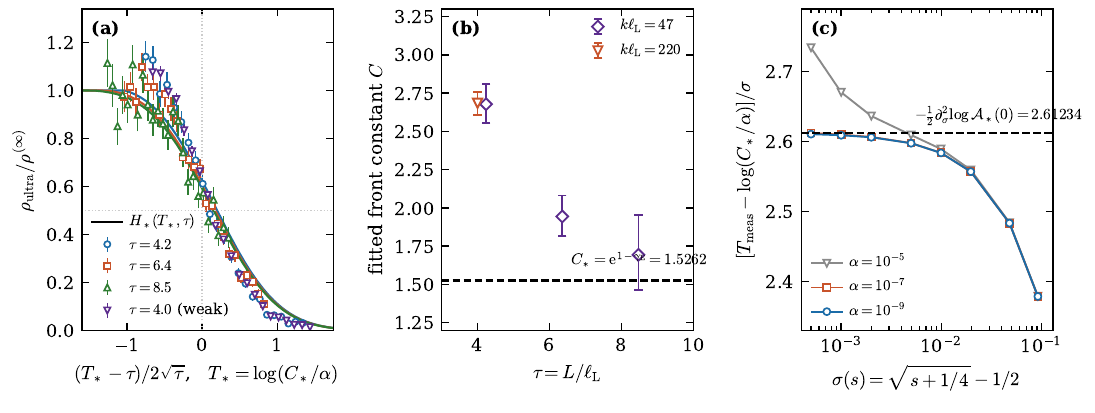}
\caption{The ultranarrow front.  (a) Measured density divided by
\eqref{eq:semi_infinite_density_main}, against the front variable, compared
with the two-term kernel \eqref{eq:two_image_kernel} (one curve per $\tau$).
Diamonds/circles/triangles: $k\ellL=47$ at $\tau=4.2,6.4,8.5$; inverted
triangles: $k\ellL=220$ at $\tau=4.0$.  (b) Front constant $C$ obtained by
fitting $H_*(\log(C/\alpha),\tau)$ to the data of (a) with $\alpha<0.3$.  The
two points near $\tau\simeq4$ come from parameter sets differing by a factor
$5$ in $k\ellL$ and coincide, showing the departure from
$C_*=\e^{1-\gamma_{\rm E}}$ (dashed) to be a function of $\tau$ alone.
(c) Independent test from the spectral problem of
Section~\ref{sec:perfect-ultranarrow}, with no disorder average.  The dashed
line is the finite-$\sigma$ correction predicted by
Eq.~\eqref{eq:numerical-amplitude-correction}.}
\label{fig:front}
\end{figure}

Figure~\ref{fig:front}(a) divides the measured density by
\eqref{eq:semi_infinite_density_main} and plots the result against the front
variable $(T_*-\tau)/2\sqrt\tau$.  The data follow the two-term kernel
\eqref{eq:two_image_kernel} over the whole crossover; in particular both terms of the kernel are needed to describe the front
correctly, raising the kernel at the front centre from $\tfrac12$ to
$H_*(\tau,\tau)=0.63,\,0.61,\,0.59$ at the three lengths.

Fitting the front position, i.e.\ replacing $C_*$ by a free constant $C$ in
$T_*=\log(C/\alpha)$, gives panel (b): $C$ falls monotonically with $\tau$
towards $C_*=\e^{1-\gamma_{\rm E}}$, reaching $1.69\pm0.25$ at $\tau=8.5$.  The
residual excess is a finite-$\tau$ effect and not a lattice artefact: the two
parameter sets, which differ by a factor $5$ in $k\ellL$, give
$C=2.68\pm0.13$ at $\tau=4.24$ and $C=2.68\pm0.08$ at $\tau=4.00$.  Its origin
is the $O(\sigma^{2})$ term in the spectral amplitude
\eqref{eq:stationary-amplitude-exact}, which is not small when the front is
controlled by $\sigma=O(\tau^{-1/2})$.  Larger
$\tau$ is inaccessible by this route, since $\Gamma_{\rm ultra}$ of
\eqref{eq:ultranarrow_scale_main} falls below the pole-resolution floor once
$\tau\gtrsim9$.

The constant is therefore determined instead from the spectral problem itself,
without relying on the disorder-averaged resonance data.  The endpoint resolvent
$\widehat H_\alpha(s)=-1/\mathscr W_\alpha(-s)$ of
\eqref{eq:ultranarrow-starting-resolvent} is evaluated by integrating the
confluent-Heun equation \eqref{eq:confluent-heun-form} numerically --- the regular
branch outward from $x=0$, the  decaying solution branch inward from $\alpha x\gg1$ ---
and $T_{\rm meas}\equiv-\log[s\widehat H_\alpha(s)]/\sigma(s)$ is compared with
$T_*=\log(C_*/\alpha)$. \underline{} Using Eq.~\eqref{eq:app-numerical-resolvent-asymptotic} and neglecting the
$O(\alpha)$ factor $e^{-\alpha/2}$ on this scale, we obtain
\begin{equation}
        \frac{T_{\rm meas}-T_*}{\sigma}
        \longrightarrow
        -\frac12\partial_\sigma^2\log\mathcal A_*(0)
        =
        \frac12\left(\frac{5\pi^2}{6}-3\right)
        =2.61234\ldots .
\label{eq:numerical-amplitude-correction}
\end{equation}
The computed values approach this limit to $2\times10^{-3}$, providing an
independent numerical check of the matching amplitude
\eqref{eq:stationary-amplitude-exact}.

\subsection{Summary}

Within the regime of validity, the closed forms of
Sections~\ref{sec:radial} and \ref{sec:perfect-ultranarrow} reproduce the lattice
data without adjustable parameters: the coefficient $1/4\pi$ of the
broad-width law to $10\%$, the $1/\Gamma\to1/\Gamma^{2}$ crossover exponent to
$0.05$, and the shape of the ultranarrow front including its image term.  The
front constant $C_*=\e^{1-\gamma_{\rm E}}$, approached but not resolved by the
Monte Carlo at $\tau\lesssim9$, is confirmed to four digits by direct solution
of the associated spectral problem.  The two approximations flagged in
Section~\ref{subsec:main-status} --- the weak-disorder continuum limit and the
generic-phase average --- are thus supported at the level of the final
predictions, and the observed departures are traced to explicitly identified
corrections rather than to either of them.

\section{Discussion}
\label{sec:discussion}

We have developed a finite-length theory of reflection resonances in a
one-dimensional weakly disordered sample, using the reflection-intensity
evolution in the presence of uniform absorption.  The main new result connects
the semi-infinite resonance density derived recently in
Ref.~\cite{FyodorovMeibohm2026} to the finite-length cutoff at ultranarrow
resonance widths.  The analysis is facilitated by a supersymmetric
factorization of the Hamiltonian associated with the Fokker--Planck evolution.
Passing to the partner problem separates the stationary state from the
positive spectrum and makes the spectral matching at the distant logarithmic
wall particularly simple.  Poisson summation then yields the finite-length
first-passage kernel, while the smooth matching amplitude fixes the order-one
displacement $C_*=\e^{1-\gamma_{\rm E}}$.

The reflection formulation also provides a direct route to the finite-length
Wigner time-delay problem.  Near unit reflection, the weak-absorption variable
$u=(1-r)/\alpha$ becomes $\GamL\tau_{\rm W}$ and obeys the
Comtet--Texier diffusion.  A nonideal contact enters through the corresponding
arbitrary-coupling phase average.  For the closed continuum sample the same
time delay has a more direct interpretation.  The combination of two exact identities
\[
        \tau_{\rm W}(E_n)=\frac{2k_n}{|\psi_n'(L)|^2},\qquad -\partial_L E_n=|\psi_n'(L)|^2,
\]
turns the complete finite-length time-delay distribution into the
boundary-force density \eqref{eq:boundary-force-density-main}.  This is the
strict one-dimensional boundary counterpart of the general single-channel
correspondence between time delays and eigenfunctions of Ossipov and Fyodorov
\cite{OssipovFyodorov2005}.

The force interpretation also suggests optical tests.  One route is continuous
spatial imaging.  Anderson-localized optical modes have been resolved directly
by near-field scanning microscopy
\cite{RiboliEtAl2011,FaggianiEtAl2016}, while near-field probes can also
convert a local modal response into a measurable resonance-frequency shift
\cite{VignoliniEtAl2010}.  In a structure with an effectively Dirichlet
boundary condition, resolving the mode sufficiently close to the boundary
would give access to the boundary slope and hence to the optical analogue of
$|\psi_n'(L)|^2$.

A second possible route is to generate many disorder realizations in a
programmable photonic potential.  Anderson localization has been observed in
optically induced disordered lattices \cite{SchwartzEtAl2007}; more directly
for this purpose, computer-controlled spatial light modulators have been used
to create reproducible ensembles of randomized photonic potentials with
tunable disorder strength and correlation length
\cite{BoguslawskiEtAl2013}.  Such a platform could in principle vary the
position of a terminal boundary as well as the disorder realization and
determine the corresponding modal frequency or propagation-constant shift.

A still closer mechanical realization is provided by disordered
optomechanical-crystal waveguides.  Disorder-induced localized optical modes
exhibit measurable optomechanical coupling, and a broad distribution of
coupling rates has already been observed
\cite{GarciaEtAl2017,ArreguiEtAl2023}.  Finally, laser-written truncated
one-dimensional arrays have demonstrated Anderson localization specifically
at a boundary \cite{SzameitEtAl2010}.  These platforms are not identical to
the scalar Dirichlet model, but together they provide practical routes for
testing the predicted finite-length distribution and scaling of the boundary
response.

A second global consequence is the direct one-dimensional derivation of the
Moldauer--Simonius sum rule.  For fixed $0<\cT<1$, integrating the width
density gives Eq.~\eqref{eq:MS-average-main}.  The result was known from
scattering theory and random-matrix theory, and has recently been derived in
a general nonlinear-$\sigma$-model framework; see
Ref.~\cite{KurilovOstrovsky2026}.  Our contribution here is a direct
microscopic derivation of this relation for the strict one-dimensional
Anderson problem.  The same calculation makes its perfect-coupling logarithmic
behaviour concrete: the transparency-generated cutoff of the broad-width
sector moves to infinity as $\cT\uparrow1$, exposing the
$1/(4\pi\Gamma^2)$ density over an ever wider interval.  The logarithmic first
moment of this inverse-square tail is exactly the Moldauer--Simonius
divergence.

The deep ultranarrow tail is instead Gaussian in
$\log[\Gamma_{\rm ultra}(L)/\Gamma]$ and is therefore of log-normal type
rather than a pure power law.  Its physical mechanism is different: once the
localization centre has reached the closed end, further reduction of the
width requires an atypically large accumulated logarithmic spatial decay
across the full sample.  Thus the same resonance distribution contains two
opposite extreme sectors with different origins: a local short-time escape
sector producing the broad $1/\Gamma^2$ law, and a whole-sample
rare-localization sector producing the ultranarrow moving front.

Within the stated regime, the numerical checks of Section~\ref{sec:numerics} test the coefficient
$1/(4\pi)$ of the broad-width $\Gamma^{-2}$ law, the crossover
scaling, and the ultranarrow kernel without adjustable parameters, while the
spectral analysis independently fixes the displacement
$C_*=\e^{1-\gamma_{\rm E}}$.

Several problems remain open even within strict one dimension.  In reflection
geometry, the short-sample regime, where narrow resonances are generated by
atypical disorder configurations, is parametrically opposite to the
long-sample regime studied here.  This regime was treated in
Ref.~\cite{FyodorovMeibohm2026} using a WKB approach.  It would be useful to
revisit the problem from a complementary Feynman--Kac path-integral
formulation leading to an instanton analysis.

For a sample open at both ends, projective Kac--Rice can still count poles
from an inverse reflection amplitude, but a logarithmic reflection observable
also sees zeros of that reflection amplitude.  Such zeros are physically
interesting in their own right: on the real axis they correspond to
reflectionless scattering, while zeros of the full scattering matrix in an
absorbing system underlie coherent perfect absorption, see e.g. discussion in
\cite{FyodorovSuwunnaratKottos2017}.  The latter phenomenon has been
demonstrated experimentally also in a disordered medium, in the form of a
``random anti-laser'' with nearly perfect absorption of appropriately shaped
incoming waves \cite{PichlerEtAl2019}.  A joint statistical treatment of
poles and zeros lies outside the present scalar density calculation.  The
phase-resolved sub-wavelength regime, the lattice band centre, and the band
edges likewise require separate scaling theories.

A particularly important direction is the extension of the present strict
one-dimensional results to higher dimensions, and especially the
characterization of resonance statistics near the Anderson localization
transition.  Despite important qualitative and analytical insights, this
problem remains much less understood than its one-dimensional counterpart;
see, for example, Refs.~\cite{Kottos2005,FyodorovSkvortsovTikhonov2024}.

Another largely open direction concerns the statistics of non-orthogonal
eigenfunctions of the effective non-Hermitian Hamiltonian
Eq.~\eqref{eq:effective_H} in Anderson-localized systems.  This situation
contrasts with that in zero-dimensional quantum chaotic systems, where
eigenfunction non-orthogonality and its physical consequences have been
studied in considerable detail; see, e.g.,
Refs.~\cite{SchomerusFrahmPatraBeenakker2000,FyodorovMehlig2002,FyodorovSavin2012,
FyodorovOsman2022}.

Finally, a largely unexplored direction is resonance scattering in
interacting disordered many-body systems.  Here by resonances we mean poles
of the scattering matrix, or equivalently complex eigenvalues of an
appropriate effective non-Hermitian Hamiltonian describing coupling of the
interacting system to external channels.  It would be particularly
interesting to understand how interaction and Anderson or many-body
localization affect the statistics of such poles and their widths.  This
problem is conceptually distinct from the ``many-body resonances'' commonly
discussed for closed interacting systems, where the term refers instead to
near-degeneracies and hybridization of many-body eigenstates.

\subsection{Towards a rigorous treatment}

The continuum derivation developed above is controlled at the level usually
adopted in theroretical physics: the approximations are explicit, their
regimes of validity are identified, and the resulting limiting problems are
analysed directly.  We have not, however, attempted to convert these
arguments into a fully rigorous asymptotic theory.  This appears to be a
realistic programme, comparable in scope to rigorous analyses of
one-dimensional closed random Schr\"odinger operators by Dumaz and
collaborators, but with the additional difficulties associated with the open
boundary and resonance condition.  Three main steps would be required.

First, one needs to justify the weak-disorder limit that takes the exact
M\"obius recursion \eqref{eq:exact-interface-mobius} to the complex reflection
diffusion \eqref{eq:main-complex-reflection-diffusion}.  This is the
reflection-coordinate analogue of the Riccati diffusion limits used in the
rigorous theory of one-dimensional localization
\cite{DumazLabbe2020Localization,DumazLabbe2024Crossover}; a related critical
scaling leads to the $\mathrm{Sch}_\tau$ process
\cite{KritchevskiValkoVirag2012}.

Second, the rapid reflection phase has to be averaged uniformly over the
radial evolution.  Appendix~\ref{app:reflection-phase-evolution} exhibits the
underlying separation of scales: the clean phase rotates with rate $2k$,
whereas the radial drift and disorder-induced diffusion evolve on the much
longer localization and absorption scales.  A rigorous treatment would have
to turn this scale separation into a uniform phase-averaging statement.

Third, the ultranarrow front requires uniform control of the spectral
asymptotics used in Sections~\ref{subsec:ultranarrow-poisson} and
\ref{subsec:ultranarrow-spectral-front}.  This includes the opening discrete
spectrum and its boundary weights, the Legendre--Bessel matching, the Poisson
summation, and the contour deformation through the front region.  One must
in particular show uniformly on the front scale that the matching errors,
including the derivatives needed when reconstructing the resonance density,
remain negligible; that the higher Poisson harmonics are exponentially
suppressed; and that higher terms in the small-$\sigma$ expansion of the
matching amplitude produce only subleading corrections.  The terms discarded
when the energy derivatives are reduced to the dominant absorption derivative
must likewise be shown to remain smaller throughout this regime.

\section*{Acknowledgements}
This paper is dedicated to Prof. Uzy Smilansky on the occasion of his 85th birthday. The influence of Uzy's contributions to subjects studied here and on the author's research interest is undeniable and gratefully acknowledged. 

The research reported in this paper was supported by grant UKRI1015, ``Non-Hermitian random
matrices: theory and applications.''  The author is grateful to Prof. Pavel Exner for discussion of results and interest in the work, and the Doppler
Institute for Mathematical Physics and Applied Mathematics in Prague for
hospitality during completion of this paper.

The author acknowledges assistance from ChatGPT (OpenAI) in improving the
organization and clarity of the presentation, performing consistency checks,
and designing numerical checks.  Claude Opus 5 (Anthropic) was used for the numerical computations in
 Section~\ref{sec:numerics}, for initial drafting of that section, for help with creating figures 
and for critical reading.  All scientific arguments, results, and conclusions
remain the responsibility of the author.

\appendix

\section{It\^o formula, complex diffusions, and conditional moments}
\label{app:complex-Ito-diffusions}

For completeness, we recall the It\^o change-of-variables formula and then
explain how the conditional moments of a complex increment determine the
drift and diffusion covariance of the corresponding process.

\subsection{It\^o formula and quadratic variation}
\label{Ito} 
Let a real process $U_s$ satisfy
\[
        \dd U_s
        =
        a(U_s,s)\,\dd s
        +
        b(U_s,s)\,\dd W_s,
\]
where $W_s$ is a standard real Brownian motion.  For a twice
differentiable function $f(u,s)$, It\^o's formula reads
\begin{equation}
\begin{aligned}
        \dd f(U_s,s)
        &=
        \left[
        \partial_s f
        +a\,\partial_u f
        +\frac12 b^2\,\partial_u^2 f
        \right]\dd s
        +
        b\,\partial_u f\,\dd W_s .
\end{aligned}
\label{eq:app-one-dimensional-Ito}
\end{equation}
The formula may be remembered in the differential form
\[
        \dd f
        =
        \partial_s f\,\dd s
        +
        \partial_u f\,\dd U_s
        +
        \frac12\partial_u^2f\,(\dd U_s)^2,
\]
together with the It\^o multiplication rules
\[
        (\dd W_s)^2=\dd s,
        \qquad
        \dd W_s\,\dd s=0,
        \qquad
        (\dd s)^2=0.
\]
Indeed, since $\dd W_s=O(\sqrt{\dd s})$, the only contribution to
$(\dd U_s)^2$ of order $\dd s$ is $b^2(\dd W_s)^2=b^2\,\dd s$; terms
containing $(\dd s)^2$ or $\dd s\,\dd W_s$ are of higher order.  The
second-derivative term is the It\^o correction.  It vanishes for a linear
change of variable, but must be retained for a nonlinear one.

More generally, suppose that
\[
        \dd X_i
        =
        a_i(\boldsymbol X,s)\,\dd s
        +
        \sum_{\mu=1}^{m}
        b_{i\mu}(\boldsymbol X,s)\,\dd W_s^{(\mu)},
        \qquad i=1,\ldots,n,
\]
with independent real Brownian motions, so that
\[
        \dd W_s^{(\mu)}\dd W_s^{(\nu)}
        =
        \delta_{\mu\nu}\,\dd s.
\]
For a smooth function $F(\boldsymbol X,s)$,
\begin{equation}
\begin{aligned}
        \dd F
        &=
        \left[
        \partial_sF
        +\sum_i a_i\partial_iF
        +\frac12\sum_{i,j}
        A_{ij}\partial_i\partial_jF
        \right]\dd s
        \\
        &\quad
        +
        \sum_{i,\mu}
        b_{i\mu}\partial_iF\,\dd W_s^{(\mu)},
        \qquad
        A_{ij}
        =
        \sum_{\mu=1}^{m}b_{i\mu}b_{j\mu}.
\end{aligned}
\label{eq:app-multidimensional-Ito}
\end{equation}
For two variables this is equivalently
\[
\begin{aligned}
        \dd F(X,Y)
        &=
        F_X\,\dd X+F_Y\,\dd Y
        +\frac12F_{XX}(\dd X)^2
        +F_{XY}\,\dd X\,\dd Y
        +\frac12F_{YY}(\dd Y)^2.
\end{aligned}
\]

\subsection{Complex conditional moments and diffusion covariance}

Let $Z=X+\ii Y$ satisfy a complex-valued It\^o stochastic differential equation driven by
independent real Brownian motions,
\begin{equation}
\label{eq:app-general-complex-Ito}
        \dd Z
        =
        b(Z,\bar Z)\,\dd s
        +
        \sum_{\mu=1}^{m}
        \sigma_\mu(Z,\bar Z)\,\dd W_s^{(\mu)}.
\end{equation}
The relevant complex quadratic variations are
\[
        (\dd Z)^2
        =
        \sum_{\mu=1}^{m}\sigma_\mu^2\,\dd s,
        \qquad
        \dd Z\,\dd\bar Z
        =
        \sum_{\mu=1}^{m}|\sigma_\mu|^2\,\dd s.
\]
For example, applying It\^o's product rule to
$r=|Z|^2=Z\bar Z$ gives
\[
\begin{aligned}
        \dd r
        &=
        \bar Z\,\dd Z
        +Z\,\dd\bar Z
        +\dd Z\,\dd\bar Z
        \\
        &=
        \left[
        2\Re(\bar Z b)
        +\sum_{\mu=1}^{m}|\sigma_\mu|^2
        \right]\dd s
        +
        2\sum_{\mu=1}^{m}
        \Re(\bar Z\sigma_\mu)\,\dd W_s^{(\mu)}.
\end{aligned}
\]
The term $\sum_\mu|\sigma_\mu|^2$ is the It\^o correction.  This is the
identity used to pass from the complex reflection diffusion
\eqref{eq:main-complex-reflection-diffusion} to the phase-resolved
intensity equation
\eqref{eq:full-intensity-equation-before-average}.

For the later change of variable $y=1-r$, one has
$\dd y=-\dd r$ because the second derivative vanishes.  By contrast, for
$x=r/(1-r)$,
\[
        \frac{\dd x}{\dd r}
        =
        \frac{1}{(1-r)^2},
        \qquad
        \frac{\dd^2x}{\dd r^2}
        =
        \frac{2}{(1-r)^3},
\]
so the quadratic-variation correction is essential in deriving
Eq.~\eqref{eq:x-radial-sde-main}.

The first conditional moment determines the complex drift:
\begin{equation}
\label{eq:app-complex-first-moment}
        \frac{\EE[\dd Z\mid Z]}{\dd s}
        =
        b(Z,\bar Z).
\end{equation}
The two second conditional moments are
\begin{equation}
\label{eq:app-complex-second-moments}
\begin{aligned}
        Q(Z,\bar Z)
        &:=
        \frac{\EE[(\dd Z)^2\mid Z]}{\dd s}
        =
        \sum_{\mu=1}^{m}\sigma_\mu^2,
        \\
        C(Z,\bar Z)
        &:=
        \frac{\EE[|\dd Z|^2\mid Z]}{\dd s}
        =
        \sum_{\mu=1}^{m}|\sigma_\mu|^2.
\end{aligned}
\end{equation}

To relate these quantities to the usual real covariance matrix, write
\[
        \dd Z=\dd X+\ii\,\dd Y.
\]
Then
\[
        (\dd Z)^2
        =
        (\dd X)^2-(\dd Y)^2
        +2\ii\,\dd X\,\dd Y, \qquad
        |\dd Z|^2=
        (\dd X)^2+(\dd Y)^2.
\]
It follows that
\begin{equation}
\label{eq:app-real-covariances-from-complex}
\begin{aligned}
        \frac{\EE[(\dd X)^2\mid Z]}{\dd s}
        &=
        \frac{C+\Re Q}{2},
        \\
        \frac{\EE[(\dd Y)^2\mid Z]}{\dd s}
        &=
        \frac{C-\Re Q}{2},
        \\
        \frac{\EE[\dd X\,\dd Y\mid Z]}{\dd s}
        &=
        \frac{\Im Q}{2}.
\end{aligned}
\end{equation}
Thus $Q$ and $C$ determine the complete real diffusion covariance matrix
\begin{equation}
\label{eq:app-real-diffusion-matrix}
        \mathsf A
        =
        \frac12
        \begin{pmatrix}
                C+\Re Q & \Im Q
                \\
                \Im Q   & C-\Re Q
        \end{pmatrix}.
\end{equation}
Together with the drift $b$, this matrix determines the diffusion generator.
Different choices of the coefficients $\sigma_\mu$ may represent the same
matrix $\mathsf A$ and therefore the same diffusion law.

A particularly important case occurs when
\begin{equation}
\label{eq:app-rank-one-condition}
        |Q|=C.
\end{equation}
Indeed,
\[
        \det\mathsf A
        =
        \frac14\bigl(C^2-|Q|^2\bigr),
\]
so Eq.~\eqref{eq:app-rank-one-condition} implies that $\mathsf A$ has rank
one.  The process may then be represented using a single real Brownian
motion,
\begin{equation}
\label{eq:app-single-real-noise}
        \dd Z
        =
        b\,\dd s+\sigma\,\dd W_s, \quad \mbox{where} \quad \sigma^2=Q,
        \,\,\,
        |\sigma|^2=C.
\end{equation}
The replacement $\sigma\mapsto-\sigma$ is immaterial, since it is equivalent
to $\dd W_s\mapsto-\dd W_s$.

For the reflection process derived in
Section~\ref{subsec:weak-disorder-moments}, the limiting second moments are
\[
        Q(R)
        =
        -\frac{D}{4k^2}(1+R)^4,
        \qquad
        C(R)
        =
        \frac{D}{4k^2}|1+R|^4.
\]
They obey $|Q(R)|=C(R)$, and the covariance therefore has rank one.  A
convenient choice of noise coefficient is
\[
        \sigma(R)
        =
        -\frac{\ii\sqrt D}{2k}(1+R)^2,
\]
for which
\[
        \sigma(R)^2=Q(R),
        \qquad
        |\sigma(R)|^2=C(R).
\]
Combining this coefficient with the limiting first conditional moment gives
the complex reflection diffusion
\eqref{eq:main-complex-reflection-diffusion}.

Strictly speaking, the convergence of the first two conditional moments
identifies the limiting diffusion provided that higher one-step moments and
possible jump contributions vanish in the continuum scaling.  In the
present construction this follows from
$\xi_n=O_{\mathbb P}(\sqrt a)$ and from the expansion of the exact
M\"obius map through terms quadratic in $\xi_n$.

\subsection{Phase evolution of the reflection amplitude}
\label{app:reflection-phase-evolution}

For $\cR=\sqrt r\,\e^{\ii\varphi}$ with $r>0$, It\^o's formula applied
locally to $\log\cR$ gives
\[
        \dd\varphi
        =
        A_\varphi(r,\varphi)\,\dd s
        +
        B_\varphi(r,\varphi)\,\dd W_s.
\]
Substituting the drift and noise coefficients from
Eq.~\eqref{eq:main-complex-reflection-diffusion} gives
\[
\begin{aligned}
        &A_\varphi(r,\varphi)=
        2k
        +\frac{\delta E}{2k}
        \left[
        2+\left(\sqrt r+\frac{1}{\sqrt r}\right)\cos\varphi
        \right]
        -\frac{\eta}{2k}
        \left(\sqrt r-\frac{1}{\sqrt r}\right)\sin\varphi
        \\
        &\quad 
        -\frac{D}{8k^2}
        \left[
        \left(r+\frac{1}{r}\right)\sin(2\varphi)
        +2\left(\sqrt r+\frac{1}{\sqrt r}\right)\sin\varphi
        \right],
        \\
        &B_\varphi(r,\varphi)=
        -\frac{\sqrt D}{2k}
        \left[
        2+\left(\sqrt r+\frac{1}{\sqrt r}\right)\cos\varphi
        \right].
\end{aligned}
\]
Thus, for $r$ away from $0$ and $\infty$, the leading phase drift is simply
the clean rotation $2k$, while the remaining
drift and noise terms have respectively the scales
$|\delta E|/k$, $\eta/k$, $D/k^2$, and $\sqrt D/k$.
The apparent singularity as $r\to0$ reflects the fact that the phase of
$\cR$ is undefined at the origin; it does not imply a singularity of the
complex reflection process itself.

\section{Exact spectral problem and boundary normalization}
\label{app:heun}

This appendix supplies the exact finite-$\alpha$ operator statements used in
the real-spectrum calculation.  The two subsections fix the self-adjoint
boundary conditions, the spectral gap, the weighted Wronskian, and the
boundary normalization without making a small-$\alpha$ approximation.  We
also record the exact boundary Laplace transform used in the numerical check
of Section~\ref{sec:numerics}.

\subsection{Self-adjoint intensity operator, partner domain, and confluent-Heun equation}
\label{app:heun-self-adjoint}

The eigenvalue equation is Eq.~\eqref{eq:ultranarrow-finite-alpha-eigenvalue},
with stationary density $\pi_\alpha(x)=\alpha e^{-\alpha x}$.  Introduce
\[
        \mathfrak p_\alpha(x)=x(1+x)\pi_\alpha(x).
\]
Then
\begin{equation}
        \cL_\alpha f
        =\frac1{\pi_\alpha(x)}
        \partial_x\!\left[\mathfrak p_\alpha(x)f'(x)\right].
\label{eq:heun-divergence-form}
\end{equation}
For functions satisfying the physical boundary conditions,
\[
\begin{aligned}
        \int_0^\infty\overline f\,\cL_\alpha g\,\pi_\alpha\,\dd x
        =-\int_0^\infty
        x(1+x)\pi_\alpha(x)\overline{f'(x)}g'(x)\,\dd x
        =\int_0^\infty
        \overline{\cL_\alpha f}\,g\,\pi_\alpha\,\dd x.
\end{aligned}
\]
Thus the physical realization of $-\cL_\alpha$ is nonnegative and
self-adjoint in $L^2(\pi_\alpha(x)\,\dd x)$.

\paragraph{Boundary at $x=0$.}
Near $\vartheta=0$, the two local solutions have the same leading
power-law behavior, with the second solution acquiring a logarithmic term.  The nonlogarithmic
solution can be normalized by $u_\lambda(0)=1$ and has
\begin{equation}
        u_\lambda(x)=1-\lambda x+O(x^2).
\label{eq:heun-regular-origin}
\end{equation}
The second linearly independent solution contains $\log x$.  Its derivative behaves as
$x^{-1}$ and therefore
\[
        \int_0 x|\phi'(x)|^2\,\dd x=\infty.
\]
Thus the second solution is not admissible. The physical boundary condition is
therefore
\begin{equation}
        \lim_{x\to0}\mathfrak p_\alpha(x)\phi_\lambda'(x)=0.
\label{eq:heun-origin-boundary-condition}
\end{equation}

\paragraph{Boundary at $x=\infty$.}
For fixed $\alpha>0$, the physical solution has
\begin{equation}
        v_\lambda(x)
        =1-\frac{\lambda}{\alpha x}+O(x^{-2}),
        \qquad x\to\infty.
\label{eq:heun-recessive-infinity}
\end{equation}
We use $v_\lambda(\infty)=1$ as its normalization.  The independent solution
grows as $e^{\alpha x}x^{-2}$ up to relative inverse-power corrections and
is not square integrable with respect to $\pi_\alpha(x)\,\dd x$.  Thus the
condition at infinity also selects a unique branch.

\paragraph{Partner domain and the gap above the stationary mode.}
The factorization $\mathcal H_-=Q_\alpha^\dagger Q_\alpha$ and the partner
$\mathcal H_+=Q_\alpha Q_\alpha^\dagger$ were introduced in
Eqs.~\eqref{eq:susy-factorization-main} and
\eqref{eq:partner-hamiltonian-main}.  The zero mode is
$\Psi_0=\sqrt{m_\alpha}$ and satisfies $Q_\alpha\Psi_0=0$.  If
$\mathcal H_-\Psi=\lambda\Psi$ with $\lambda>0$, then
\[
        \mathcal H_+(Q_\alpha\Psi)=Q_\alpha \mathcal H_-\Psi
        =\lambda Q_\alpha\Psi,
\]
and $Q_\alpha\Psi\neq0$.  Conversely, $Q_\alpha^\dagger$ maps a positive
partner eigenstate back to $\mathcal H_-$.  Hence the positive spectra coincide.

Near $\vartheta=0$, we have $ U_{+,\alpha}(\vartheta)
        =\frac{3}{4\vartheta^2}+O(1)$, so the local behaviors are $\vartheta^{3/2}$ and $\vartheta^{-1/2}$; only
the former is square integrable.  At infinity the term
$\alpha^2\sinh^2\vartheta/16$ is confining.  The partner therefore has the
standard half-line self-adjoint realization with discrete simple spectrum.
For every nonzero $\chi$ in its quadratic-form domain,
\begin{equation}
\begin{aligned}
        \langle\chi,\mathcal H_+\chi\rangle
        =\frac14\|\chi\|^2
        +\int_0^\infty|\chi'(\vartheta)|^2\,\dd\vartheta
        +\int_0^\infty
        \left[
        \frac{3}{4\sinh^2\vartheta}
        +\frac{\alpha^2}{16}\sinh^2\vartheta
        \right]|\chi(\vartheta)|^2\,\dd\vartheta
        >\frac14\|\chi\|^2.
\end{aligned}
\label{eq:app-partner-gap-form}
\end{equation}
Because every excited eigenvalue of $\mathcal H_-$ occurs in $\mathcal H_+$,
\begin{equation}
        \lambda_n(\alpha)>\frac14,
        \qquad n\geq1,
        \qquad \alpha>0.
\label{eq:app-excited-gap}
\end{equation}
This proves the inequality used in
Eq.~\eqref{eq:heun-discrete-spectrum}.  The supersymmetry here belongs to the
auxiliary radial spectral problem; no physical supersymmetry of the Anderson
Hamiltonian is implied.

\paragraph{Confluent-Heun form.}
With $z=-x$, Eq.~\eqref{eq:ultranarrow-finite-alpha-eigenvalue} becomes
\begin{equation}
        \frac{\dd^2\phi_\lambda}{\dd z^2}
        +\left[\alpha+\frac1z+\frac1{z-1}\right]
        \frac{\dd\phi_\lambda}{\dd z}
        +\frac{\lambda}{z(z-1)}\phi_\lambda=0.
\label{eq:confluent-heun-form}
\end{equation}
It has regular singular points at $z=0$ and $z=1$ and an irregular singular
point at infinity, and is therefore a confluent-Heun equation.  We use the
explicit differential equation rather than a convention-dependent Heun
notation.

\subsection{Weighted Wronskians, quantization, and boundary normalization}
\label{app:heun-wronskian}

Let $u_\lambda$ be the solution regular at $x=0$, normalized by
$u_\lambda(0)=1$, and let $v_\lambda$ be the  solution decaying at infinity,
normalized by $v_\lambda(\infty)=1$.  Their weighted Wronskian is
\begin{equation}
        \mathscr W_\alpha(\lambda)
        =\mathfrak p_\alpha(x)
        \left[u_\lambda(x)v_\lambda'(x)
        -u_\lambda'(x)v_\lambda(x)\right].
\label{eq:heun-weighted-wronskian}
\end{equation}
The differential equation gives
$\partial_x\mathscr W_\alpha(\lambda)=0$, so this quantity is independent of
the matching point.  An eigenvalue occurs precisely when the two solutions
are linearly dependent:
\begin{equation}
        \mathscr W_\alpha(\lambda_n)=0.
\label{eq:heun-quantization}
\end{equation}

\paragraph{Boundary resolvent and normalized spectral weight.}
Away from the spectrum, the Green kernel is
\[
        G_\alpha(x,x_0;\lambda)
        =-\frac{u_\lambda(x_<)v_\lambda(x_>)}
        {\mathscr W_\alpha(\lambda)},
        \qquad
        x_<=\min\{x,x_0\},\quad x_>=\max\{x,x_0\}.
\]
The sign follows by integrating the resolvent equation across $x=x_0$ and
using the derivative jump determined by $\mathfrak p_\alpha(x_0)$.  Taking
$x=0$ and then $x_0\to\infty$ gives
\begin{equation}
        G_\alpha(0,\infty;\lambda)
        =-\frac1{\mathscr W_\alpha(\lambda)}.
\label{eq:heun-endpoint-resolvent}
\end{equation}
At an eigenvalue $\lambda_n$, the two solutions are proportional and define
the same eigenfunction.  Let $\phi_n$ denote this eigenfunction normalized by
\[
        \int_0^\infty
        \phi_n(x)^2\pi_\alpha(x)\,\dd x=1.
\]
Since $u_{\lambda_n}(0)=1$ and $v_{\lambda_n}(\infty)=1$,
\[
        u_{\lambda_n}(x)
        =\frac{\phi_n(x)}{\phi_n(0)},
        \qquad
        v_{\lambda_n}(x)
        =\frac{\phi_n(x)}{\phi_n(\infty)}.
\]
Differentiating the Wronskian identity at $\lambda=\lambda_n$ then gives
\begin{equation}
        \phi_n(0)\phi_n(\infty)
        =\frac1{\partial_\lambda\mathscr W_\alpha(\lambda_n)}.
\label{eq:heun-endpoint-product}
\end{equation}
This is the exact boundary coefficient used in the real spectral sum.

\paragraph{Exact partner Wronskian relation.}
Let
\[
        \Psi_u=\sqrt{m_\alpha}\,u_\lambda,
        \qquad
        \Psi_v=\sqrt{m_\alpha}\,v_\lambda,
\]
and, for $\lambda\neq0$,
\[
        \widetilde\Psi_u=\frac1\lambda Q_\alpha\Psi_u,
        \qquad
        \widetilde\Psi_v=\frac1\lambda Q_\alpha\Psi_v.
\]
The ordinary Schr\"odinger Wronskian of $\Psi_u$ and $\Psi_v$ equals
Eq.~\eqref{eq:heun-weighted-wronskian}.  For two solutions of
$\mathcal H_-f=\lambda f$ and $\mathcal H_-g=\lambda g$,
\begin{equation}
        W[Q_\alpha f,Q_\alpha g]=\lambda W[f,g].
\label{eq:app-Darboux-Wronskian}
\end{equation}
Hence
\begin{equation}
        \widetilde{\mathscr W}_\alpha(\lambda)
        :=W[\widetilde\Psi_u,\widetilde\Psi_v]
        =\frac{\mathscr W_\alpha(\lambda)}{\lambda},
\label{eq:app-partner-Wronskian}
\end{equation}
and, at a positive eigenvalue,
\begin{equation}
        \frac1{\partial_\lambda
        \widetilde{\mathscr W}_\alpha(\lambda_n)}
        =\frac{\lambda_n}
        {\partial_\lambda\mathscr W_\alpha(\lambda_n)}.
\label{eq:app-partner-residue}
\end{equation}

\paragraph{Transition kernel.}
The reduced propagator is
\begin{equation}
        h_\alpha(x,\tau\mid x_0)
        =\sum_{n=0}^\infty
        e^{-\lambda_n\tau}\phi_n(x)\phi_n(x_0).
\label{eq:heun-transition-kernel}
\end{equation}
It is symmetric and normalized with respect to the stationary measure,
\[
        h_\alpha(x,\tau\mid x_0)=h_\alpha(x_0,\tau\mid x),
        \qquad
        \int_0^\infty\pi_\alpha(x)
        h_\alpha(x,\tau\mid x_0)\,\dd x=1.
\]
Taking the closed-end limit gives
\begin{equation}
        h_\alpha(x,\tau\mid\infty)
        =\sum_{n=0}^\infty
        e^{-\lambda_n\tau}\phi_n(x)\phi_n(\infty).
\label{eq:heun-entrance-kernel}
\end{equation}
The $n=0$ contribution equals one.

For the boundary value
$H_\alpha(\tau)=h_\alpha(0,\tau\mid\infty)$, the same Green kernel gives the
exact Laplace transform
\begin{equation}
        \widehat H_\alpha(s)
        :=\int_0^\infty e^{-s\tau}H_\alpha(\tau)\,\dd\tau
        =G_\alpha(0,\infty;-s)
        =-\frac1{\mathscr W_\alpha(-s)},
        \qquad \Re s>0.
\label{eq:ultranarrow-starting-resolvent}
\end{equation}
 Eq.~\eqref{eq:ultranarrow-starting-resolvent} provides a
convenient exact normalization for the numerical spectral test.  Writing
$s=\sigma(1+\sigma)$, the same Legendre--Bessel coefficients used in
Section~\ref{sec:perfect-ultranarrow} give
\begin{equation}
        \widehat H_\alpha(s)
        =\frac1s e^{-\alpha/2}\mathcal A_*(\sigma)
        e^{-\sigma T_0}
        \left[1+O(e^{-c_0T_0})\right],
        \qquad
        \sigma=\sqrt{s+\frac14}-\frac12.
\label{eq:app-numerical-resolvent-asymptotic}
\end{equation}
This compact continuation is used only to interpret the numerical endpoint
Wronskian in Fig.~\ref{fig:front}(c).

\section{Spectral representation of the finite-length Wigner-delay distribution}
\label{app:wigner-delay}

This appendix records the spectral step behind
Eq.~\eqref{eq:delay-exact-density-main}.  Starting from the finite-length
Fokker--Planck equation \eqref{eq:delay-fokker-planck-main}, put $z=\log u$.
The backward generator is
\[
        \cL_{\rm d}
        =\partial_z^2+(e^{-z}-1)\partial_z,
\]
with stationary weight $m_{\rm d}(z)=e^{-z-e^{-z}}$.  The ground-state
transformation gives the Morse Hamiltonian
\begin{equation}
        \mathcal H_-^{\rm d}
        =-\partial_z^2
        +\frac14-e^{-z}+\frac14e^{-2z},
\label{eq:app-delay-morse}
\end{equation}
whereas its supersymmetric partner is
\begin{equation}
        \mathcal H_+^{\rm d}
        =-\partial_z^2+\frac14+\frac14e^{-2z}.
\label{eq:app-delay-liouville}
\end{equation}
The linear exponential term has cancelled.  The partner continuum is therefore
expressed through $K_{\ii q}(e^{-z}/2)$, and intertwining back gives the
Whittaker functions $W_{1,\ii q}(1/u)$.  With the stationary mode restored,
this yields precisely Eq.~\eqref{eq:delay-exact-density-main}.

For $\tau\gg1$, the small-$q$ part gives
\begin{equation}
        P(u,\tau)
        =\frac1{u^2}e^{-1/u}
        +\frac{2\sqrt\pi}{u\tau^{3/2}}
        e^{-\tau/4-1/(2u)}
        W_{1,0}\!\left(\frac1u\right)
        \left[1+O(\tau^{-1})\right].
\label{eq:app-delay-long-finite-correction}
\end{equation}

To compare with the notation of Refs.~\cite{ComtetTexier1997,TexierComtet1999},
let $\lambda_{\rm CT}$ denote their high-energy localization length and
$Z=2k\tau_{\rm W}$.  In the white-noise convention used here,
\[
        \gamma=\frac{1}{2\ellL},
        \qquad
        \lambda_{\rm CT}=\gamma^{-1}=2\ellL,
\]
so that
\begin{equation}
        u=\frac{Z}{\lambda_{\rm CT}},
        \qquad
        \tau=\frac{L}{\ellL}=\frac{2L}{\lambda_{\rm CT}}.
\label{eq:app-delay-CT-map}
\end{equation}
Equation~\eqref{eq:delay-exact-density-main} is then the finite-$L$ Whittaker
representation of Comtet and Texier.  Likewise,
Eq.~\eqref{eq:delay-exponential-functional-main} becomes
\[
        \tau_{\rm W}
        \overset{\rm d}{=}
        \frac1{k\gamma}
        \int_0^{\gamma L}e^{-2s+2W_s}\,\dd s.
\]
Their finite-size log-normal tail takes the form, up to algebraic prefactors,
\begin{equation}
        P(u,\tau)
        \sim
        \exp\!\left[-\frac{(\log u)^2}{4\tau}\right],
        \qquad
        e^{\sqrt{\tau/2}}\ll u\ll e^{\tau/2}.
\label{eq:app-delay-lognormal-window}
\end{equation}

\end{document}